\documentclass{jfm}
\usepackage{lineno,hyperref}
\usepackage{graphicx}
\usepackage{mathtools}
\usepackage{epstopdf, epsfig}
\usepackage{tikz}
\usepackage{booktabs}
\usepackage{amsmath}
\usepackage{multirow}
\usepackage{comment}
\usepackage{natbib}
\usepackage[]{color}
\usepackage[]{xcolor}
\usepackage{todonotes}
\usepackage{placeins}
\usepackage{relsize}
\usepackage{soul}
\usepackage{rotating}
\usepackage{array}

\usepackage{microtype}

\newcolumntype{x}[1]{>{\centering\arraybackslash\hspace{0pt}}p{#1}}

\allowdisplaybreaks[1]

\shorttitle{Roughness for transition delay in high-speed boundary layers}
\shortauthor{R. Jahanbakhshi and T. A. Zaki}

\title{Optimal two-dimensional roughness for transition delay in high-speed boundary layer}

\author{Reza Jahanbakhshi\aff{1,2} 
 \and Tamer A. Zaki\aff{1}   \corresp{\email{t.zaki@jhu.edu}}}

\affiliation{\aff{1}Department of Mechanical Engineering, Johns Hopkins University,\\ Baltimore, MD 21218, USA
\aff{2}Department of Aerospace, Physics \& Space Sciences, Florida Institute of Technology, \\ Melbourne, FL 32901, USA}
\begin{document}

\maketitle

\begin{abstract}
The influence of surface roughness on transition to turbulence in a Mach 4.5 boundary layer is studied using direct numerical simulations.  
Transition is initiated by the nonlinearly most dangerous inflow disturbance, which causes the earliest possible breakdown on a flat plate for the prescribed inflow energy and Mach number.
This disturbance is primarily comprised of two normal second-mode instability waves and an oblique first mode.
When localized roughness is introduced, its shape and location relative to the synchronization points of the inflow waves are confirmed to have a clear impact on the amplification of the second-mode instabilities.
The change in modal amplification coincides with the change in the height of the near-wall region where the instability wave-speed is supersonic relative to the mean flow;
the net effect of a protruding roughness is destabilizing when placed upstream of the synchronization point and stabilizing when placed downstream.
Assessment of the effect of the roughness location is followed by an optimization of the roughness height, abruptness and width with the objective of achieving maximum transition delay. 
The optimization is performed using an ensemble-variational (EnVar) approach, while the location of the roughness is fixed upstream of the synchronization points of the two second-mode waves.
The optimal roughness disrupts the phase of the near-wall pressure waves, suppresses the amplification of the primary instability waves, and mitigates the nonlinear interactions that lead to breakdown to turbulence. The outcome is a sustained non-turbulent flow throughout the computational domain.
\end{abstract}

\section{Introduction}
\label{sec:Introduction}

When high-speed transitional boundary layers encounter surface roughness, the resulting interaction is difficult to anticipate.  
Depending on the location and shape of the roughness, and the state of the boundary layer, transition to turbulence may be promoted or delayed. 
Even in the latter case, a longer sustained laminar flow may be a result of taming a particular route to turbulence, but other instability waves may become more amplified.  
Nonetheless, the potential thermal and mechanical benefits of a sustained laminar state in flight are such that roughness has previously been explored as an effective strategy to delay transition in high-speed boundary layers \citep{Fujii2006,Marxen2010,Riley2014,Fong2015,Zhao2019}.
In the present effort, we examine the influence of roughness location and shape on laminar-to-turbulence transition in a Mach 4.5 zero-pressure-gradient boundary layer using direct numerical simulations, and explain the observed shifts in transition location.
We subsequently perform an optimization of the roughness parameters in order to achieve the longest possible delay of breakdown to turbulence within our simulation domain.

\subsection{Stability of high-speed boundary layers}

The work by \citet{Lees1946investigation} extended the Rayleigh inflection-point criterion to compressible flows.  At the generalized inflection point, $ \partial_y ( \bar{\rho} \; \partial_y \bar{u} ) $ vanishes, where $\bar{\rho}$ and $\bar{u}$ are the base-state density and streamwise velocity and $\partial_y \equiv \partial / \partial y$ is the derivative in the wall-normal direction.  The key implication is that a compressible, zero-pressure-gradient boundary layer can be inviscidly unstable, unlike the incompressible counterpart which has a viscous Tollmien-Schlichting instability.   
\citet{Mack1969,Mack1984} performed extensive linear stability computations of the compressible boundary layer, and identified an infinite sequence of inflectional instabilities at high Mach number, the first two of which are now known as the first and second Mack modes.
This spectrum of discrete, unstable modes was also identified by \citet{smith1990inviscid} and \citet{cowley1990instability} using asymptotic analysis in the limit of infinite Mach number. The higher order modes are reported to be unstable over relatively small ranges of high frequencies.

Mack's first-mode reaches its maximum energy near the generalized inflection point of the boundary-layer profile, or the local maximum of $\bar{\rho} \; \partial_y \bar{u}$.
Mack's higher-modes, which only exist at high Mach number, are rooted in acoustic waves that are trapped inside the boundary layer near the wall in a region where the phase-speed of the wave $c$ is locally supersonic relative to the mean flow $\bar{u}$, or $c - \bar{u} \geq a$ where $a$ is the local speed of sound.
While most of these modes are inviscid, the analysis by \citet{smith1989first} showed that the first modes are of a viscous-inviscid kind; Directed outside of the local wave-Mach-cone direction, i.e.\,mode angles higher than $\tan^{-1}(\sqrt{M^2_{\infty} - 1})$ where $M_\infty$ is the free-stream Mach number, these modes exhibit a triple-deck structure.

Mack's modes are potential precursors of transition in compressible boundary layers, and both the first- and second-mode waves have been observed in various experiments, see e.g.\, \citep{Kendall1975,Lysenko1984,Stetson1992,Laurence2016,Casper2016,Kegerise2016,Zhu2018,Liu2019}.
At high-subsonic and moderate-supersonic speeds, boundary-layer transition in low-disturbance environments occurs as a result of excitation and amplification of instabilities that resemble Mack's first-mode waves.
As Mach number increases, the generalized inflection point moves to the outer region of the boundary layer and the growth rate of first-mode instabilities becomes smaller than the second-mode instabilities.  The latter dominate transition in high-Mach-number, supersonic and hypersonic boundary layers.
According to inviscid linear stability \citep{Mack1984}, the growth rate of the second-mode exceeds that of the first-mode, for an adiabatic flat-plate boundary layer, at $M_\infty \approx 4$, where $M_\infty$ is the free-stream Mach number.
For cooled boundary layers, the second-mode could become the dominant instability at even lower Mach numbers.

More recent work has pointed out that extending the terminology first- and second-model to viscous flows may not be pertinent \citep{Tumin2007,Fedorov2011b,Fedorov2011}.
Instead, the discrete modes were distinguished as follows: The slow mode, or mode S, has a phase speed that approaches the slow acoustic wave $c/u_\infty = 1 - 1 /M_{\infty}$ in the limit $\alpha \delta \ll 1$, where $\alpha$ is the streamwise wavenumber of the wave and $\delta$ is the local boundary layer thickness.  This limit is achieved, for exmaple, near the leading edge when $\alpha$ is finite and $\delta \to 0$.  
In the same limit the fast mode,  or mode F, has a phase speed that approaches the fast acoustic wave $c/u_\infty = 1 + 1 /M_{\infty}$.
\citet{Fedorov2011b} argued that, in terms of the spatial stability of an adiabatic wall at a finite Reynolds number, there only exists one unstable discrete mode, Mode S, which exhibits features of the inviscid first-mode or second-mode instabilities depending on frequency and Reynolds number.  
Mode F can also become unstable, for example in a cold-wall boundary layer.
As the Reynolds number increases downstream of the leading edge, the phase speed of mode S increases and that of mode F decreases until they synchronize.
The synchronization point is an important modulator for many stability features of high-speed boundary layers \citep{Fedorov2011b,Fedorov2011,Fong2015,Zhao2018,Park2019,Dong2021,Jahanbakhshi2021}.
For example, \citet{Park2019} examined the sensitivity of the linear stability of high-speed boundary layers to the distortions to the base velocity and temperature profiles;
they showed that the sensitivities of modes S and F to the distortions increase with Reynolds number, but near the synchronization point there is a sudden drop and jump, respectively, in the sensitivities.
Despite the arguments by Fedorov and Tumin, the terminology ``first-mode" and ``second-mode" remain widely adopted by the high-speed boundary-layer research community, and is therefore retained herein.

\subsection{Effects of roughness on transition to turbulence}

Introducing isolated or distributed roughness elements can promote breakdown to turbulence, relative to a smooth surface, by increasing the amplification rate of existing instabilities or spurring new one. Examples include the generation of wakes and unstable shear layers downstream of tall roughness elements \citep{Ergin2006}, and formation of streamwise vorticity behind shorter roughness elements that can initiate stationary crossflow instabilities in three-dimensional boundary layers \citep{Radeztsky1999}.
Our interest is, however, in carefully designed roughness elements that can delay transition.

Depending on the shape and location, roughness contributes to (i) receptivity \citep{Liu2020,Dong2020} and/or (ii) local scattering of perturbations, by roughness-induce mean-flow distortion, non-homogeneous forcing and non-parallel effects \citep{Xu2016,Wu2016,Dong2021}. When the roughness is small compared to the local boundary layer height, these effects can be studied within the framework of triple-deck theory \citep{Stewartson1969flow,Smith1973laminar,Wu2016,Dong2021}.
Using a large-Reynolds-number asymptotic analysis, \citet{Dong2020} showed that the distortion of a small free-stream acoustic wave by the curved wall of an isolated surface elements of height $h \ll \delta$ contributes to receptivity, and the amplitude of the resulting eigenmode scales with $\mathcal{O} ( h / \delta )$;  In addition, the interactions between the roughness-induced mean-flow distortion and the acoustic wave leads to receptivity that scales as $\mathcal{O} ( [h/\delta] Re_\delta^{-1/3} )$, where $Re_\delta$ is the Reynolds number based on $\delta$.  In another study, \citet{Liu2020} performed DNS and asymptotic analysis at moderate and large Reynolds numbers, respectively, to show that the amplitude of the excited viscous Mack first mode for the strong receptivity regime scales as $\mathcal{O} ( [h/\delta] Re_x^{1/4} )$, where $Re_x$ is the Reynolds number based on the distance from the leading edge.

Two notable early experimental studies that reported transition delay on a roughed-wall are the works by \citet{James1959} and \citet{Holloway1964}. 
The former examined 2D-roughness elements in free flight-tests with $2.8 < \textrm{Ma}_{\infty} < 7$, and the latter tested spherical roughness elements mounted on a flat-plate in a Mach 6 wind-tunnel.
More recent experiments in which roughness-induced delay of transition was observed were conducted by \citet{Fujii2006}, \citet{Bountin2013} and \citet{Fong2015}.
The measurements by \citet{Fujii2006} were on a Mach-7 5-degree half-angle sharp cone on which either a wavy two-dimensional or spherical roughness elements were mounted.
He found that, at high stagnation temperature and pressure conditions, transition was delayed when the wavelength of wavy-wall roughness is similar to the unstable second-mode (approximately $2 \delta$); spherical roughness elements had little effect on the location of transition to turbulence.
At low stagnation temperature and pressure conditions, on the other hand, both the wavy wall and spherical roughness elements promoted transition to turbulence relative to a smooth wall.
\citet{Bountin2013} examined the effects of a wavy wall on stability of a Mach-6 boundary-layer. They observed flow over the shallow-grooved plate was stabilized in a high-frequency band and destabilized at low frequencies, emphasizing that roughness must be carefully selected depending on the flow regime taking into account potential environmental disturbance spectra that may force the boundary layer.
\citet{Fong2015} studied a Mach-6 flow over a flared cone with initial half-angle of 2 degrees with six equi-spaced, two-dimensional elliptical roughness elements.
They reported that second-mode instabilities can be damped if the roughness is placed downstream of the synchronization point of the fast and slow second modes.
However, it is noteworthy that in this experiment the roughness had an unanticipated effect of promoting the amplification of the first-mode instability.

Computational and theoretical studies have also examined the influence of roughness on high-speed boundary layer stability \citep{Duan2010, Marxen2010, Riley2014, Groskopf2016, Zhao2019,Dong2021,haley2023roughness}.
We first consider the impact of two-dimensional modifications of the surface.  
Using direct numerical simulations (DNS), \citet{Duan2010} studied a Mach 5.92 boundary layer on a flat plate with an isolated two-dimensional elliptic bump, and drew similar conclusions as \citet{Fong2015}.
The DNS by \citet{Marxen2010} examined the impact of two-dimensional hyperbolic-shaped isolated roughness on a flat-plate boundary layer at free-stream Mach 4.8.
Depending on modal frequency, the roughness element can either amplify or damp the disturbance waves.
\citet{Riley2014} investigated the effect of two-dimensional compliant panels (convex or concave panel buckling) on boundary-layer stability for Mach 4 flow over a wedge.
They used linear stability theory and the parabolized stability equations, and showed that placing panels near the leading edge of the wedge promotes naturally occurring high-frequency disturbances. However, placement near the trailing edge enhanced stability.
Most recently, \citet{Zhao2019} used the harmonic linearized Navier-Stokes equations to evaluate the effect of a two-dimensional hump or indentation on a flat wall on boundary-layer stability. They confirmed that the synchronization points of the instability waves are critical modulators of the impact of roughness.
The earlier referenced study by \citet{Dong2021}, which developed large-Reynolds-number asymptotic theory for the impact of localised roughness on first and second Mack modes, attributed the dominant roughness effect to the interaction of the oncoming perturbation with the mean-flow distortion in the main layer and the inhomogeneous forcing from the curved wall.
Most recently, \citet{haley2023roughness} investigated how a single roughness strip and an array of six sequential strips influence stability of second modes on a Mach-8 straight blunt cone. For both cases, they observed stabilization of high-frequency and destabilization of low-frequency modes.
As for three-dimensional roughness elements, \citet{Groskopf2016} studied their impact on a Mach-4.8 boundary layer over a flat plate.  The roughness parameters included the spanwise width to streamwise length ratio, height, and skewing angle with respect to the oncoming flow.
The authors observed that, in the wake of obliquely placed elements, strong low-speed streaks are generated due to the induced cross flow. Oblique placement was also associated with larger amplification of instabilities.

\subsection{Objectives}

The above discussion highlights both the wealth of discoveries from experimental, theoretical and numerical studies of roughness in transitional high-speed flows, and also the challenge in anticipating the impact of roughness on transition.
The outcome depends on the roughness parameters, including shape and location, and on the details of the flow configuration. A choice that does not guarantee robust transition delay can, therefore, lead to an undesirable, potentially catastrophic outcome especially given the uncertainty in the environmental conditions relevant to high-speed flights.
In the present work, we examine the key roughness parameters that impact transition to turbulence in a flat-plate boundary layer at Mach 4.5.  
We then optimize these parameters to achieve maximum transition delay within our computational domain.
The inflow condition in our simulations is the nonlinearly most dangerous disturbance at the prescribed level of inlet energy, which was previously computed for the same configuration \citep{Jahanbakhshi2019}.
The delay of the associated transition mechanism using surface roughness is analysed in detail.

This paper is organized as follows: Computational aspects, inflow condition, and the roughness geometry are introduced in \S\ref{sec:DNS}, while validation of the computational model is provided in Appendix \ref{Appendix_A}.
The results and discussions are reported in \S\ref{sec:Results}, and are followed by a summary in \S\ref{sec:Summary}.

\section{Computational framework}
\label{sec:DNS}

\begin{figure}
\centerline{%
\includegraphics[trim=0 0 0 0, clip,width=0.999\textwidth] {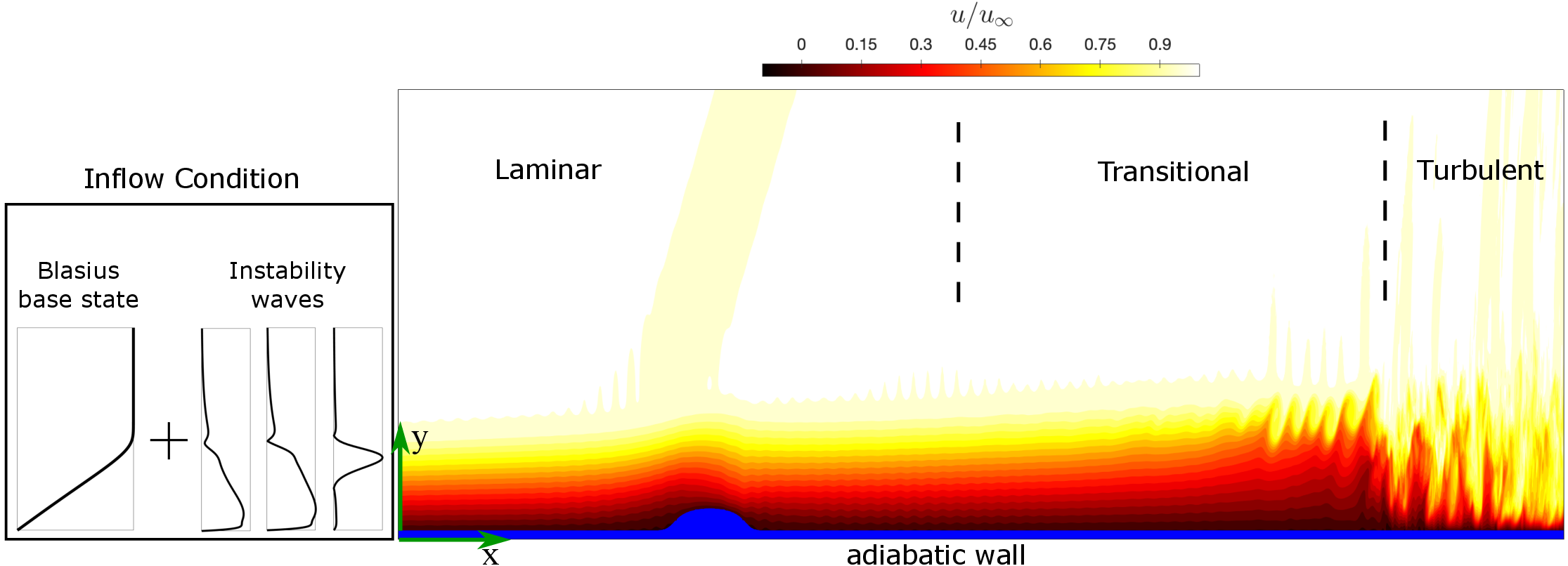}%
}%
\caption{Schematic of transitional boundary-layer flow over a flat plate with an isolated, protruding roughness element. Contours show the streamwise velocity $u$  normalized by its free-stream value, $u_\infty$.}
\label{FIG:Schematic}
\end{figure}

Direct numerical simulations are performed to study the influence of surface roughness on transitional high-speed boundary layers, and specifically the capacity to delay breakdown to turbulence.
The flow satisfies the compressible Navier-Stokes equations, and a sample flow configuration is provided in figure \ref{FIG:Schematic}.  The contours show the streamwise velocity of a boundary layer over a protruding roughness.  Throughout this work, flow variables are non-dimensionalized using the free-stream velocity $\tilde{u}_\infty$, density $\tilde{\rho}_\infty$, temperature $\tilde{T}_\infty$, and viscosity $\tilde{\mu}_\infty$, where $\tilde{\bullet}$ represents dimensional quantities.  Lengths are normalized using the Blasius scale at the inflow, $\tilde{l} = \sqrt{\tilde{\mu}_\infty \tilde{x}_0 / \tilde{\rho}_\infty \tilde{u}_\infty}$, where $\tilde{x}_0$ is the distance of the inflow plane from the virtual boundary-layer origin.  In terms of these reference scales, we define the Reynolds number $Re_l \equiv \tilde{\rho}_\infty \tilde{u}_\infty \tilde{l} / \tilde{\mu}_\infty$, which is equivalent to $\sqrt{ Re_{x_0} } = \sqrt{ \tilde{\rho}_\infty \tilde{u}_\infty \tilde{x}_0 / \tilde{\mu}_\infty } $.  Both values are equal to the non-dimensional location of the inflow plane, $x_0 = \tilde{x}_0 / \tilde{l} = \sqrt{ Re_{x_0} } = Re_l $, while the Reynolds number based on the streamwise distance is given by $Re_x = Re_l~x$.

In non-dimensional form, the compressible Navier-Stokes equations are,  
\begin{equation}
		\frac{\partial \boldsymbol{q} }{\partial t} + \nabla \cdot \left( \boldsymbol{f}_{i} - \boldsymbol{f}_{v} \right) = 0, 
	\label{Eq:Navier_Stokes}
\end{equation}
where $\boldsymbol{q} = \begin{bmatrix} \rho & \rho \boldsymbol{u} & E \end{bmatrix}^{\top}$ is the state vector, $\boldsymbol{f}_{i} = \begin{bmatrix} \rho \boldsymbol{u} & \rho \boldsymbol{u} \boldsymbol{u} + p \boldsymbol{I} & \boldsymbol{u} ( E + p ) \end{bmatrix}^{\top}$ represents inviscid fluxes, $\boldsymbol{f}_{v} = \begin{bmatrix} 0 & \boldsymbol{\tau} & ( \boldsymbol{u} \cdot \boldsymbol{\tau} - \boldsymbol{\theta} ) \end{bmatrix}^{\top}$ are the viscous fluxes, 
and $\left(\bullet\right)^{\top}$ denotes the transpose.  
In the equations, $\rho$ is the density, $\boldsymbol{u} = \begin{bmatrix} u & v & w \end{bmatrix}^{\top}$ is the three-dimensional velocity vector, $p$ is the pressure, $E = \rho e + 0.5 \rho \boldsymbol{u} \cdot \boldsymbol{u}$ is the total energy, $e$ is the specific internal energy, $\boldsymbol{\tau}$ is the viscous stress tensor, $\boldsymbol{\theta}$ is the heat-flux vector, and $\boldsymbol{I}$ is the unit tensor.
The system of equations is closed by assuming a calorically perfect gas whose thermodynamic properties are related by,
\begin{equation}
		p = \left( \gamma - 1 \right) \rho e \;\;\;\; \textrm{and} \;\;\;\; T = \gamma ( \gamma - 1 ) M_\infty^2 e,
\label{Eq:State_Equation}
\end{equation}
where $T$ is the temperature, $\gamma$ is the ratio of specific heats, $M_\infty = \tilde{u}_\infty \left[ (\gamma-1) \tilde{c}_p \tilde{T}_\infty \right]^{-1/2}$ is the free-stream Mach number, and $\tilde{c}_p$ is the specific heat at constant pressure.
The viscous stress and heat flux are modelled as,
\begin{equation}
	\boldsymbol{\tau} = \frac{\mu}{Re_l} \left[ \nabla \boldsymbol{u} + ( \nabla \boldsymbol{u})^{\top} + \left( \frac{\mu_b}{\mu} - \frac{2}{3} \right) ( \nabla \cdot \boldsymbol{u} ) \boldsymbol{I} \right]  \;\;\;\; \textrm{and} \;\;\;\; \boldsymbol{\theta} = - \frac{ \mu \nabla T }{(\gamma-1) M_\infty^2 Re_l Pr},
\end{equation}
respectively, where $\mu$ is the dynamic shear viscosity, $\mu_b$ is the bulk viscosity, and $Pr$ is the Prandtl number.

The direct numerical simulations adopt a finite-difference discretization of the governing Navier-Stokes equations on a structured Cartesian grid.
A sixth-order-accurate central difference scheme in the split form by \citet{Ducros2000} is adopted for the inviscid fluxes, and is replaced by a fifth-order-accurate WENO scheme with Roe flux splitting near shocks.
The viscous fluxes are computed using a conservative discretization that has the resolution characteristics of a sixth-order scheme.
Time is advanced by a fourth-order accurate Runge-Kutta method.
In order to simulate complex geometries, a cut-stencil method \citep{Greene2016} is implemented that changes the discretization of the governing equations near the body and applies the boundary conditions just at the interface between the solid and the fluid.
The method generates precise locations for the body.
Validation of the original algorithm for Cartesian geometries is available in the literature \citep[see e.g.][]{Larsson2009,Johnsen2010,Kawai2012,Volpiani2018}, and the newly implemented cut-stencil method is discussed in more details in Appendix~\ref{Appendix_A}.

The operating gas is air for which Prandtl number is $Pr = 0.72 $ and the ratio of specific heats is $\gamma = 1.4$. The Mach number in the free-stream is $ M_{\infty} = 4.5$. Sutherland's law \citep{Sutherland1893} models the temperature dependence of the dynamic viscosity, and Stokes' hypothesis relates the dynamic and bulk viscosity coefficients.
The streamwise position of the inflow plane is $x_0 = \sqrt{Re_{x_0}} = 1800$, which was selected based on the transition Reynolds numbers in high-altitude flight tests being $\sqrt{Re_{x_{tr}}} > 2000$ for $M_\infty > 4$ \citep{Harvey1978,Schneider1999}.
At the inlet plane, the Blasius base-state is prescribed along with a superposition of linear instability waves. The amplitudes and relative phases of all the modes were optimized in an earlier study \citep{Jahanbakhshi2019} such that they lead to the earliest possible breakdown to turbulence on an adiabatic flat plate (see also \S\ref{sec:InflowModes}). Periodic boundary conditions are enforced in the homogeneous spanwise direction, convective outflow is prescribed at the right and top boundaries, and the bottom boundary is a no-slip adiabatic wall.

\begin{figure}
\centerline{%
\includegraphics[trim=0 0 0 0, clip,width=0.99\textwidth] {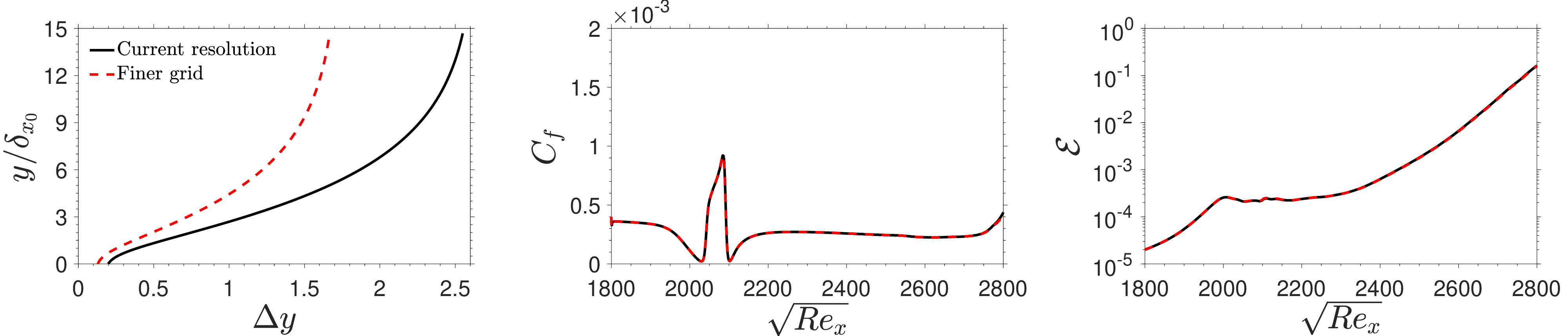}%
\put(-385.0,70.0){$(a)$}
\put(-252.0,70.0){$(b)$}
\put(-125.0,70.0){$(c)$}
}
\vspace{2mm}
\centerline{%
\includegraphics[trim=0 355 0 0, clip,width=0.99\textwidth] {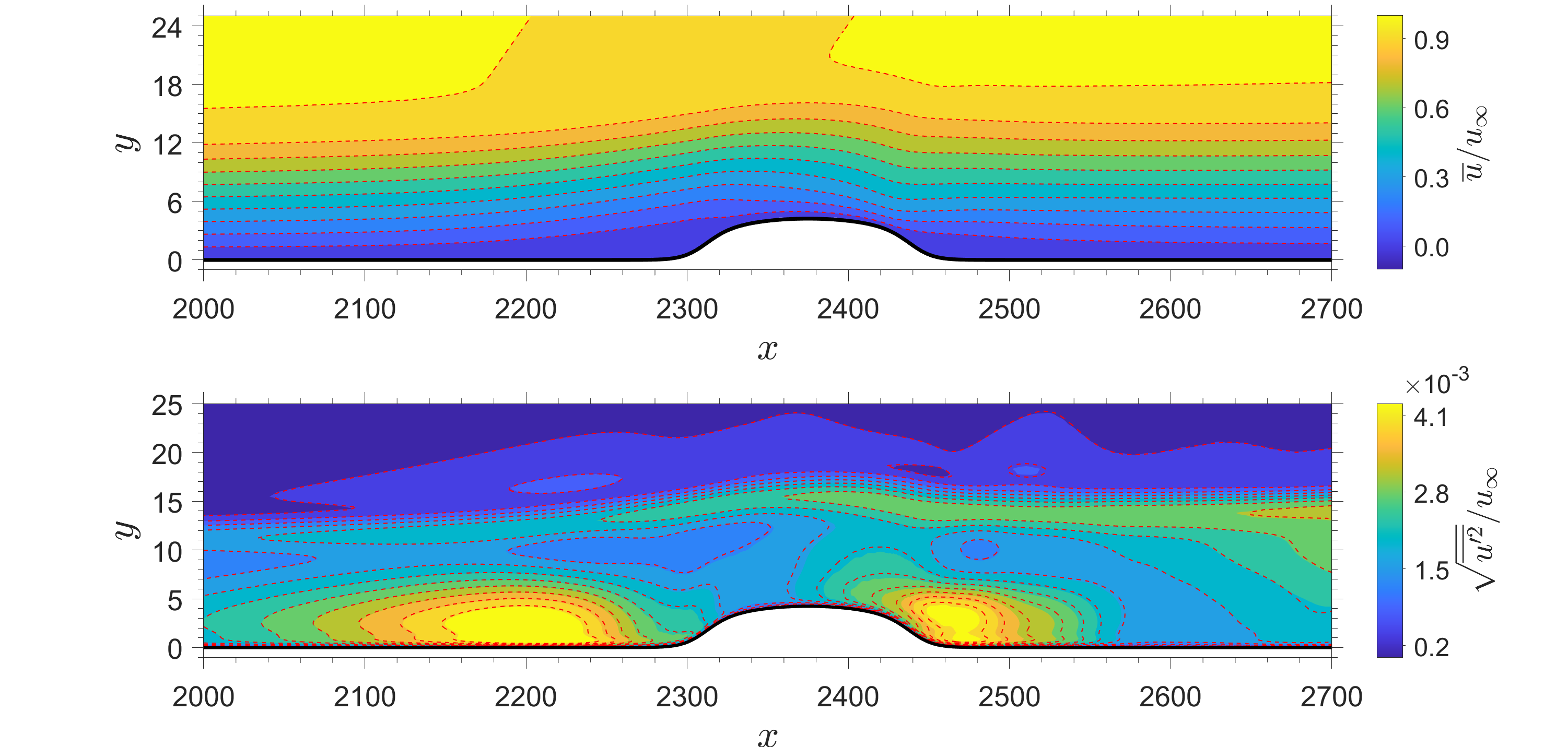}%
\put(-360.0,80.0){$(d)$}
}
\caption{$(a)$ Wall-normal grid-size and $(b-d)$ sample flow over the optimal roughness for delaying transition.  $(b)$ Downstream behaviour of the skin-friction coefficient. $(c)$ Downstream evolution of the disturbance energy. $(d)$ Contours of averaged streamwise velocity in the vicinity of the roughness; colours are from the herein adopted resolution $(N_x, N_y, N_z) = (2985, 189, 151)$, and dashed-lines are from the finer grid $(N_x, N_y, N_z) = (3978, 279, 201)$.}
\label{FIG:GridIndependence}
\end{figure}

The extents of the computational domain in the streamwise, wall-normal and spanwise directions are, respectively, $L_x = 2984$, $L_y = 204$ and $L_z = 150$.
The domain is discretized using a uniform grid in the horizontal directions with $N_x = 2985$ and $N_z = 151$ points.
A hyperbolic tangent stretching of the grid with $N_y = 189$ points is used to discretize the wall-normal direction. The wall-normal grid spacing that was adopted throughout this work is reported in figure \ref{FIG:GridIndependence}(a), as well as a finer grid that was used to verify grid independence. At the inlet plane, the main grid has fifty-four points within the boundary layer.
Figures \ref{FIG:GridIndependence}($b$,$c$) provide evidence of grid independence by comparing the main and finer grids. The figures report skin-friction coefficient, 
\begin{equation}
    C_f \equiv \frac{ \tau_{\textrm{wall}}}{ 0.5 \rho_{\infty} u_{\infty}^2 }, 
\end{equation}
and the disturbance energy, 
\begin{equation}
    \mathcal{E} \equiv \frac{1}{2} \int_{0}^{L_y} \left( \overline{\rho} \Bigl\{ \overline{\boldsymbol{u}^{\prime} \cdot \boldsymbol{u}^{\prime}} \Bigl\} + \frac{\overline{\rho} \overline{T}}{(\gamma-1) \gamma M_\infty^2} \Biggl\{ \frac{\overline{\rho^{\prime 2}}}{\overline{\rho}^2} + \frac{\overline{T^{\prime 2}}}{\overline{T}^2} \Biggl\} \right) dy,
\end{equation}
as a function of $\sqrt{Re_x}$.  Over-line denotes averaging in time and in the homogeneous spanwise direction, and the primed variables are the perturbations with respect to this average.
As demonstrated in figure \ref{FIG:GridIndependence}, a finer resolution does not yield any perceptible changes in the skin friction, the disturbance energy, or the mean streamwise-velocity contours (we also confirmed grid-independence of the spanwise and time-averaged variance).  Additionally, we verified that the instability modes prescribed at the inlet and their nonlinear interactions, leading to transition, are all fully resolved, and that our predictions of transition is grid independent.

\subsection{Inflow instability modes}
\label{sec:InflowModes}

The inflow disturbance is synthesized as a superposition of linear stability eigenmodes of the local boundary-layer profile, which span the relevant range of frequencies $\omega$ and spanwise wavenumbers $\beta_z$, 
\begin{equation}
    \boldsymbol{q}^{\prime}\left(x_0, y, z, t\right) = \sum_{\omega, \beta_z}
\Re \{ \breve{\boldsymbol{q}}_{\left< \omega , \beta_z \right>}(y) \exp( \alpha x_0 - i ( \beta_z z + \omega t) )\}.
\end{equation}
The eigenmodes at each $\langle \omega, \beta_z\rangle$ pair are obtained by substituting the ansatz $\breve{\boldsymbol{q}}(y) \exp( \alpha x - i ( \beta_z z + \omega t) )$ in the linear perturbation equations, and the resulting spatial eigenvalue problem is solved for the spectrum of eigenfunctions $\breve{\boldsymbol{q}}(y)$ and associated complex eigenvalues $\alpha = \alpha_r + i \alpha_i$.  Only the linearly most unstable wave, which in the present study is the slow mode, was retained at each $\left<\omega, \beta_z\right>$ pair. 
The linear-stability operators and solution algorithm are standard \citep[see e.g.][for details]{Park2019}.

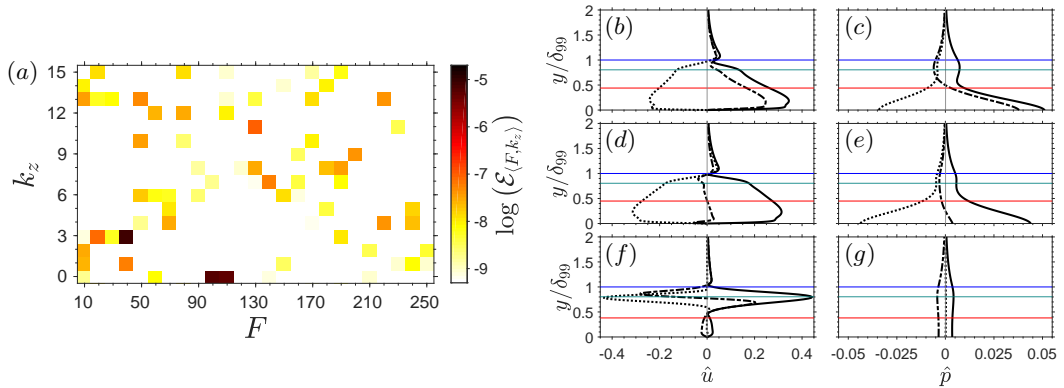
\begin{figure}
\centerline{%
\includegraphics[trim=0 -60 0 0, clip,width=0.5\textwidth] {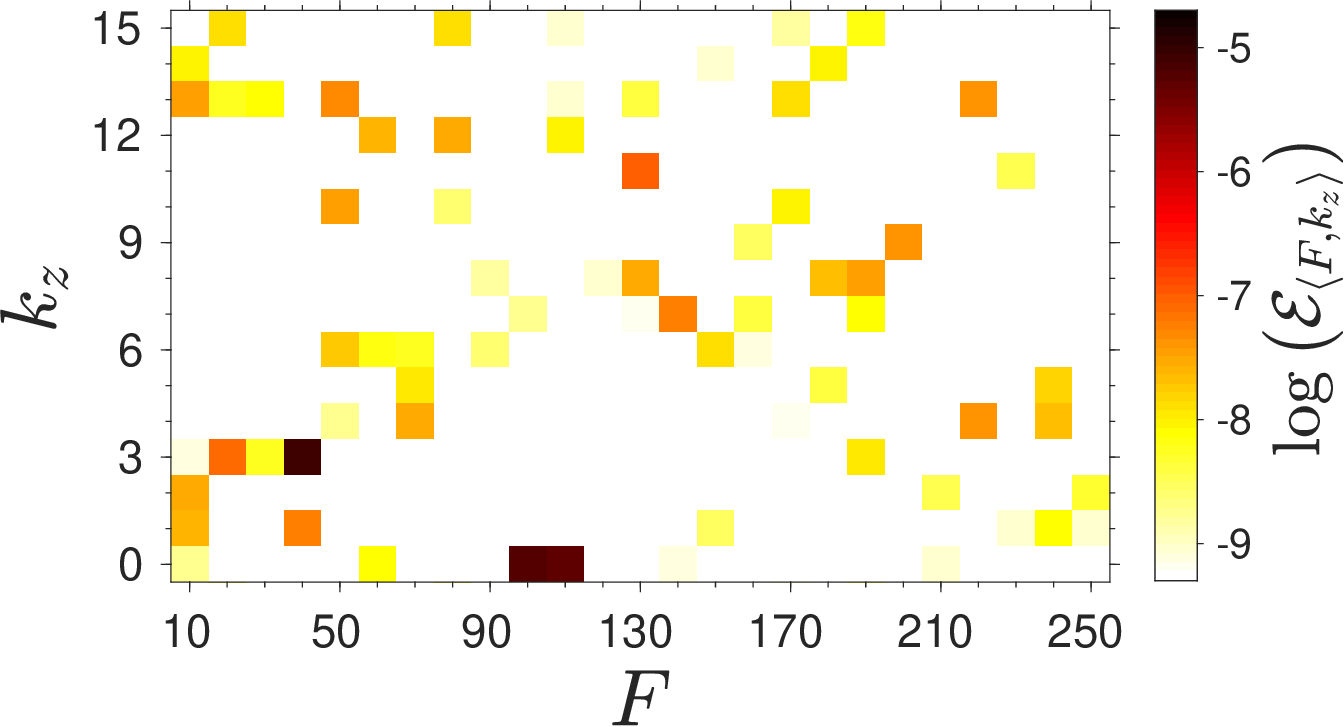}%
\hspace{2.00mm}
\begin{tikzpicture}
\node[anchor=south west,inner sep=0] at (0,0) {\includegraphics[trim=0 0 0 0, clip,width=0.5\textwidth] {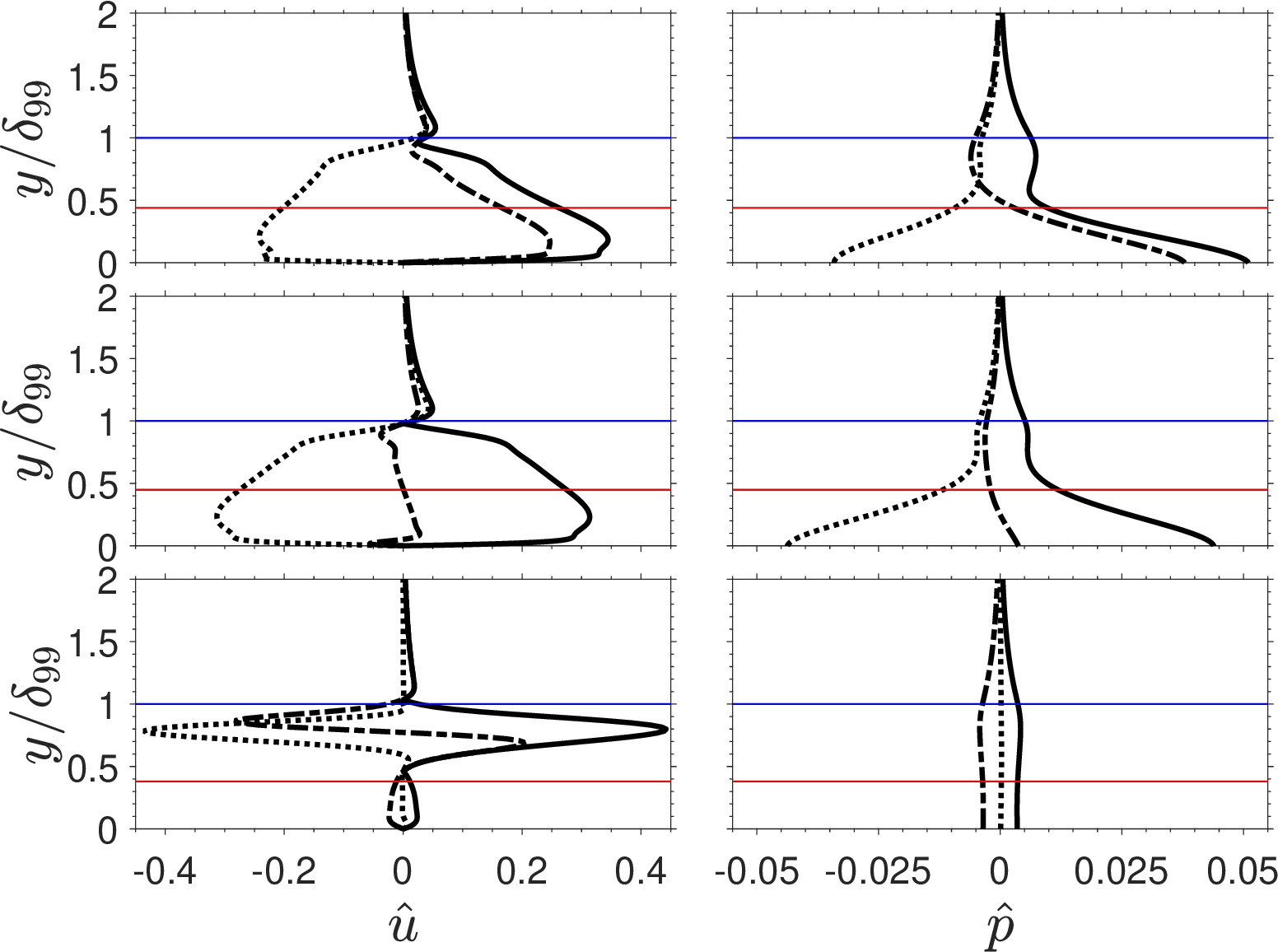}};
\draw[gray, thin, opacity=0.8] (2.135,0.65) -- (2.135,1.98);
\draw[gray, thin, opacity=0.8] (2.135,2.14) -- (2.135,3.48);
\draw[gray, thin, opacity=0.8] (2.135,3.64) -- (2.135,4.98);
\draw[gray, thin, opacity=0.8] (5.287,0.65) -- (5.287,1.98);
\draw[gray, thin, opacity=0.8] (5.287,2.14) -- (5.287,3.48);
\draw[gray, thin, opacity=0.8] (5.287,3.64) -- (5.287,4.98);
\draw[teal, thin, opacity=0.8] (0.72,1.186) -- (3.55,1.186);
\draw[teal, thin, opacity=0.8] (0.72,2.680) -- (3.55,2.680);
\draw[teal, thin, opacity=0.8] (0.72,4.174) -- (3.55,4.174);
\draw[teal, thin, opacity=0.8] (3.87,1.186) -- (6.70,1.186);
\draw[teal, thin, opacity=0.8] (3.87,2.680) -- (6.70,2.680);
\draw[teal, thin, opacity=0.8] (3.87,4.174) -- (6.70,4.174);
\end{tikzpicture}
\put(-395.0,115.0){$(a)$}
\put(-170.0,132.0){$(b)$}
\put(-80.0,132.0){$(c)$}
\put(-170.0,89.5){$(d)$}
\put(-80.0,89.5){$(e)$}
\put(-170.0,47.0){$(f)$}
\put(-80.0,47.0){$(g)$}
}%
\caption{The inflow instability waves. $(a)$ The spectral energy. $(b)-(g)$ profiles of the important modes: $(b,c)$ Mode $ \left< 110 , 0 \right>$, $(d,e)$ mode $ \left< 100 , 0 \right>$, and $(f,g)$ mode $ \left< 40 , 3 \right>$. (Thick) Magnitude; (dashed-dotted) real part; (dotted) imaginary part of the mode shapes. The Horizontal thin lines (top to bottom) mark the 99\% boundary-layer thickness, the generalized inflection point, and the relative sonic line $c - \bar{u} = a$.}
\label{FIG:InflowSpectra}
\end{figure}

For a prescribed level of the total disturbance energy, \citet{Jahanbakhshi2019} optimized the amplitudes and phases of the inflow instability waves in order to achieve the earliest possible transition to turbulence on a flat plate. The spectral makeup of the resulting nonlinearly most dangerous inflow disturbance is provided in figure \ref{FIG:InflowSpectra}$a$, as a function of the normalized frequency and integer spanwise wavenumber, 
\begin{subequations}
\begin{gather}
F \equiv \frac{\omega}{\sqrt{Re_{x_0}}} \times 10^6 \quad \textrm{ and }   \quad   k_{z} \equiv \frac{\beta_{z} L_z}{2 \pi}.            \tag{\theequation a,b}
\end{gather}
\label{Eq:Normalized_Freq_wavenum}
\end{subequations}
The modal energy, $\mathcal{E}_{\left< F , k_z \right>}$, is defined as,
\begin{equation}
\mathcal{E}_{\left< F , k_z \right>} = \frac{1}{2}
	\int_{0}^{L_y} \left( \overline{\rho} \bigl\{ \boldsymbol{\hat{u}}^* \boldsymbol{\hat{u}} \bigl\} + \frac{\overline{\rho} \overline{T}}{(\gamma-1) \gamma M_\infty^2} \Biggl\{ \frac{ \hat{\rho}^* \hat{\rho} }{\overline{\rho}^2} + \frac{ \hat{T}^* \hat{T} }{\overline{T}^2} \Biggl\} \right)_{\left< F , k_z \right>} dy ,
\label{Eq:FourierTransform_Energy_Mode_FKz}
\end{equation}
where hatted variables are the Fourier coefficients in frequency-spanwise wavenumber space, and star denotes the complex-conjugate transpose.
At the inlet plane, the total disturbance energy is $ \sum_{F,k_z} \mathcal{E}_{\left< F , k_z \right>} = 2 \times 10^{-5}$.

\begin{table}
\begin{center}
\def~{\hphantom{0}}
\begin{tabular}{ccccccc}
Mode & & $\omega$ & & $\beta_z$ & & $ \alpha $ \\ [3pt]
$\left<F,k_z\right>$ & & & & & & $\left( \alpha_r, \alpha_i \right)$ \\ [3pt]
\hline
\\
$\left<110,0\right>$ & & $0.198$ & & $0$ & & $\left( 0.0039, 0.2166 \right)$ \\
$\left<100,0\right>$ & & $0.180$ & & $0$ & & $\left( 0.0004, 0.1950 \right)$ \\
$\left<40,3\right>$ & & $0.072$ & & $0.1257$ & & $\left( 0.0011, 0.0840 \right)$ \\
\hline
\end{tabular}
\caption{Parameters of the inlet instability waves. Definitions are provided in \S\ref{sec:InflowModes}.}
\label{TABLE:Mode_Parameters}
\end{center}
\end{table}

Figure \ref{FIG:InflowSpectra}$a$ shows that the majority of the total energy, approximately 95\%, is assigned to only three inlet waves: $\left< 110 , 0 \right>$, $\left< 100 , 0 \right>$ and $\left< 40 , 3 \right>$.
The parameters for these three waves are provided in table \ref{TABLE:Mode_Parameters}; our simulations resolve each wavelength with 29 to 75 grid points in the streamwise direction, and 50 to 151 points in the span.
The streamwise velocity and pressure of the associated eigenfunctions are plotted in figures \ref{FIG:InflowSpectra}$b$ to \ref{FIG:InflowSpectra}$g$. The pressure profiles in figures \ref{FIG:InflowSpectra}$c$ and \ref{FIG:InflowSpectra}$e$, for modes $\left< 110 , 0 \right>$ and $\left< 100 , 0 \right>$, have a single zero crossing in the wall-normal direction, and are second-mode instabilities.
\citet{Jahanbakhshi2021} decomposed the momentum-density vector of these waves into acoustic, entropic and solenoidal components, and confirmed that both $\left< 110 , 0 \right>$ and $\left< 100 , 0 \right>$ have acoustic characteristics, in agreement with the interpretation by \cite{Mack1984}.
The associated pressure perturbations reflect back and forth between the wall and the sonic line of the relative flow.
Mode $\left< 40 , 3 \right>$ with no zero crossings is a first-mode, vortical instability.
Each of these three modes plays an important role in causing the earliest transition location, at the present total level of energy. For example, removing one of the modes and redistributing its energy to the other instability waves leads to a downstream shift in transition onset.

Figures \ref{FIG:InflowSpectra}$b,d,f$ show that the second-mode instabilities reach their maximum values below the relative sonic line close to the wall, while the first-mode instability is appreciable beyond the relative sonic line, close to boundary-layer edge.
The difference in the wall-normal dependence of the two classes of instabilities makes the second-mode waves more susceptible to control strategies that are applied at the wall, e.g.\,short roughness elements or wall heating/cooling. 

\begin{figure}
\centerline{%
\includegraphics[trim=0 0 0 0, clip,width=0.999\textwidth] {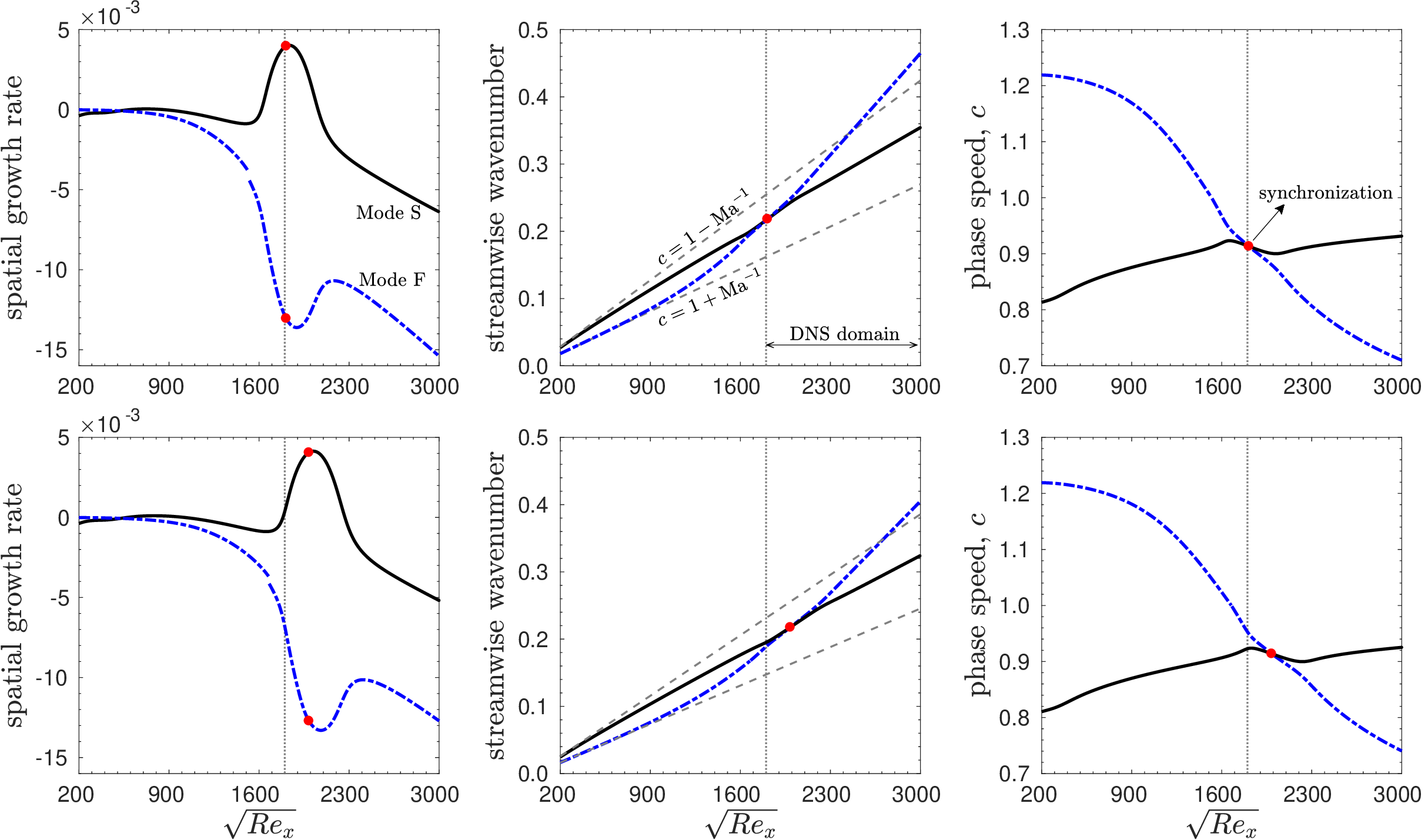}%
\put(-360.0,209.0){$(a)$}
\put(-230.0,209.0){$(c)$}
\put(-100.0,209.0){$(e)$}
\put(-360.0,99.0){$(b)$}
\put(-230.0,99.0){$(d)$}
\put(-100.0,99.0){$(f)$}
}%
\caption{$(a-b)$ Spatial growth rate, $\alpha_r$, $(c-d)$, streamwise wavenumber, $\alpha_i$, and $(e-f)$ phase speed, $\omega / \alpha_i$, of Modes S (solid line) and F (dashed-dotted lines) corresponding to the instabilities $\left< 110 , 0 \right>$, $(a-c-e)$, and $\left< 100 , 0 \right>$, $(b-d-f)$. The red dots indicate the synchronization point, the dotted lines mark the start of the computational domain in our DNS, and the dashed lines represent the phase speed of acoustic waves. Reproduction of figure 5 by \citet{Jahanbakhshi2021}.}
\label{FIG:SynchronizationInfo}
\end{figure}
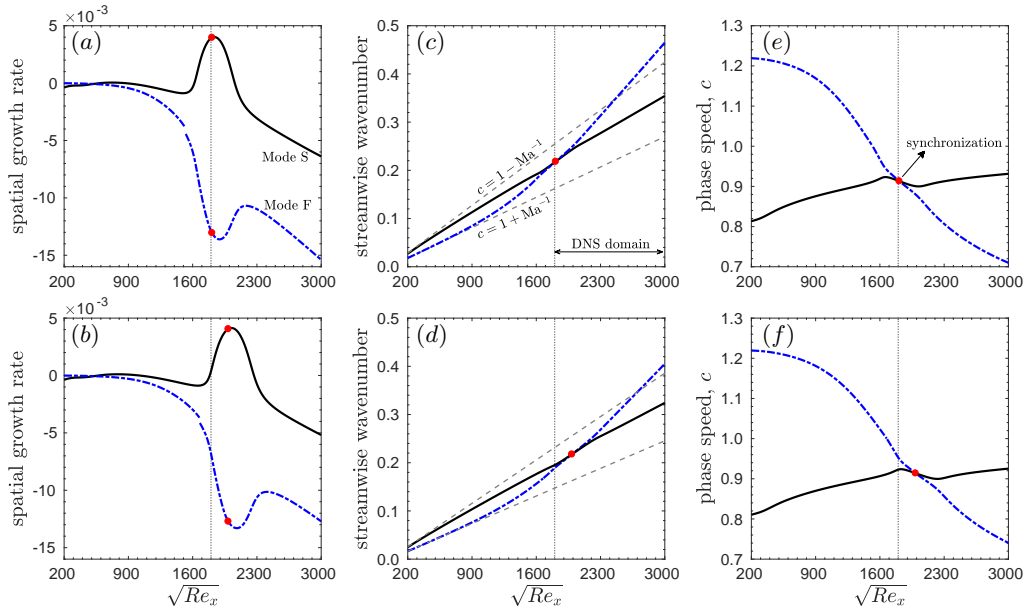

The synchronization of the fast and slow modes has a significant impact on the instability waves. Figure \ref{FIG:SynchronizationInfo} reports the spatial growth rates, streamwise wavenumbers, and phase speeds of modes F and S at $\left< F , k_z \right> = \left< 110 , 0 \right>$ and $\left< 100 , 0 \right>$.
These results are obtain using parallel, spatial, linear-stability theory and confirm typical characteristics of slow and fast modes for a high-Mach-number flow over an adiabatic flat-plate \citep{Fedorov2011b}. The growth rates of modes F, in panels $(a)$ and $(b)$, show a discontinuity at the location where the phase speed approaches unity, which corresponds to crossing the continuous branch of entropy and vorticity modes. Panels $(c)$ and $(d)$ show that at low Reynolds number modes S and F follow the slow and fast acoustic branches, $c = 1 \mp 1/M_{\infty}$,  of the spectrum.  Panels $(e)$ and $(f)$ show the evolution of the phase speeds, which approach one another with Reynolds number until they synchronize.
The Reynolds numbers at synchronization are $\sqrt{Re_x} = 1811$ for mode $\left< F , k_z \right> = \left< 110 , 0 \right>$ and $\sqrt{Re_x} = 1988$ for mode $\left< 100 , 0 \right>$.

\subsection{Geometry of roughness elements}
\label{sec:RoughnessGeometry}

The wall geometry $\hat{h}_r(x)$ is defined over three streamwise segments: $x < X_0$, $X_0 \leq x \leq X_N$ and $ x > X_N $. In the first and last segments, the wall is flat, and therefore $\hat{h}_r = 0 $.
In the range $X_0 \leq x \leq X_N$, a two-dimensional roughness element is defined, 
\begin{equation}
    \hat{h}_r = \frac{H_r}{2} \left[ \tanh \left( L_r \left[ x - X_r + W_r \right] \right) - \tanh \left( L_r \left[ x - X_r - W_r \right] \right) \right] \sin \left( n\pi \frac{ x - X_0}{X_N - X_0} \right),
\label{Eq:RoughnessGeometry}
\end{equation} 
where $H_r \geq 0$, $W_r \geq 0$, $L_r \geq 0$, $X_r = \frac{X_0 + X_N}{2}$, and $n \in \mathbb{Z} $ are the geometrical parameters that, respectively, determine the maximum height, streamwise extent, abruptness, center (location of maximum height), and streamwise integer wavenumber of the roughness.
Note that the surface function is not $C^2$ continuous at $x = X_0$ and $X_N$, which can trigger numerical instability.
In order to obtain a surface function that is $C^2$ at all $x$, we reconstruct the above surface topography using a fourth-order spline in which the first and final knots are at $x = X_0$ and $x = X_N$.
This spline reconstruction is given by $\boldsymbol{h}_{r} = \boldsymbol{B}_{4} [ ( \boldsymbol{B}^{\top}_4 \boldsymbol{B}_4 )^{-1} \boldsymbol{B}^{\top}_4 \hat{\boldsymbol{h}}_{r} ] $, where $\boldsymbol{B}_4 \in \mathbb{R}^{N_x \times N_c}$ is a matrix whose columns are B-splines of order 4, and $N_c$ is the number of control points.
In a discretized domain, $\boldsymbol{h}_{r} \equiv h_r (x) $ and $\hat{\boldsymbol{h}}_{r}  \equiv \hat{h}_r (x) $ are vector quantities in $\mathbb{R}^{N_x \times 1}$ space.
The size of the control points, $N_c$, is set such that the residuals function $(\boldsymbol{h}_{r} - \hat{\boldsymbol{h}}_{r})^{\top}(\boldsymbol{h}_{r} - \hat{\boldsymbol{h}}_{r}) < 10^{-5}$.

\begin{table}
\begin{center}
\def~{\hphantom{0}}
\begin{tabular}{c|cccc|cccc}
Case & X1p1 & X1p3 & X1m1 & X1m3 & X2p1 & X2p3 & X2m1 & X2m3 \\ [3pt]
\hline
$n$ & $+1$ & $+3$ & $-1$ & $-3$ & $+1$ & $+3$ & $-1$ & $-3$ \\
$(X_0, X_N)$ & \multicolumn{4}{c|}{$(2000, 2250)$} & \multicolumn{4}{c}{$(2250, 2500)$} \\
$(H_r, W_r, L_r)$ & \multicolumn{4}{c|}{$(3.0,72.5,0.06)$} & \multicolumn{4}{c}{$(3.0,72.5,0.06)$} \\
\hline
\end{tabular}
\caption{Geometrical parameters for examining effects of location and shape of the roughness on transition to turbulence. The boundary-layer thickness in the reference, flat-plate case is $\delta_{99} = 14.7$ at X1 and $\delta_{99} = 15.6$ at X2.}
\label{TABLE:RoughnessCases}
\end{center}
\end{table}

\begin{figure}
\centerline{%
\includegraphics[trim=0 0 0 0, clip,width=0.999\textwidth] {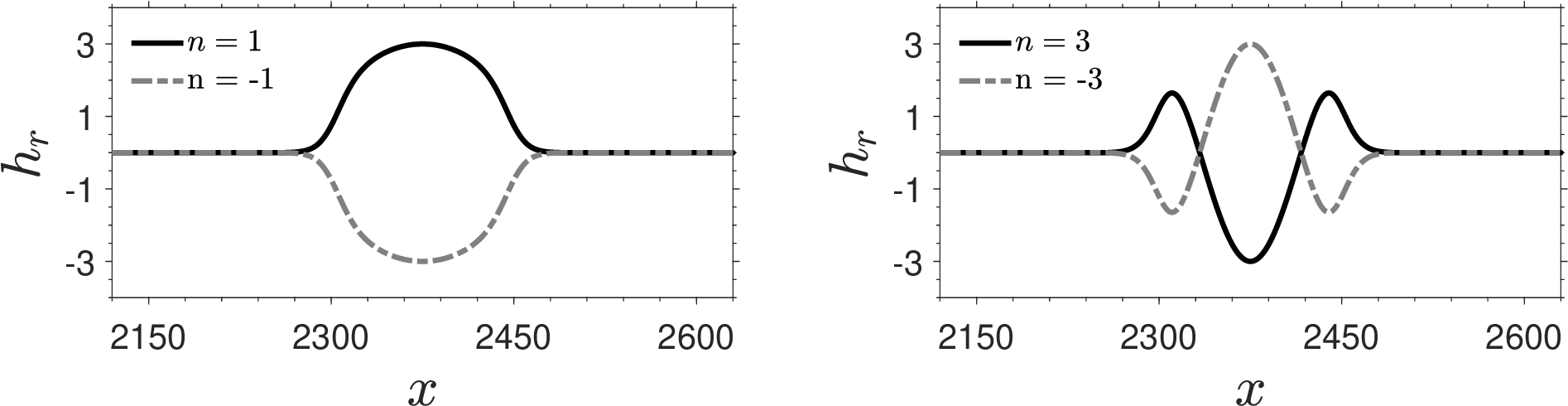}%
\put(-385.0,90.0){$(a)$}
\put(-183.0,90.0){$(b)$}
}%
\caption{Shape of the roughness elements positioned between $X_0 = 2250$, $X_N = 2500$. $(a)$ $n = \pm 1$ and $(b)$ $n = \pm 3$. The remaining roughness parameters are reported in table \ref{TABLE:RoughnessCases}.}
\label{FIG:RoughnessShapes}
\end{figure}

The geometrical parameters of eight roughness elements are summarized in table \ref{TABLE:RoughnessCases}.  These surface topographies will be adopted in numerical simulations in order to examine the effect on transition, and the results will be discussed in \S\ref{sec:InitialGuess}.  The eight configurations involve two roughness locations and four wave-numbers: 
The designation `X1' references cases with roughness elements positioned between $X_0 = 2000$ and $X_N = 2250$, while `X2' has $X_0 = 2250$ and $X_N = 2500$. 
These locations were informed by the synchronization locations of the two most energetic inflow second-mode instabilities.
The position X1 is post-synchronization of mode $\left< 110 , 0 \right>$ and pre-synchronization of mode $\left< 100 , 0 \right>$, whereas X2 is located post-synchronization of both second-modes.
The streamwise extent of these roughness elements, $X_N - X_0 = 250$, is much longer than the streamwise wavelength of the three dominant inlet instability waves (see table \ref{TABLE:Mode_Parameters}).
The designations `p1' and `p3' reference roughness elements with positive values of $n = 1$ and $n = 3$, respectively, while `m1' and `m3' refer to negative values of $n$.
Figure \ref{FIG:RoughnessShapes} shows the dependence of the surface geometry on the four values of $n$, at the upstream location.
As this figure shows, the main feature of the surfaces defined by $n = 1$ and $n = -3$ is their protrusion above the flat wall, whereas roughness elements with $n = -1$ and $n = 3$ are primarily indentations, or cratering of the surface.

\section{Results and discussion}
\label{sec:Results}

In this section, we examine of the effects of roughness parameters on the location where the flow transitions to a turbulent state.  Specifically, we examine the influence of the roughness location, streamwise integer wave-number, maximum height, width and abruptness, $\{X_r, n, H_r, W_r, L_r\}$ in (\ref{Eq:RoughnessGeometry}). 
The first two geometrical parameters are examined in \S\ref{sec:InitialGuess}, while the remaining three are optimized in \S\ref{sec:OptimalRoughness} to achieve maximum transition delay in our computational domain.

\subsection{Effects of location and shape of the roughness on transition}
\label{sec:InitialGuess}
While a general surface topography that delays transition can be sought by optimization, we consider localized roughness since the associated impact on the flow is less ambiguous to analyze and due to practical considerations.  The optimization can be performed for all the roughness parameters.  However, we will first demonstrate the impact of the roughness location and general shape, $\{X_r, n\}$, on transition.  This initial set of simulations will serve two roles:  Firstly, the simulations will be analyzed to determine the key impact of roughness on transition dynamics.  Secondly, based on these simulations, we will select the initial location and shape of the roughness whose parameters $\{H_r, W_r, L_r\}$  we will further optimize.  

\begin{figure}
\centerline{%
\includegraphics[trim=150 0 150 0, clip,width=0.9999\textwidth] {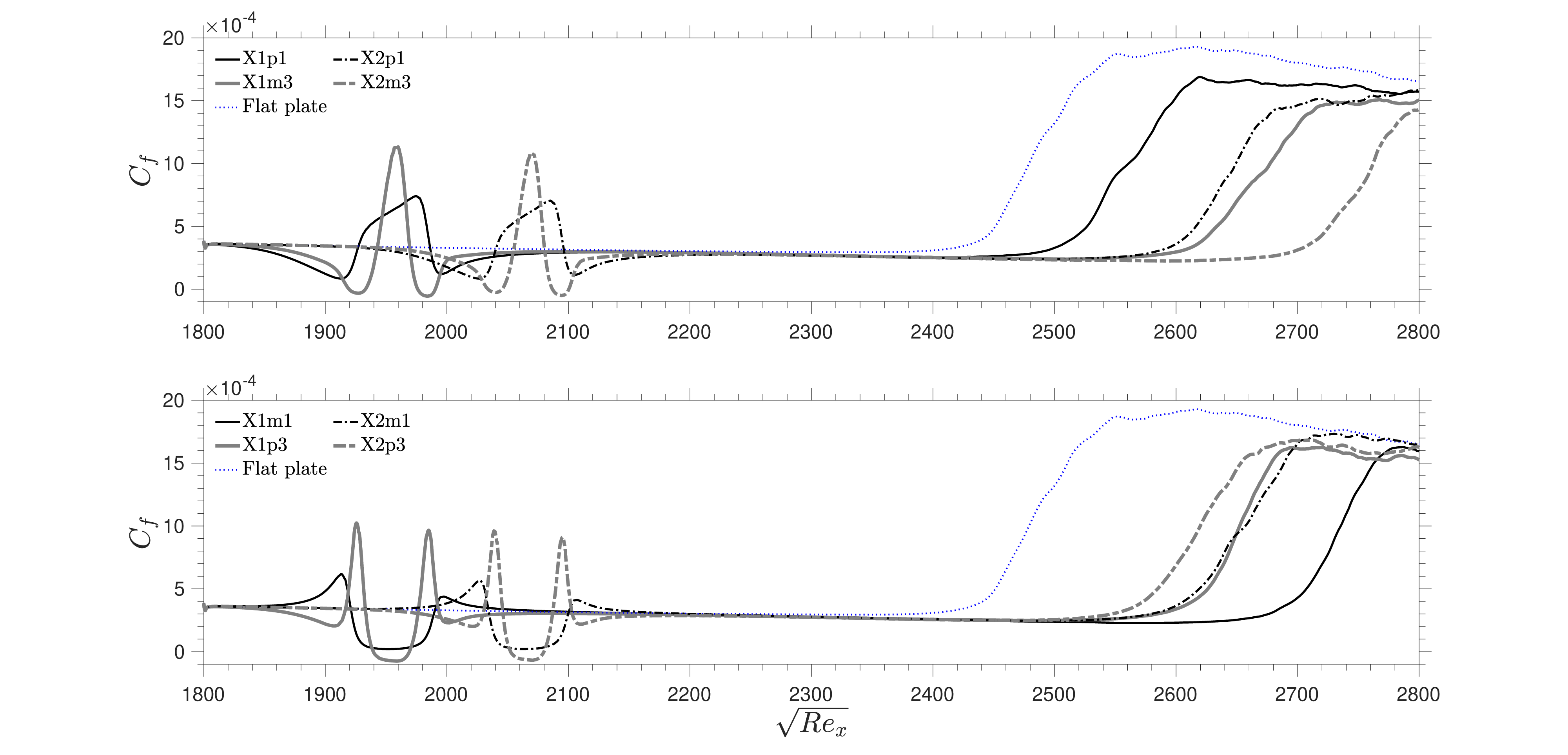}%
\put(-387.0,205.0){$(a)$}
\put(-387.0,98.0){$(b)$}
}%
\caption{Skin-friction coefficients for cases where the primary feature of the roughness is $(a)$ protrusion and $(b)$ indentation. The geometric parameters of each roughness are provided in table \ref{TABLE:RoughnessCases}. The results for transition on a flat plate are included for reference.}
\label{FIG:CfCurvesPriliminary}
\end{figure}

For each of the roughness configurations listed in table \ref{TABLE:RoughnessCases}, simulations of transition were performed when the inflow disturbance is the superposition of instability waves summarized in \S\ref{sec:InflowModes}.  
The skin-friction curves from these simulations are reported in figure \ref{FIG:CfCurvesPriliminary}.
Panel $(a)$ highlights the cases in which the roughness can mainly be considered as a protrusion, whereas panel $(b)$ shows the indentation cases.
It is evident that both the location and shape of the roughness appreciably affect the evolution of the incoming boundary-layer disturbances as manifested by the observed shift in the location of transition to turbulence.  
Figure \ref{FIG:CfCurvesPriliminary}$a$ shows that the examined protrusions can more effectively delay transition when they are placed at X2, i.e.~post-synchronization of the second modes. 
The present simulations are therefore consistent with previous studies that noted the importance of the position of roughness relative to the synchronization point of the oncoming disturbance waves \citep{Fong2015,Zhao2018,Dong2021}.
The increase in the streamwise wavenumber from one to three further delays transition, at either locations of the roughness. 
Compared to $n = 1$, the roughness with $n= - 3$ is a more slender protrusion into the flow that has a relatively larger impact on the boundary-layer thickness due to the cratering that precedes the peak.
The results in figure \ref{FIG:CfCurvesPriliminary}$a$ are specific to roughness elements whose primary feature is a protrusion.
When the primary feature is an indentation, or cratering, transition is similarly delayed relative to the flat plate.  However, some of the trends are reversed as shown in figure \ref{FIG:CfCurvesPriliminary}$b$: the most appreciable delay in transition takes place when the roughness location is upstream, at X1, and the streamwise wavenumber $n = -1$ is more effective compared to $n = 3$.

The shift in transition location reported in figure \ref{FIG:CfCurvesPriliminary} should be viewed in the context of the effect of each roughness element on the near-wall flow.
In hypersonic boundary layers, an important modulator of the amplification of second-mode instabilities is the thickness of the near-wall region where the instability phase-speed is supersonic relative to the flow.
In this region the pressure waves reflect at the wall and at the relative sonic line ($c - \bar{u} = a$) where the waves change from compression to expansion and vice versa \citep{Morkovin1987}.
These waves are typically phase-tuned with the harmonic vorticity and temperature waves that are traveling along the relative sonic line.
If the thickness of the relative supersonic region changes due to surface modifications, e.g.~roughness elements or localized cooling/heating, the growth rate of the second-mode instabilities is altered.
In case of roughness elements, parameters such as location relative to the synchronization point, streamwise wave-number, height compared to the relative sonic line, and abruptness/width relative to wavelength of instabilities, all affect the outcome.

\begin{figure}
\centerline{%
\includegraphics[trim=270 85 300 90, clip,width=0.5\textwidth] {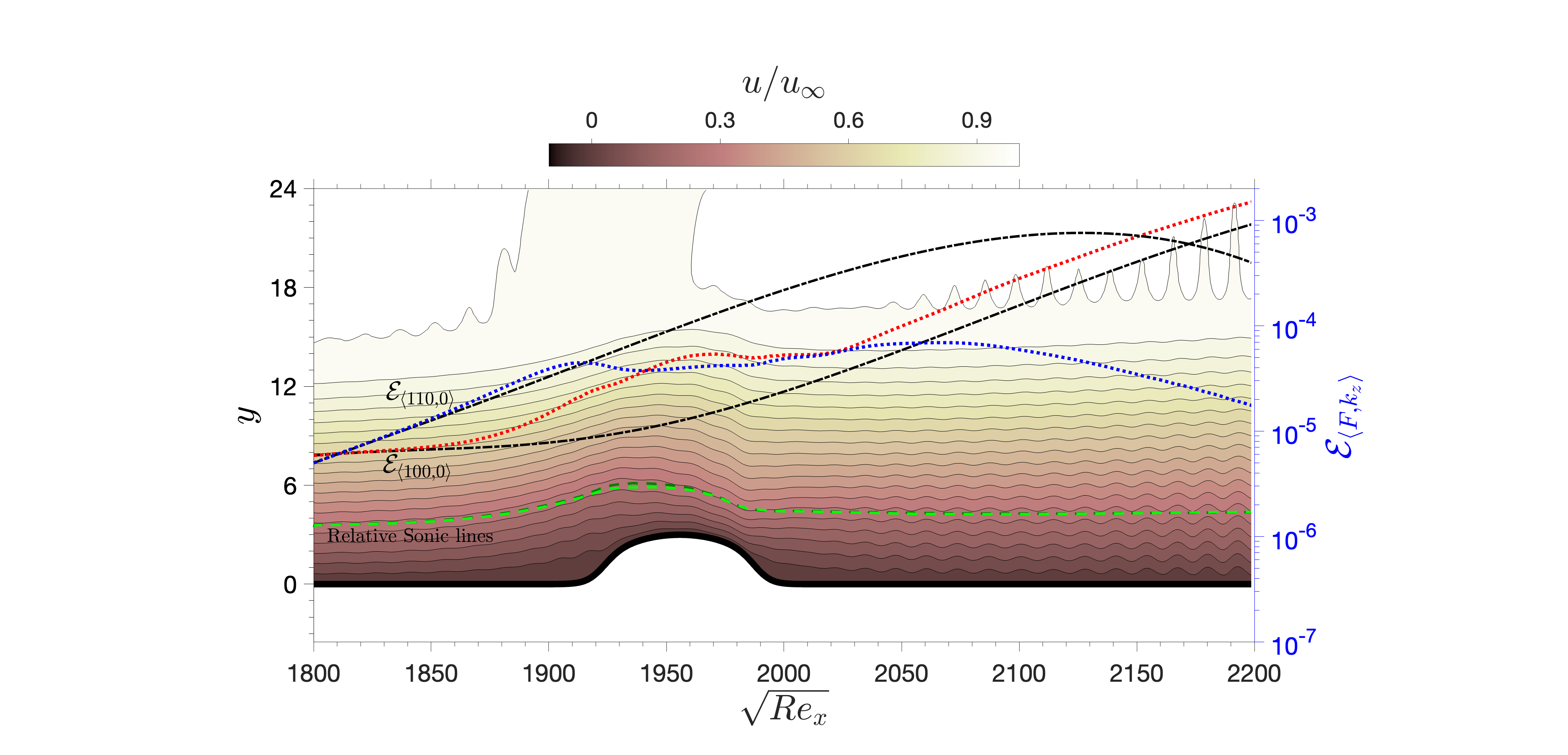}%
\hspace{4.0mm}
\includegraphics[trim=320 85 250 90, clip,width=0.5\textwidth] {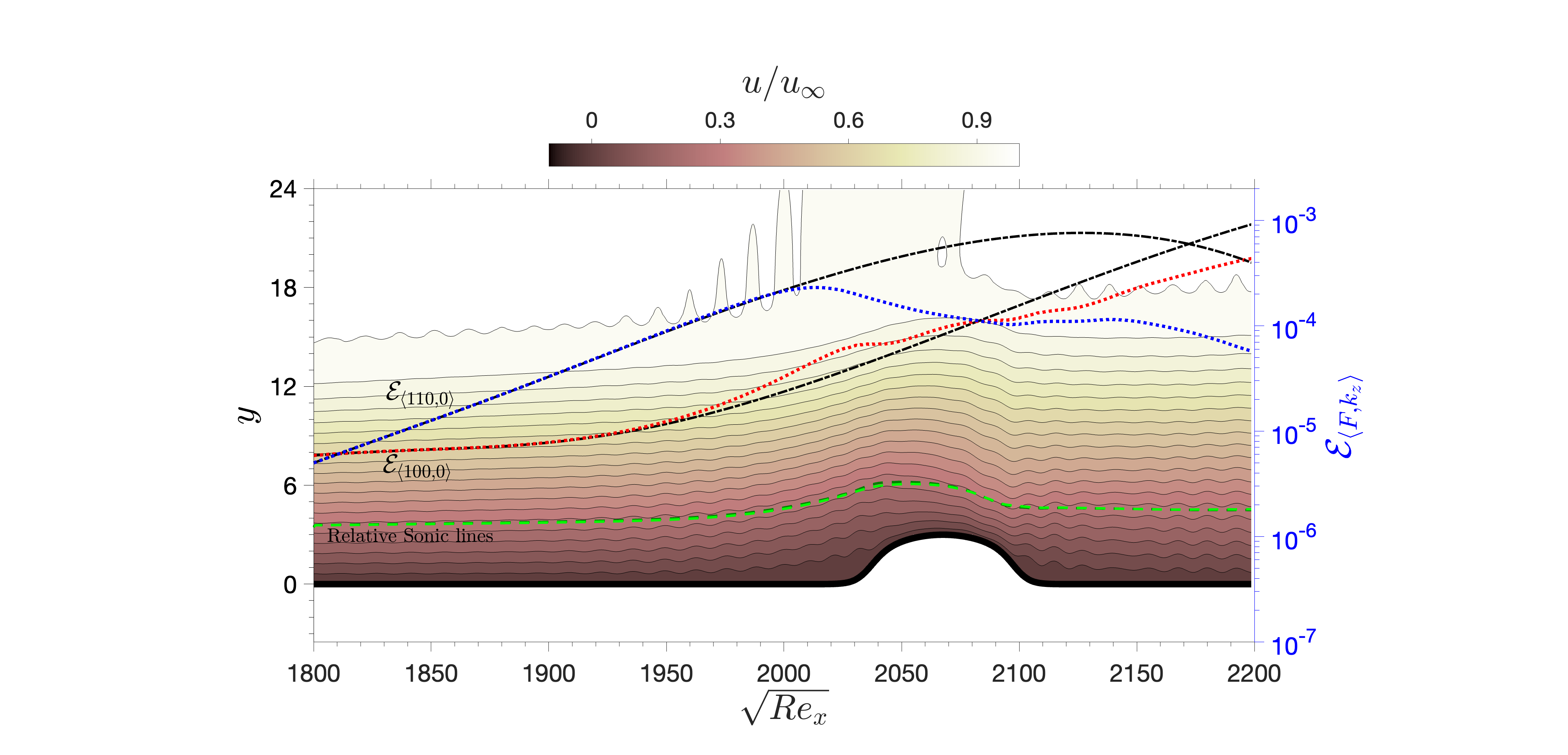}%
\put(-405.0,85.0){$(a)$}
\put(-205.0,85.0){$(b)$}
}%
\vspace{4.0mm}
\centerline{%
\includegraphics[trim=270 30 300 210, clip,width=0.5\textwidth] {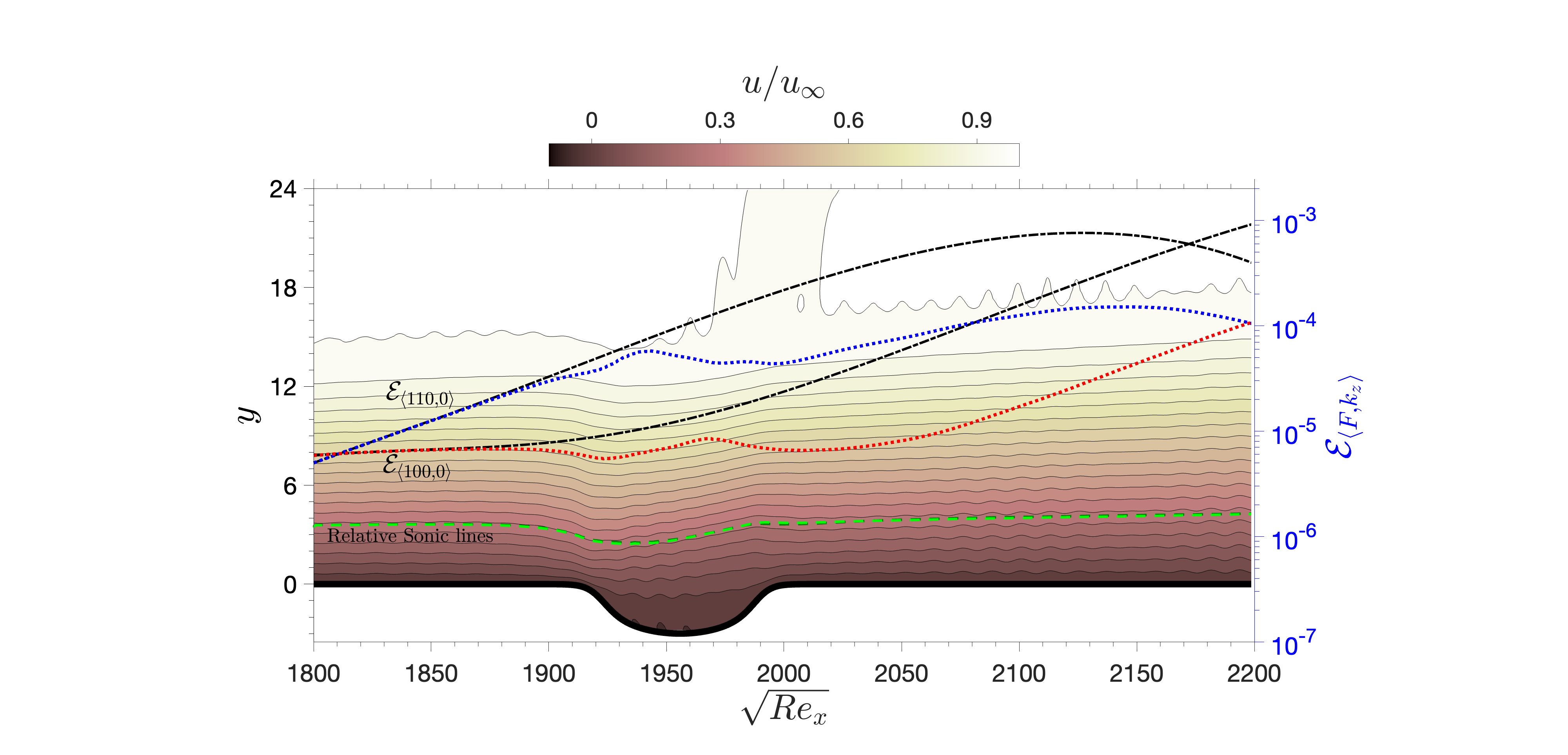}%
\hspace{4.0mm}
\includegraphics[trim=320 30 250 210, clip,width=0.5\textwidth] {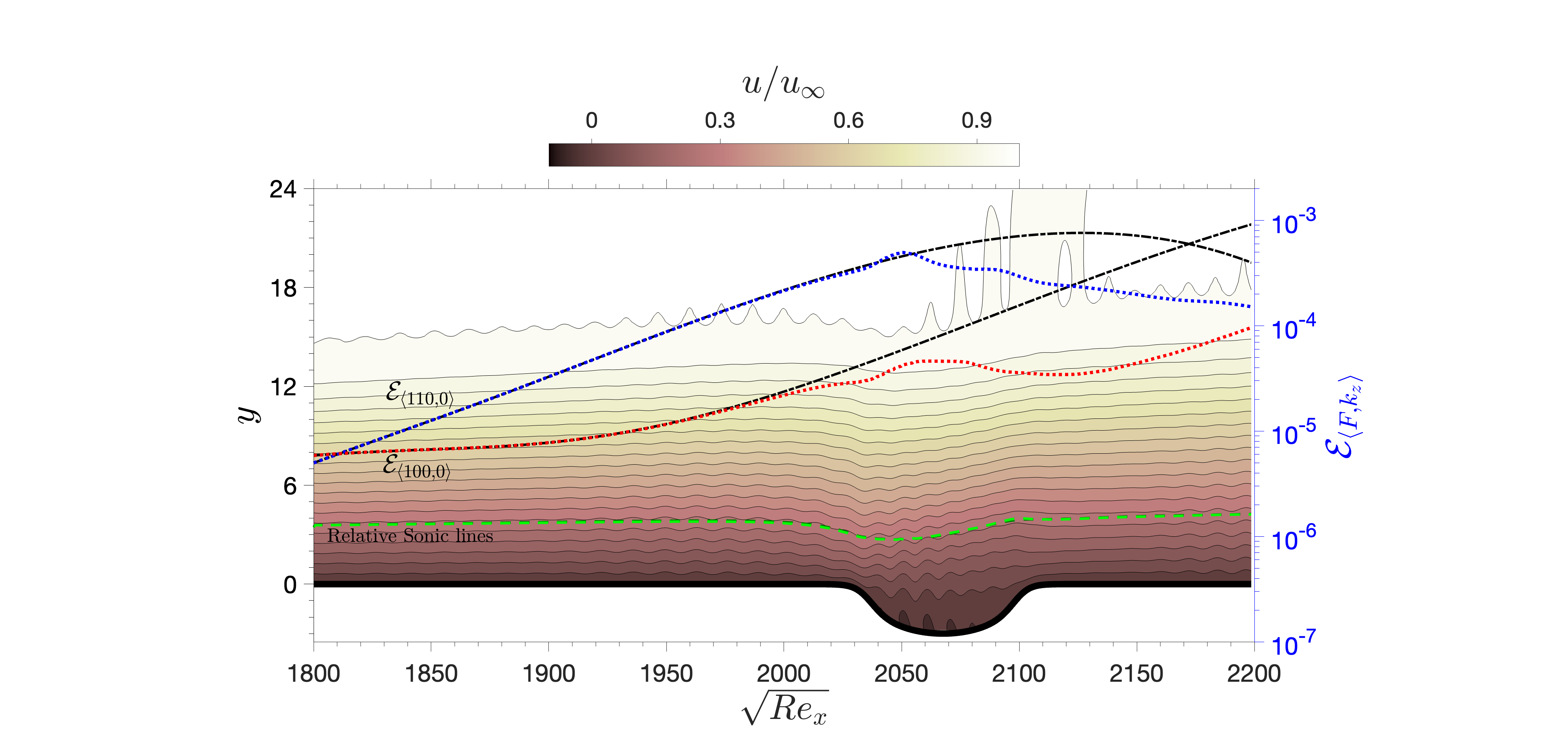}%
\put(-405.0,92.0){$(c)$}
\put(-205.0,92.0){$(d)$}
}%
\caption{Contours of instantaneous streamwise velocity for $(a)$ X1p1, $(b)$ X2p1, $(c)$ X1m1, and $(d)$ X2m1. Right axis is the energy of modes $\left< 110 , 0 \right>$ and $\left< 100 , 0 \right>$ for each case (dotted blue and red lines) compared to the reference flat-plate case (dashed-dotted black lines). The relative sonic lines of modes $\left< 100 , 0 \right>$ and $\left< 110 , 0 \right>$ are marked by dark and light green dashed lines, and are mostly overlapping.}
\label{FIG:X12Bump_UContoursE100_Z75}
\end{figure}

When the primary feature of the roughness is a protrusion, modification of the relative supersonic region pre-synchronization can have an appreciable net destabilizing effect. This trend is reversed when the modification to the relative supersonic region takes place post-synchronization; the net effect on the instability waves becomes stabilizing.
This trend is captured in the growth rate of mode $\left< 100 , 0 \right>$ for which the locations X1 and X2 are pre and post its synchronization location ($\sqrt{Re_x} = 1988$).
Figures \ref{FIG:X12Bump_UContoursE100_Z75}(a,b) quantify this behaviour, where the growth rate of mode $\left< 100 , 0 \right>$ is reported for protrusions at X1p1 and at X2p1 (the figures also shows the growth rate of mode $\left< 110 , 0 \right>$).
The contours are iso-levels of the instantaneous streamwise velocity.
Upstream of the roughness, there is a zone where the iso-lines diverge away from the wall, specifically where the relative supersonic region thickens; Atop the roughness, the iso-lines converge and thus the relative-supersonic region thins; Finally within a third zone downstream of the roughness, the iso-lines recover their natural boundary-layer spreading.
The wave is initially destabilized before it reaches the roughness, followed by a stabilization region on top of the roughness, and finally a re-adjustment zone across which the instability essentially recovers its amplification rate for an undisturbed boundary layer.
When the protrusion is positioned upstream of the synchronization point of mode $\left< 100 , 0 \right>$, the net effect was a higher modal energy of this wave.
Placement of the protruding roughness downstream of the synchronization point is shown in panel $(b)$. The final energy attained by mode $\left< 100 , 0 \right>$ in this case is lower than the upstream placement of this roughness element.  
The above dependence of the amplitude of mode $\left< 100 , 0 \right>$ on the roughness location may seem at odds with transition being delayed relative to the flat plate, for both roughness locations (figure \ref{FIG:CfCurvesPriliminary}). For the explanation, it is important to recall that both roughness configurations are downstream of synchronization of the other dominant second-mode wave, namely $\left< 110 , 0 \right>$.  As a result, while this mode also exhibits the three-stage stabilization/destabilization/readjustment behaviour across the roughness, it is ultimately stabilized in both cases (figures \ref{FIG:X12Bump_UContoursE100_Z75}$(a)$ and \ref{FIG:X12Bump_UContoursE100_Z75}$(b)$).
Comparison of the DNS results with computations of modal evolution using linearized-Navier-Stokes equations (not shown here) confirmed that the destabilization zones of modes $\left< 100 , 0 \right>$ and $\left< 110 , 0 \right>$ are primarily caused by the roughness-induced base-flow distortion, while the stabilization and re-adjustment zones are strongly influenced by the interaction of the instability waves with the finite slope of the roughness geometry.

When the main roughness feature is an indentation (figures \ref{FIG:X12Bump_UContoursE100_Z75}$(c,d)$), the relative supersonic region initially thins, then thickens, and finally thins again, with an associated stabilization/destabilization/stabilization and re-adjustment of the  second-mode waves.
Roughness location at either X1 or X2 leads to a net stabilization for both second-mode waves.  For mode $\left< 100 , 0 \right>$ this outcome is noteworthy because the synchronization location of this mode is downstream of X1 and upstream of X2.  According to \citet{Dong2021} (see their figures 16 and 18), indentations have a similar albeit weaker scattering effect as protrusions, and hence one expects that the instability wave is destabilized when an indentation is introduced pre-synchronization and stabilized when the indentation is downstream of synchronization.  In contrast to that work, in our nonlinear simulations where the roughness is relatively short in the streamwise direction, mode  $\left< 100 , 0 \right>$ is stabilized in both configurations.

Similar to earlier works, the herein considered roughness parameters were guided by knowledge of (a) the importance of the synchronization point, (b) the sensitivity of the modal amplification to distortions in the base flow and (c) the non-parallel effects induced by the roughness. The roughness parameters were not, however, optimized to guarantee a particular outcome.  We next consider such optimization; specifically we seek an optimal roughness geometry that can lead to sustained laminar flow throughout the entire computational domain in our simulations.

\subsection{Optimal protruding roughness for transition delay}
\label{sec:OptimalRoughness}

In this section, we present the nonlinear optimization of the roughness height, width and abruptness, and examine the flow field associated with the optimal roughness.

\subsubsection{Ensemble-variational optimization of the roughness}
\label{sec:EnVarOptimization}

The base design that is adopted as a starting point of the optimization has streamwise wave-number $n = 1$, which corresponds to a simple protrusion.
This choice is primarily motivated by its simplicity to aid the interpretation of the impact on the flow, and ease of manufacturing relative to cratering of the surface.
Figure \ref{FIG:CfCurvesPriliminary}$a$ showed that protrusions at X2 are more effective in delaying transition. Therefore, $X_0 = 2250$ and $X_N = 2500$ are selected as the start and end positions of the roughness element.

The optimization is performed using an ensemble-variational (EnVar) approach.
Starting from the initial design, here a roughness with 
$\{H_r,  W_r, L_r\}=\{0.22\delta_{x_0}, 5.37\delta_{x_0}, 0.81 \delta^{-1}_{x_0} \}$ where $\delta_{x_0} = 13.5$ is the boundary-layer thickness at the inlet plane,
EnVar updates the estimate of the control vector, $ \boldsymbol{c} = [ H_r \;\; W_r \;\; L_r ]^{\top}$, at the end of each iteration using the gradient of the cost function; the gradient is evaluated from the outcomes of an ensemble of possible solutions.  
The cost function is the integrated skin-friction coefficient along the plate which, once minimized, ensures the farthest possible downstream location of transition to turbulence.
The iterative optimization is halted once the identified roughness can maintain a laminar state throughout the entire computational domain.
It should be noted that the solution to the above nonlinear optimization is not unique and, similar to other gradient-based methods, the reported results are only guaranteed to be a local optimum for a given choice of the initial guess.  However, as long as the discovered roughness design accomplishes the objective of the optimization, it is deemed successful. 

The choice of the EnVar technique is justified because the low-dimensional nature of the control vector.  In addition, unlike adjoint methods which may place limits on the time horizon of the flow solution \citep{Zaki2021}, EnVar is perfectly suited for long-time integration and evaluation of statistical cost functions \citep{Mons2019,Mons2021}.  EnVar has successfully been adopted in high-speed boundary layers for assimilation of measurements into flow simulations \citep{Buchta2021,Buchta2022} and optimization \citep{Jahanbakhshi2021}. The reader is referred to those studies for the details of the algorithm.  

The optimization procedure seeks the roughness parameters that minimize the cost function
\begin{equation}
\mathcal{J} = \underbrace{ \frac{1}{2} \Vert \boldsymbol{c} - \boldsymbol{c}^{(e)} \Vert_{\boldsymbol{B}^{-1}}^2 }_{\mathcal{J}_p} + \underbrace{ \frac{1}{2} \Vert \mathcal{G} ( \boldsymbol{q} ) \Vert_{\boldsymbol{R}^{-1}}^2 }_{ \mathcal{J}_o }. 
\label{Eq:CostFunction}
\end{equation}
The first term $\mathcal{J}_p$ is a regularization which ensures that the optimal $\boldsymbol{c}$ at the end of each iteration does not deviate appreciably from the previous estimate $\boldsymbol{c}^{(e)}$. The second term, $\mathcal{J}_o$, is the objective function that is formulated by defining $ \mathcal{G} \left( \boldsymbol{q} \right) = C_f \left({ dx / L_x }\right)^{\frac{1}{2}} $ where the skin-friction coefficient is $C_f = \tau_{wall} / ( \frac{1}{2} \rho_{\infty} u_{\infty}^2 )$.  
In equation (\ref{Eq:CostFunction}), $\boldsymbol{B}$ and $\boldsymbol{R}$ are the co-variance matrices of the prior term and the observed skin friction, respectively.
In order to ensure that the maximum slope of the predicted roughness at the end of each iteration of the EnVar algorithm is appropriately resolved by the grid, a constraint is introduced in the optimization procedure.
Specifically, we require that $| d \boldsymbol{h}_r / dx | \leq s $ where $s$ is a pre-determined limit on the slope of the roughness which is chosen based on the grid resolution, $s \sim \mathcal{O}(\Delta y_{\text{near wall}} / \Delta x)$.

\begin{figure}
\centerline{%
\includegraphics[trim=0 0 0 0, clip,width=0.50\textwidth] {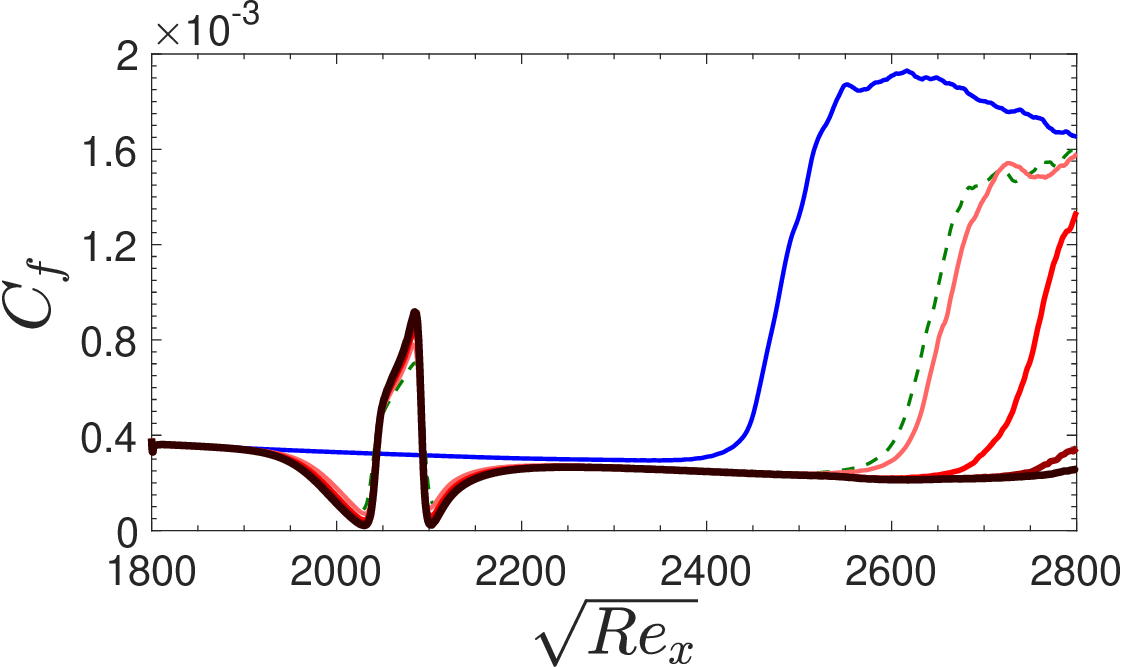}%
\includegraphics[trim=-20 0 0 0, clip,width=0.50\textwidth] {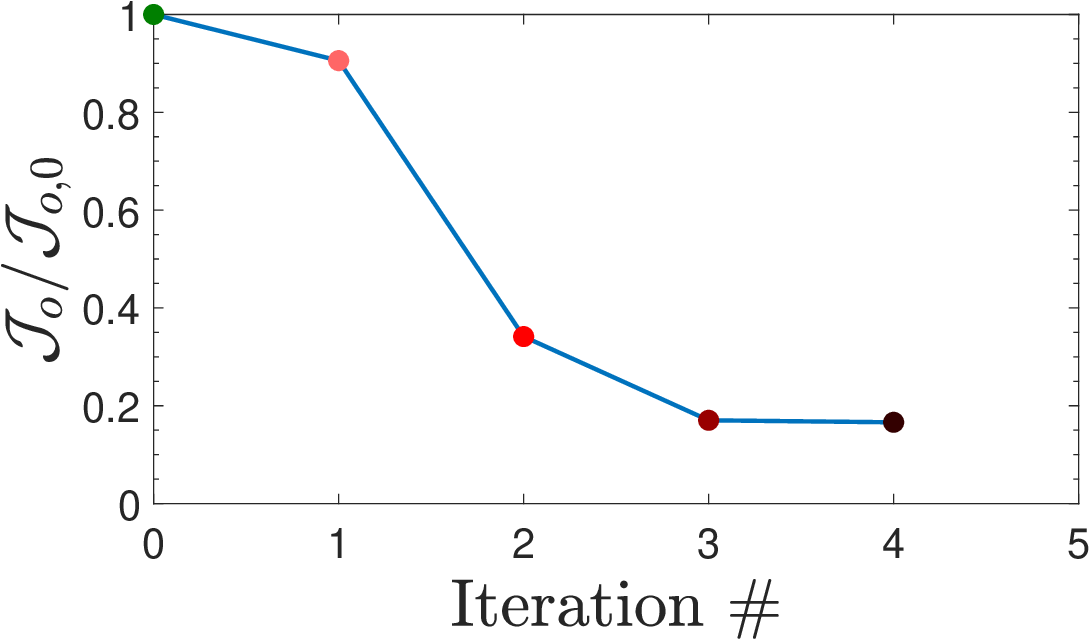}%
\put(-385.0,100.0){$(a)$}
\put(-185.0,100.0){$(b)$}
\put(-261.0,50.0){\scriptsize \begin{turn}{77} \textit{Flat plate} \end{turn} }
\put(-240.0,60.0){ \begin{turn}{-50} $ \xrightarrow{ \textit{EnVar search} } $ \end{turn} }
}%
\vspace{2mm}
\centerline{%
\includegraphics[trim=0 0 0 0, clip,width=0.60\textwidth] {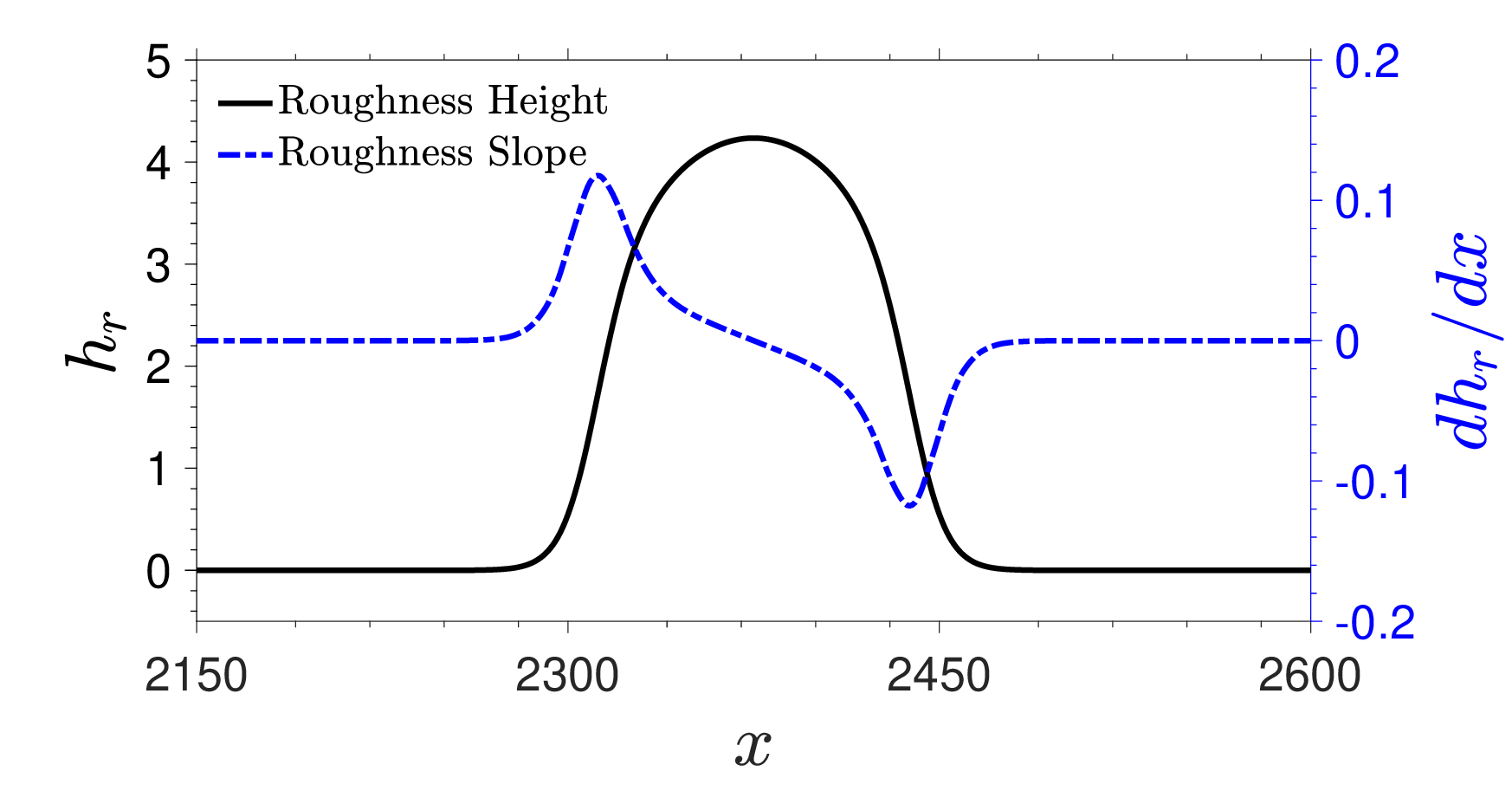}%
\put(-225.0, 105.0){$(c)$}
}%
\caption{The outcomes of EnVar iterations. $(a)$ Skin friction of the reference flow over a flat plate and the predictions at consecutive iterations versus $\sqrt{Re_x}$. $(b)$ Cost function of each iteration normalized by the initial guess. $(c)$ The geometry of the optimal roughness, after the fourth EnVar iteration, versus streamwise distance $x$.}
\label{FIG:EnVarOutCome}
\end{figure}

The skin friction and normalized cost function, $C_f$ and $\mathcal{J}_o/\mathcal{J}_{o,0}$, associated with the mean control vector $\boldsymbol{c}$ at the end of each EnVar iteration are reported in figures \ref{FIG:EnVarOutCome}($a$,$b$).
These plots demonstrate that the location, where the boundary layer transitions to a turbulent state, shifts downstream after consecutive iteration.  In other words, the optimization procedure is effective at updating the roughness parameters in a manner to reduce the cost function (\ref{Eq:CostFunction}) and delay breakdown to turbulence.
At the end of iteration \# 4, the boundary layer is laminar throughout the computational domain.

The height and slope of the optimal roughness after the fourth EnVar iteration are depicted in figure \ref{FIG:EnVarOutCome}($c$).
The associated geometric parameters are $\{H_r, W_r, L_r\} = \{0.31 \delta_{x_0}, 4.84 \delta_{x_0}, 0.89 \delta^{-1}_{x_0} \}$.
Therefore, compared to the initial guess, the optimal roughness is taller, more slender and more abrupt.
Despite the larger height, the optimal roughness is still below the relative sonic line inside the boundary layer. As for the width $2 W_r$, it is approximately 4.1 times the streamwise wavelength of mode $\left< 100, 0 \right>$ and 4.5 times the wavelength of mode $\left< 110, 0 \right>$ (see table \ref{TABLE:Mode_Parameters}).
In order to provide an appreciation for the physical size of the roughness elements, we can convert the current roughness parameters to dimensional quantities.  For this purpose, we adopt the reference scales from the experiments by \citet{Kendall1975} which examined the stability of a flat-plate boundary layer at the same free-stream Mach number and temperature.
The optimal dimensional parameters of our roughness then become $\{H_r, W_r, L_r\} = \{1.1 \; \textrm{mm}, 16.3 \; \textrm{mm}, 0.264 \; \textrm{mm}^{-1} \}$, which are physically and practically relevant.
For example, in the experiment by \citet{Fong2015} for a Mach-6 boundary layer, the height and width of the roughness were set to $H_r = 0.665 \; \textrm{mm}$ and $W_r = 2.66 \; \textrm{mm}$, respectively; \citet{Bountin2013} set the roughness height and width in their experiment for a Mach-6 boundary layer to $H_r = 1.8 \; \textrm{mm}$ and $W_r = 12 \; \textrm{mm}$; \citet{Fujii2006}'s Mach-7 experiments included roughness height in the range $0.5 \; \textrm{mm}$ to $0.9 \; \textrm{mm}$ and roughness width in the range $0.5 \; \textrm{mm}$ to $4 \; \textrm{mm}$.

\subsubsection{Effect of optimal roughness on transition mechanism}
\label{sec:OptimalTransitionMechanism}

\begin{figure}
\centerline{%
\includegraphics[trim=0 0 0 0, clip,width=0.99\textwidth] {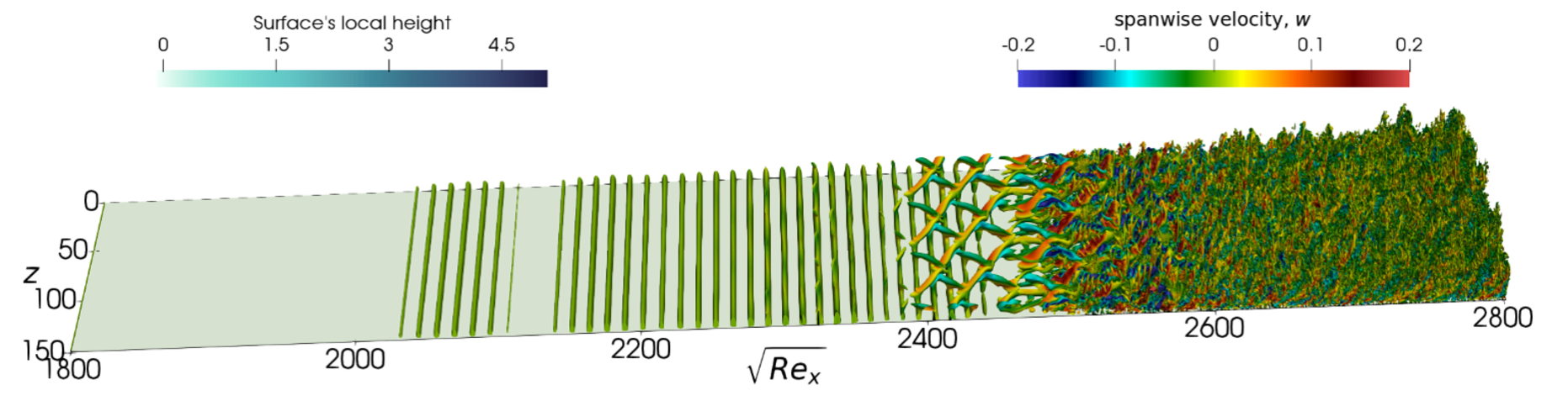}%
\put(-390.0,50.0){$(a)$}
}%
\centerline{%
\includegraphics[trim=0 0 0 55, clip,width=0.99\textwidth] {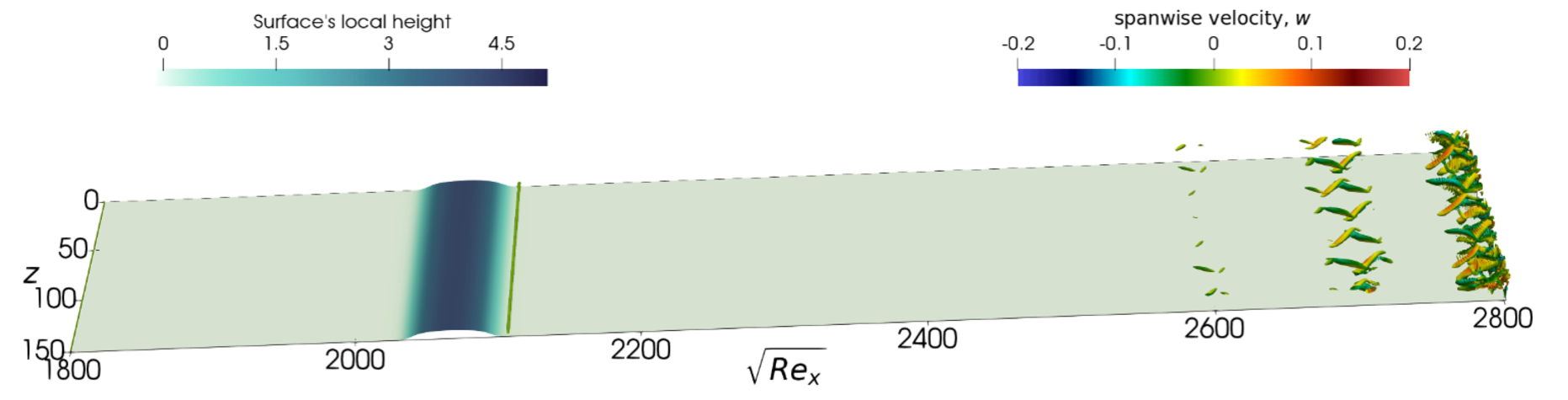}%
\put(-390.0,50.0){$(b)$}
}%
\caption{Isosurfaces of $Q$-criteria for $(a)$ flow over a flat plate and $(b)$ the optimal roughness.}
\label{FIG:FlatPlate_X2OptBump_Q}
\end{figure}

Two instantaneous flow fields are contrasted in  figure \ref{FIG:FlatPlate_X2OptBump_Q}: the structures are iso-surfaces of the second invariant of the velocity-gradient tensor, and are coloured by the spanwise velocity perturbations.  Figure \ref{FIG:FlatPlate_X2OptBump_Q}(a) is the reference simulation over a flat plate, while figure \ref{FIG:FlatPlate_X2OptBump_Q}(b) shows the outcome of the optimization procedure.  The visual difference highlights the efficacy of the optimized roughness in suppressing the transition precursors, and delaying the onset of turbulence.

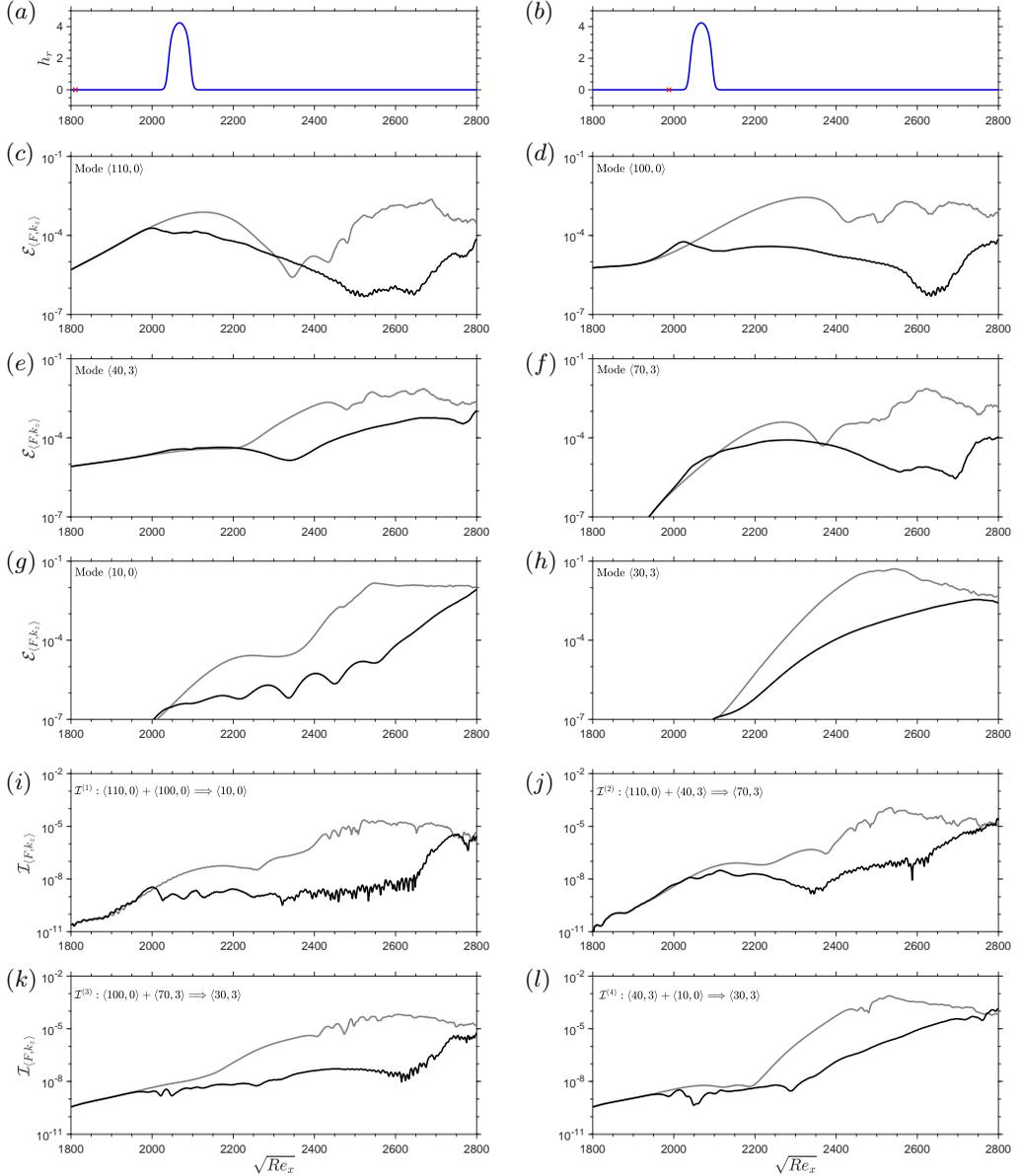
\begin{figure}
\centerline{%
\includegraphics[trim=120 480 190 490, clip,width=0.99\textwidth] {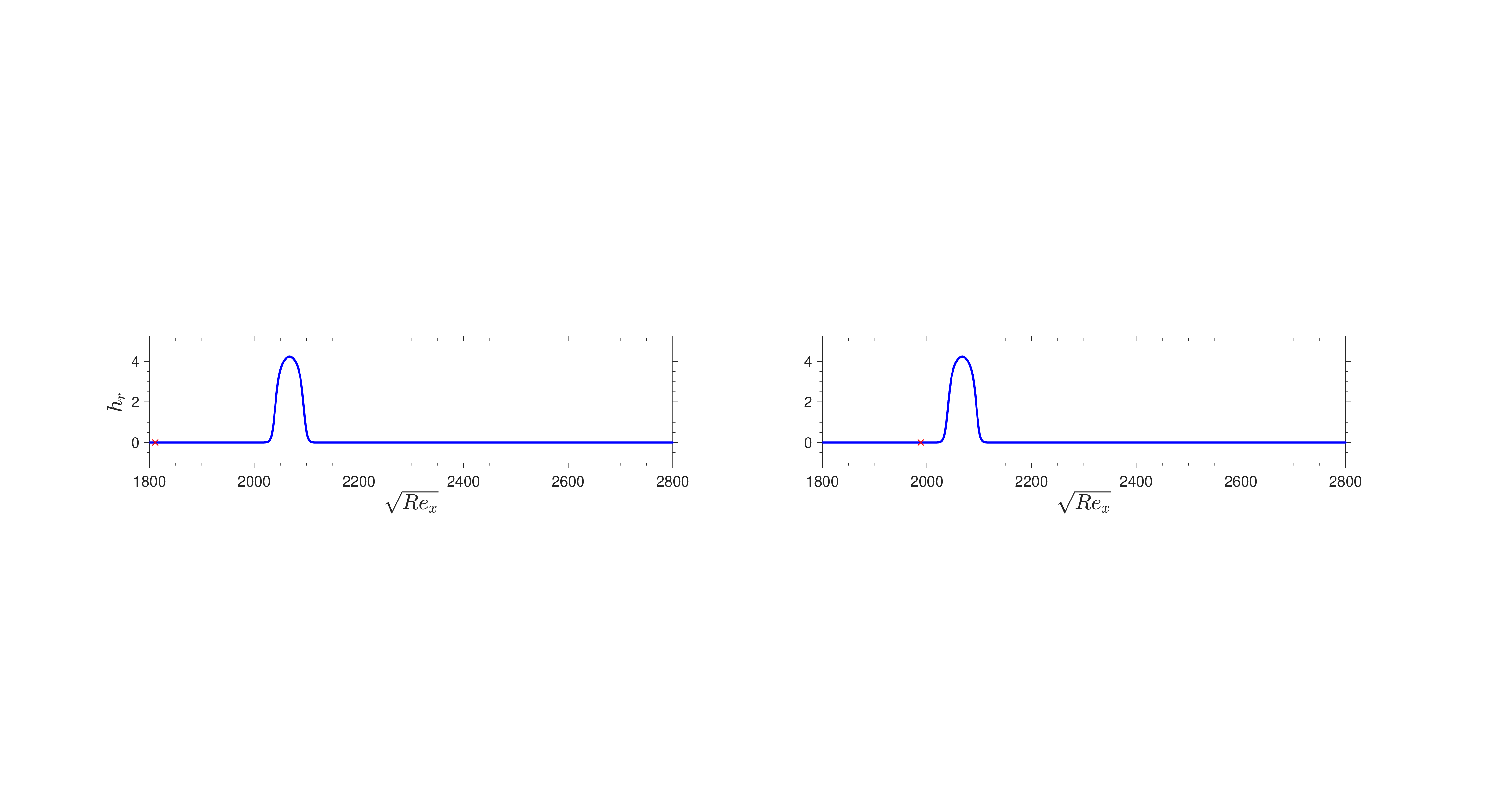}%
\put(-385.0,40.0){$(a)$}
\put(-187.0,40.0){$(b)$}
}%
\vspace{2.0mm}
\centerline{%
\includegraphics[trim=120 70 190 0, clip,width=0.99\textwidth] {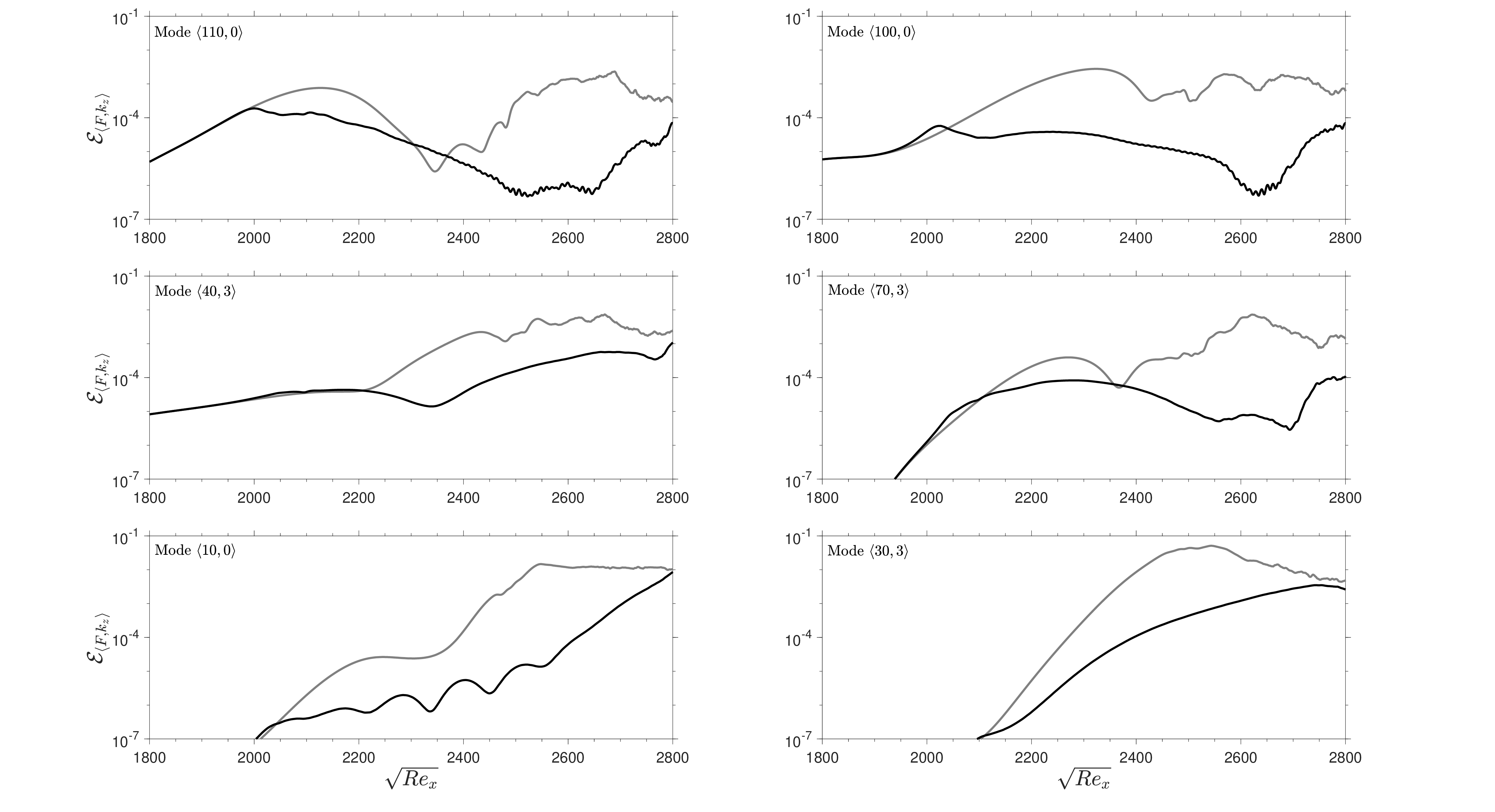}%
\put(-385.0,220.0){$(c)$}
\put(-187.0,220.0){$(d)$}
\put(-385.0,140.0){$(e)$}
\put(-187.0,140.0){$(f)$}
\put(-385.0,65.0){$(g)$}
\put(-187.0,65.0){$(h)$}
}%
\vspace{2.0mm}
\centerline{%
\includegraphics[trim=120 120 190 220, clip,width=0.99\textwidth] {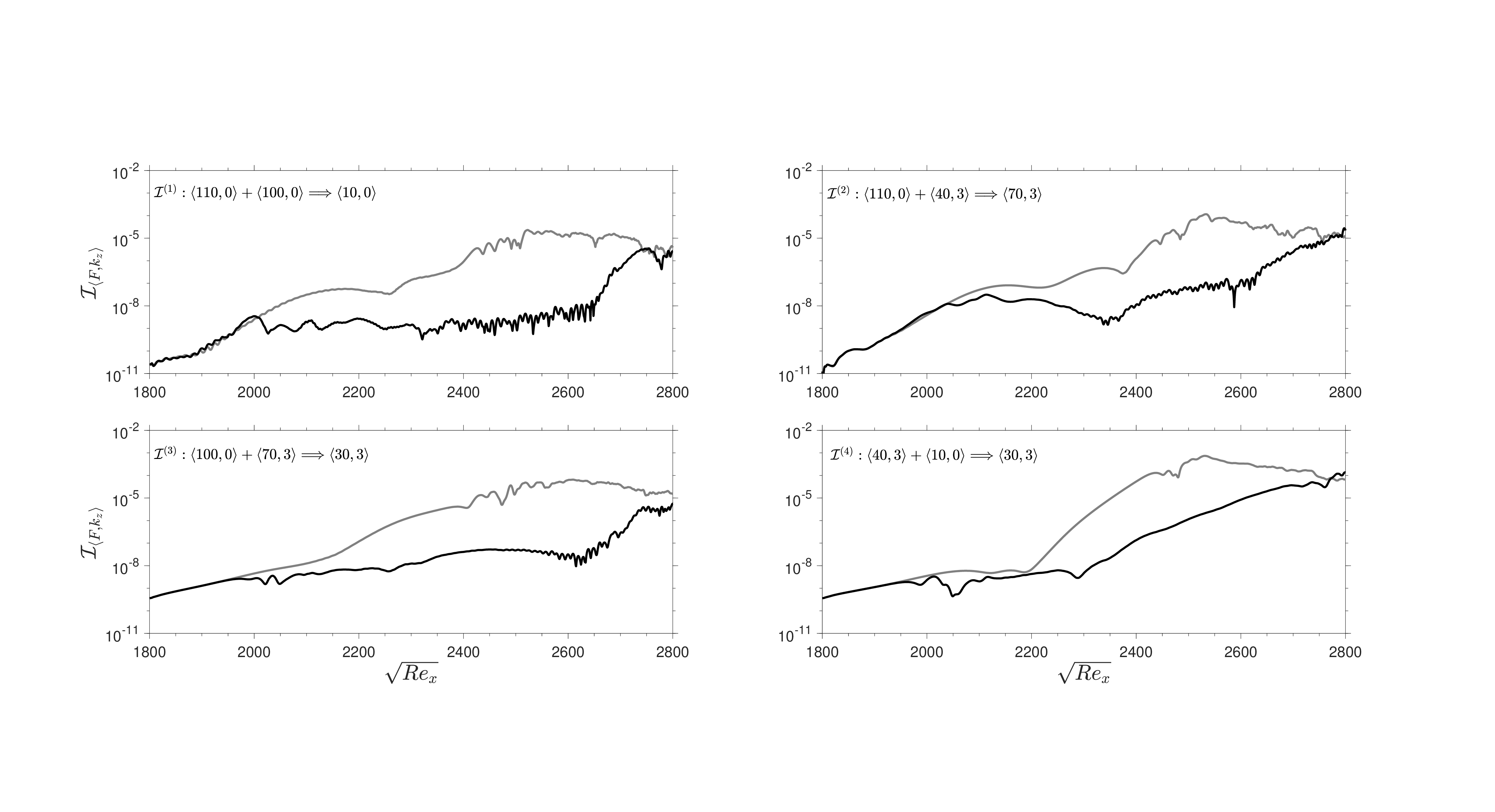}%
\put(-385.0,160.0){$(i)$}
\put(-187.0,160.0){$(j)$}
\put(-385.0,85.0){$(k)$}
\put(-187.0,85.0){$(l)$}
}%
\caption{$(a,b)$: The optimal protruding roughness at the end of iteration \# 4, with `x' marking the synchronization Reynolds numbers of modes $\left< 110,0 \right>$ and $\left< 100,0 \right>$, respectively. ($c$-$h$): Spectral energy, $\mathcal{E}_{\left< F , k_z \right>}$, for selected instability modes versus streamwise coordinate. ($i$-$l$) Modal nonlinear energy-transfer coefficient, computed for key triad interactions. Gray lines are reference curves for flat-plate case; black lines are the results of the boundary layer over the optimal roughness.}
\label{FIG:EnergySpectralDensity}
\end{figure}

The impact of the optimized roughness element on the downstream development of key instability waves and nonlinear energy exchanges is examined in figure \ref{FIG:EnergySpectralDensity}.  
In the first two panels $(a,b)$, we reproduce the shape of the optimized roughness and mark the synchronization locations of the slow and fast modes for waves ($a$) $\left< 110, 0 \right>$ and ($b$) $\left< 100, 0 \right>$; the roughness location is downstream of both synchronization points.
Figures \ref{FIG:EnergySpectralDensity}($c$-$h$) compare the downstream development of the spectral energy, $\mathcal{E}_{\left< F , k_z \right>}$, over a flat plate (gray lines) and in presence of the optimal roughness (black lines). The remaining panels ($i$-$l$) report the nonlinear energy transfer among wave triads $\bigl\{ \left< F , k_{z} \right>, \left< F_1 , k_{z,1} \right> , \left< F_2 , k_{z,2} \right> \bigl\}$, which is evaluated using 
\begin{equation}
	\mathcal{I}_{\left< F , k_{z} \right>} =  \sum_{\pm F , \pm k_{z}} \int_{0}^{L_y} \left\vert \boldsymbol{\hat{\mathcal{A}}}^*_{\left< F_1 , k_{z,1} \right>} \boldsymbol{\hat{\mathcal{B}}}_{\left< F_2 , k_{z,2} \right>}
											+ \boldsymbol{\hat{\mathcal{A}}}^*_{\left< -F_2 , -k_{z,2} \right>} \boldsymbol{\hat{\mathcal{B}}}_{\left< -F_1 , -k_{z,1} \right>} \right\vert dy
\label{Eq:NonLinearEnergyTransfer}
\end{equation}
where $F = F_2 - F_1$ and $k_z = k_{z,2} - k_{z,1}$ \citep{Cheung2010,Jahanbakhshi2019}.
The symbols $\boldsymbol{\hat{\mathcal{A}}}$ and $\boldsymbol{\hat{\mathcal{B}}}$ are the Fourier coefficients of 
$\boldsymbol{\mathcal{A}} =
	\begin{bmatrix}
		\rho u^{\prime\prime} u^{\prime\prime} & \rho u^{\prime\prime} v^{\prime\prime} & \rho u^{\prime\prime} w^{\prime\prime}
	\end{bmatrix}^{\top} $
and 
$\boldsymbol{\mathcal{B}} =
	\begin{bmatrix}
		u^{\prime\prime} & v^{\prime\prime} & w^{\prime\prime}
	\end{bmatrix}^{\top}$ 
where the double-prime denotes fluctuations with respect to the Favre (density-weighted) average.
The mathematical definition of $\mathcal{I}_{\left< F , k_z \right>}$ does not depend on the ordering of the modes within the triad, and hence this expression only measures the energy exchange and not the direction of the transfer.
In other words, the designation $\left< F_1 , k_{z,1} \right> + \left< F_{2} , k_{z,2} \right> \Rightarrow \left< F , k_z \right>$ in figures \ref{FIG:EnergySpectralDensity}($i$-$l$) merely identifies the triad and bears no directional significance, although the outcome of the exchange can be gleaned, with caution, by simultaneously considering the spectra of the individual waves (panels $c$-$f$).

To establish the background against which the influence of roughness is examined, we summarize the transition mechanism in the flat-plate case with the aid of the spectra in figures \ref{FIG:EnergySpectralDensity}($c$-$h$).  
The three most energetic inflow modes, $\{\left< 110 , 0 \right>, \left< 100 , 0 \right>, \left< 40 , 3 \right>\}$, initially amplify and then spur other waves via nnonlinear interactions.  The pair of second modes participate in $\mathcal{I}^{(1)}$ (figure \ref{FIG:EnergySpectralDensity}($i$)) and generate $\left< 10 , 0 \right>$.  Another triad, $\mathcal{I}^{(2)}$ in figure \ref{FIG:EnergySpectralDensity}($j$), activates mode $\left<70 , 3 \right>$ which is visible as oblique structures in figure \ref{FIG:FlatPlate_X2OptBump_Q}$(a)$ near $\sqrt{\textrm{Re}_x} \approx 2300$.
The last two interactions, $\mathcal{I}^{(3)}$ and $\mathcal{I}^{(4)}$ in figures \ref{FIG:EnergySpectralDensity}($k$-$l$), involve both the inflow and newly formed waves, and spur the formation of mode $\left< 30 , 3 \right>$. This mode amplifies faster than any other wave, forms the $\Lambda$-shaped structures in figure \ref{FIG:FlatPlate_X2OptBump_Q}$(a)$ in the range $2400 < \sqrt{\textrm{Re}_x} < 2550$, and is the ultimate cause of breakdown to turbulence.

Comparing the reference case with the optimal roughness in figures \ref{FIG:EnergySpectralDensity}($c$-$l$), both modes $\left<110,0\right>$ and $\left<100,0\right>$ are attenuated in the latter configuration as the roughness element is approached (see panels ($c$-$d$)).
The effect is more pronounced for mode $\left<100,0\right>$ whose synchronization point is much closer to the roughness.  The subsequent nonlinear interactions that involve these two instabilities, especially interactions $\mathcal{I}^{(1)}$ and $\mathcal{I}^{(3)}$ involving mode $\left<100,0\right>$, are also mitigated.  The outcome is an appreciable weakening of the nonlinearly generated modes that cause breakdown to turbulence.

Another observation from figure \ref{FIG:EnergySpectralDensity} is that the first-mode wave $\left<40,3\right>$ is negligibly destabilized as a result of the introduced roughness.
This trend is different from previous efforts \citep[see e.g. ][]{Fong2015, Bountin2013}, and is due to the smaller height of the roughness in our simulation.
For comparison, the height of the roughness in the experiments by \citet{Fong2015} and by \citet{Bountin2013} was approximately equal to $0.5 \delta_{X_r}$, while in our analysis the height is $H_r = 0.23 \delta_{X_r}$.
Mode $\left<40,3\right>$ is appreciable between the relative sonic line and the boundary layer edge, and this region is minimally affected by the presence of the roughness which is positioned below the sonic line.

The results presented in figure \ref{FIG:EnergySpectralDensity} highlight that roughness can have a direct, local effect on the instability waves. For example, in figures \ref{FIG:EnergySpectralDensity}($c$-$d$), the growth rates of modes $\left< 100 , 0 \right>$ and $\left< 110 , 0 \right>$ are significantly altered in the region $2000 < \sqrt{Re_x} < 2100$, as these instabilities approach and travel over the roughness.
As noted in \S\ref{sec:InitialGuess}, the dominant local roughness effects in our configuration are the roughness-induced mean-flow distortion and non-parallel effects in regions where where $d h_r / dx$ is large.
In particular, the non-parallel effects of the optimal roughness (which is taller, more slender and more abrupt) is more pronounced compared to roughness X2p1.
In addition to these local effects, the influence of roughness on transition precursors can persist downstream. For example, figure \ref{FIG:EnergySpectralDensity}($e$) shows that the growth rate of mode $\left< 40 , 3 \right>$ in the roughed-wall case reduces dramatically for $\sqrt{Re_x} > 2200$.
This change can be traced back to the local-effects on the amplitude of modes $\left< 100 , 0 \right>$ and $\left< 110 , 0 \right>$: stabilization of these waves by the roughness mitigates the subsequent nonlinear interactions (figures \ref{FIG:EnergySpectralDensity}($i$-$l$)).  
Specifically, over a flat plate, mode $\left< 40 , 3 \right>$ is activated near $\sqrt{Re_x} \approx 2200$ in a triad interaction $\mathcal{I}^{(4)}$ which involves $\left< 10 , 0 \right>$ and $\left< 30 , 3 \right>$ (figure \ref{FIG:EnergySpectralDensity}($l$)); In the rough-wall case, this triad-interaction is delayed and muted, due to a delay in generation of modes $\left< 10 , 0 \right>$ and $\left< 30 , 3 \right>$ from the preceding interactions.  In light of these observations, we turn our attention to the local effect of the optimal roughness on the second-mode instabilities $\left< 100 , 0 \right>$ and $\left< 110 , 0 \right>$.

\subsubsection{Local effect of optimal roughness on second-mode instabilities}
\label{sec:LocalEffects}

\begin{figure}
\centerline{%
\includegraphics[trim=100 130 50 190, clip,width=0.99\textwidth] {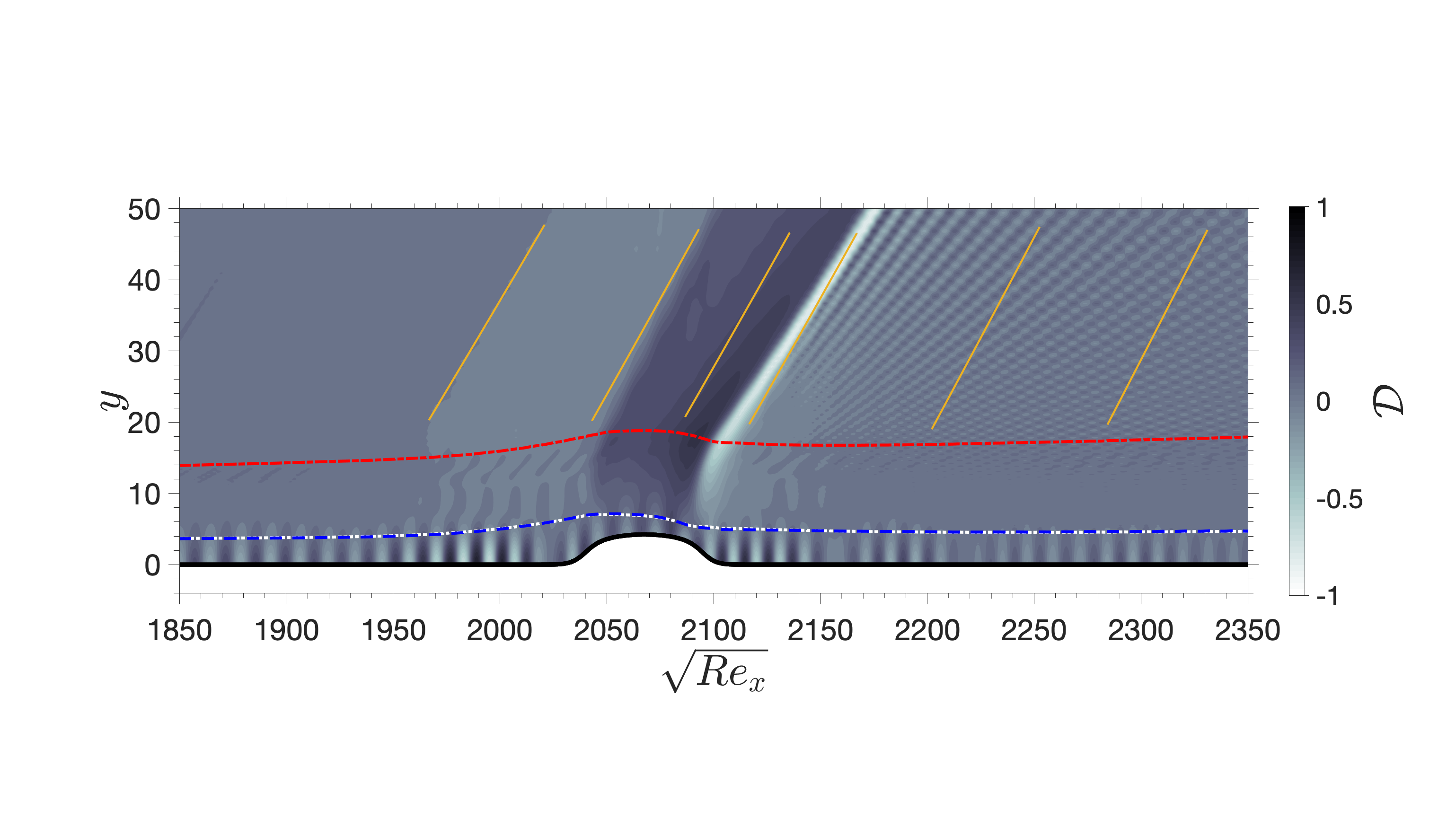}%
}%
\caption{Contours of $\mathcal{D}$ defined by equation (\ref{Eq:NumericalSchlieren}). Dark colors are expanded regions ($\mathcal{D} > 0$, $\nabla \cdot \boldsymbol{u} > 0$), and light colors mark compressed zones ($\mathcal{D} < 0$, $\nabla \cdot \boldsymbol{u} < 0$). Red dashed-dotted line is the boundary-layer edge, $\delta_{99}$; yellow lines are the local Mach lines; and white dotted and blue dashed lines mark the relative sonic line, where $c - \bar{u} = a$, for modes $\left< 100 , 0 \right>$ and $\left< 110 , 0 \right>$, respectively.}
\label{FIG:X2OptBump_midPlane_Schlieren}
\end{figure}

\begin{figure}
\centerline{%
\includegraphics[trim=90 190 170 190, clip,width=0.99\textwidth] {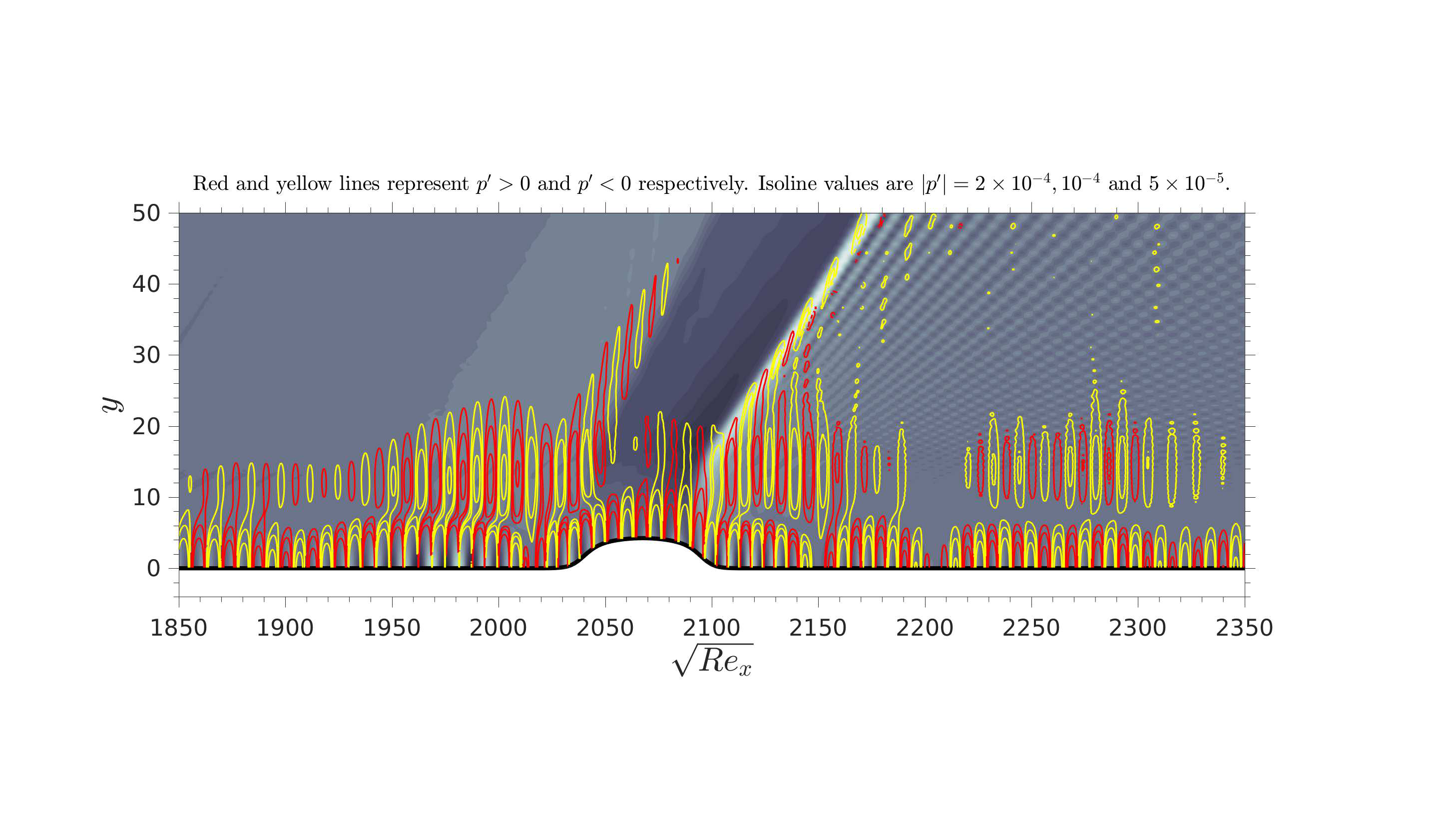}%
\put(-385.0,133.0){$(a)$}
\put(-338.0,55.0){$\overbrace{}^{\textrm{t1U}}$}
\put(-341.0,24.0){$\underbrace{}_{\textrm{t1L}}$}
\put(-220.0,105.0){$\nwarrow$}
\put(-350.0,115.0){$p^{\prime}$ propagating into the free stream}
}%
\centerline{%
\includegraphics[trim=90 190 170 190, clip,width=0.99\textwidth] {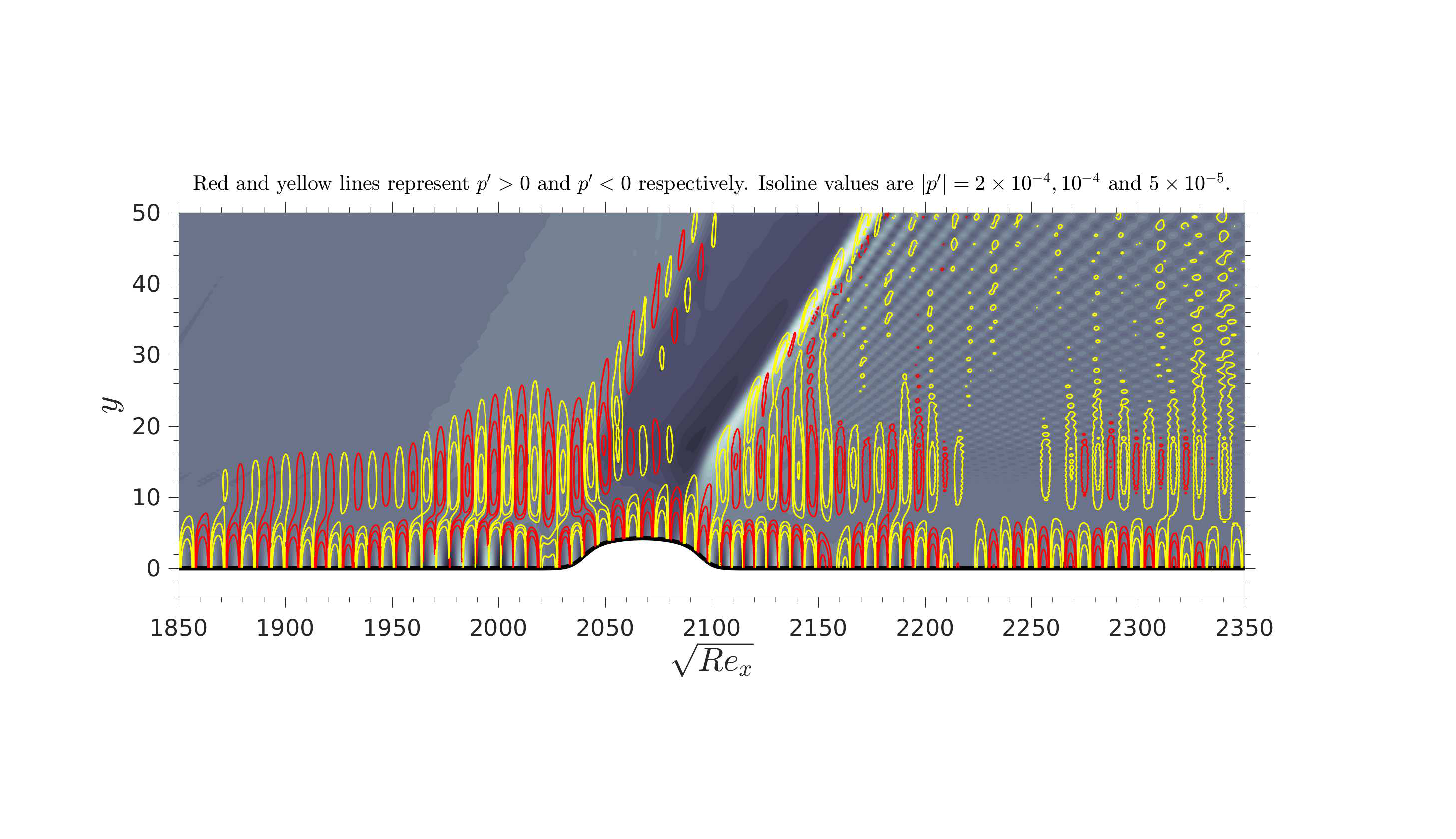}%
\put(-385.0,133.0){$(b)$}
\put(-238.0,79.0){$\overbrace{}^{\textrm{t2U}}$}
\put(-243.0,24.0){$\underbrace{}_{\textrm{t2L}}$}
\put(-219.0,107.0){$\nwarrow$}
\put(-350.0,115.0){$p^{\prime}$ propagating into the free stream}
}%
\centerline{%
\includegraphics[trim=90 150 170 190, clip,width=0.99\textwidth] {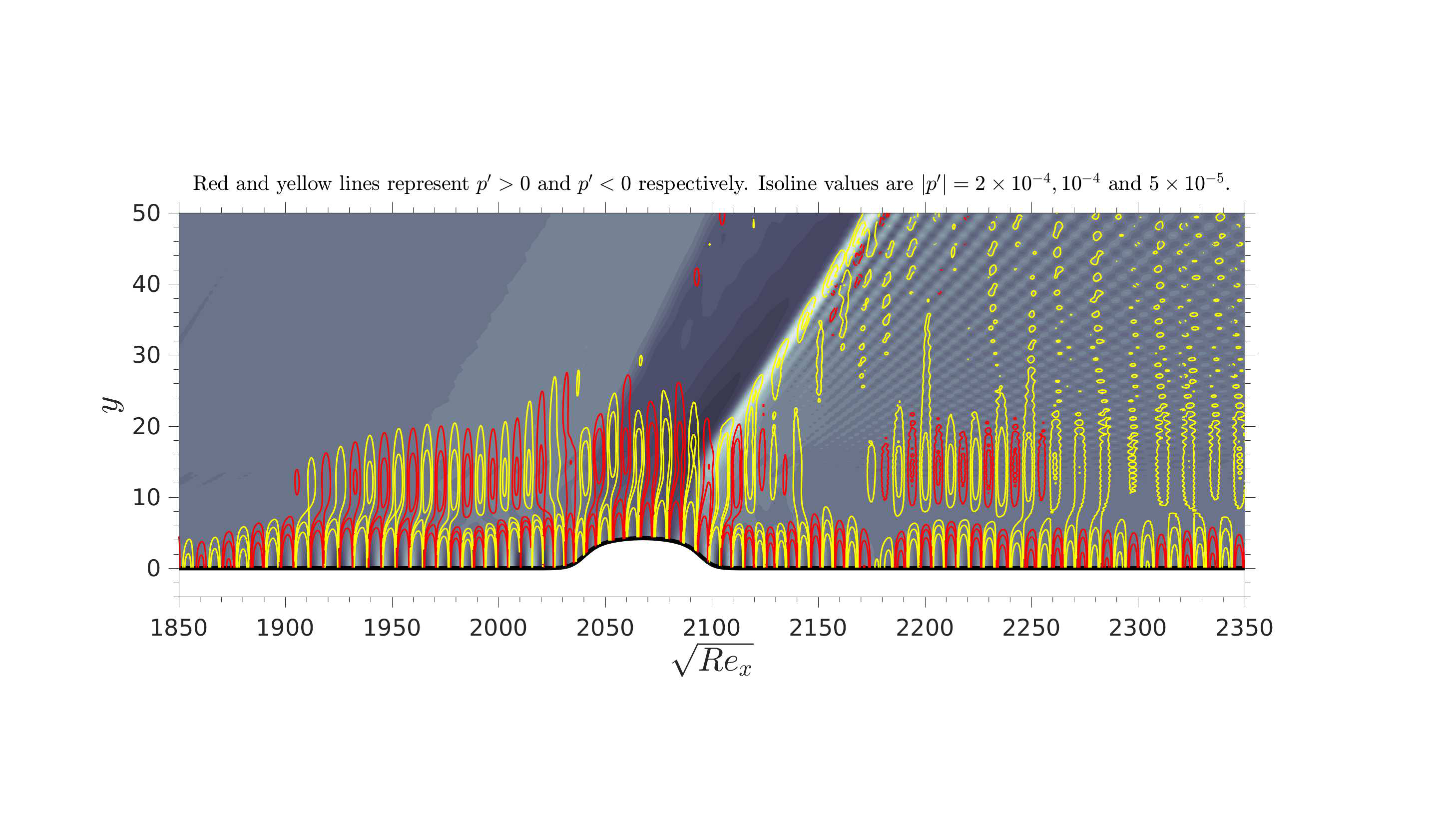}%
\put(-385.0,147.0){$(c)$}
\put(-206.0,90.0){$\overbrace{}^{\textrm{t3U}}$}
\put(-206.0,46.0){$\underbrace{}_{\textrm{t3L}}$}
}%
\caption{Snapshots of $\mathcal{D}$ and pressure fluctuations, $p^{\prime}$, at $(a)$ $t = 0$, $(b)$ $t = 418.6$ and $(c)$ $t = 575.4$. Red and yellow lines are $p^{\prime} > 0$ and $p^{\prime} < 0$ respectively. Pressure fluctuations iso-lines are $\{0.5, 1, 2\} \times 10^{-4}$ (from outer to inner lines).}
\label{FIG:X2OptBump_midPlane_SchlierenPressure}
\end{figure}
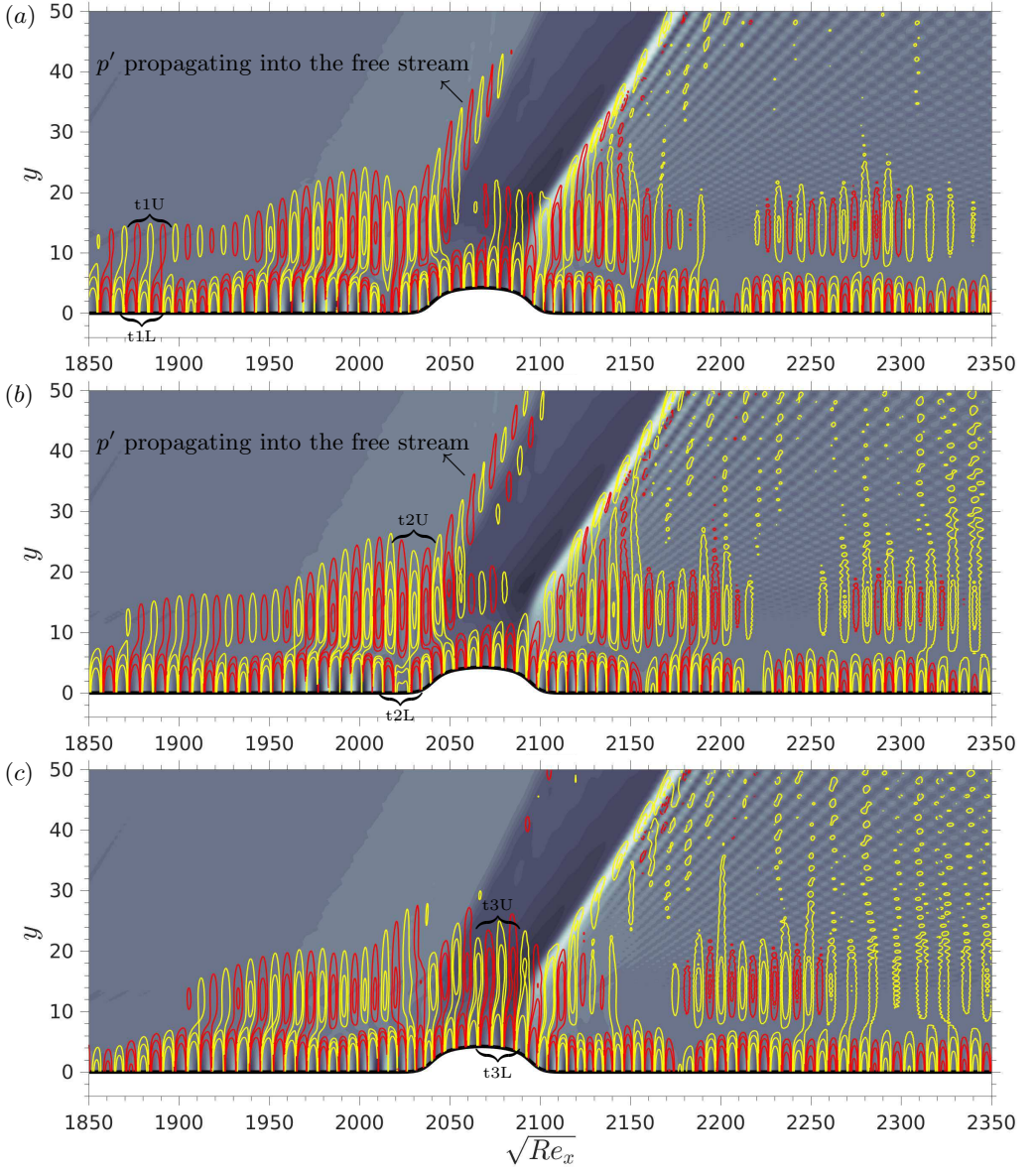

Figures \ref{FIG:X2OptBump_midPlane_Schlieren} and \ref{FIG:X2OptBump_midPlane_SchlierenPressure} provide flow visualizations in order to aid the physical interpretation of how the roughness modifies the near-wall region and alters the stability behaviour of the second modes.
These figures correspond to the simulation of the boundary layer over the optimal roughness after the final iteration of the EnVar procedure.
Figure \ref{FIG:X2OptBump_midPlane_Schlieren} shows contours of 
\begin{equation}
    \mathcal{D} = \tanh{ \left( \xi_0 \nabla \cdot \boldsymbol{u} \right) } ,
\label{Eq:NumericalSchlieren}
\end{equation}
where $\xi_0$ is a constant that adjusts the background color. A few additional lines are marked on the figure:  The boundary-layer edge is identified as the location where $\bar{u} = 0.99 u_\infty$;  The relative sonic lines are defined as the heights at which $c - \bar{u} = a$, where $c$ is the phase speed of either mode $\left< 100 , 0 \right>$ or ${\left< 110 , 0 \right>}$.
Below the relative sonic line, the alternating compressing and expanding acoustic waves travel supersonically relative to the mean flow.
Inviscid, linear perturbation theory predicts that the effect of the wall roughness propagates to infinity with constant strength along the lines $x - [M^2_\infty - 1]^{{1}/{2}} y =  \textrm{constant}$, which represent the local Mach lines \citep{Liepmann2001elements}.
In figure \ref{FIG:X2OptBump_midPlane_Schlieren}, several Mach lines are plotted with slope $[M_{y = 35}^2 - 1]^{-{1}/{2}}$ where $M_{y = 35}$ is the Mach number at $y = 35$.  These lines agree with the prediction from the theory, and qualitatively capture the free-stream compression zone upstream the roughness element. On top of the roughness, the slope of the compression zone progressively deviates from the theory, and is followed by an expansion zone.  
The deviation from theory is expected in light of its restrictive assumptions, for example it assumes that $ \theta_{max} \sqrt{M_\infty^2 - 1} \ll 1$ where $\theta_{max}$ is the maximum inclination angle of the roughness \citep{Liepmann2001elements}, while the geometrical features of the optimal roughness in our simulation yield $ \theta_{max} \sqrt{M_\infty^2 - 1} = 0.5 H_r L_r \sqrt{M_\infty^2 - 1} \approx 0.6 $.
Figure \ref{FIG:X2OptBump_midPlane_Schlieren} also shows that, in the zone upstream of the roughness, the near-wall relative supersonic region becomes thicker whereas above the protrusion this region becomes thinner.
The trapped acoustic waves reflect back and forth at the wall and the relative sonic line. At the latter location, the waves change from compression to expansion and vice versa.
As the height of the relative supersonic zone changes, the periods between consecutive reflections at the relative sonic line change, which disrupts the amplification of the instability mode.

Figure \ref{FIG:X2OptBump_midPlane_SchlierenPressure} shows three instantaneous fields, which highlight the pressure fluctuations, $p^\prime = p - \bar{p}$ inside the boundary layer.
While the visualized $p^\prime$ is the outcome of a superposition of different instability waves, rather than a single one, the figure is helpful in explaining the behaviour of the instabilities in response to the roughness element.
Two separate regions in the wall-normal profile of $p^\prime$ can be identified in this figure:
(i) near-wall (below the relative sonic line) $p^\prime$ contours are caused by the acoustic component of the instability waves and (ii) $p^\prime$ above the relative sonic line where the vortical and thermal components of the instability wave are prominent.
Panels $(a)$ to $(c)$ track a series of compression-expansion-compression in $p^\prime$, and the three figures correspond to three time instances, \{t1, t2, t3\}.   As we discuss the wall-normal profile of $p^\prime$, our focus will be on the phase relation below (region denoted $L$) and above (region denoted $U$) the relative sonic line.
At t1 (panel $(a)$), the the wall-normal profiles of $p'$ as we traverse from t1L to t1U have have a fixed phase relation that is initially preserved with downstream distance. In this configuration, they are phase-tuned. 
At time t2 (panel $(b)$), the $p^\prime$ packet has traveled to the marked location, and the change in phase along the wall-normal profile of $p'$ across the sonic line has changed appreciably, by approximately $180^\circ$. We will refer to this process as phase-detuning.  
At time t3 (panel $(c)$), the wall-normal profile of $p'$, crossing from t3L to t3U, recovers the original phase relation.
Another interesting observation from panels $(a)$ and $(b)$ is the propagation of pressure fluctuations, $p^\prime$, along the Mach lines emanating above the roughness, into the free stream.  This process is absent at t3 (panel $(c)$), and has a low normalized frequency on the order of $F=10$.
The most energetic mode at this frequency is $\left< 10 , 0 \right>$, which is generated by non-linear interaction of modes $\left< 110 , 0 \right>$ and $\left< 100 , 0 \right>$, and whose wavelength is
$\lambda_{\left< 10 , 0 \right>} / 2 W_r  = 2.5$ relative to the roughness streamwise extent.

\begin{figure}
\centerline{%
\includegraphics[trim=0 0 0 0, clip,width=0.99\textwidth] {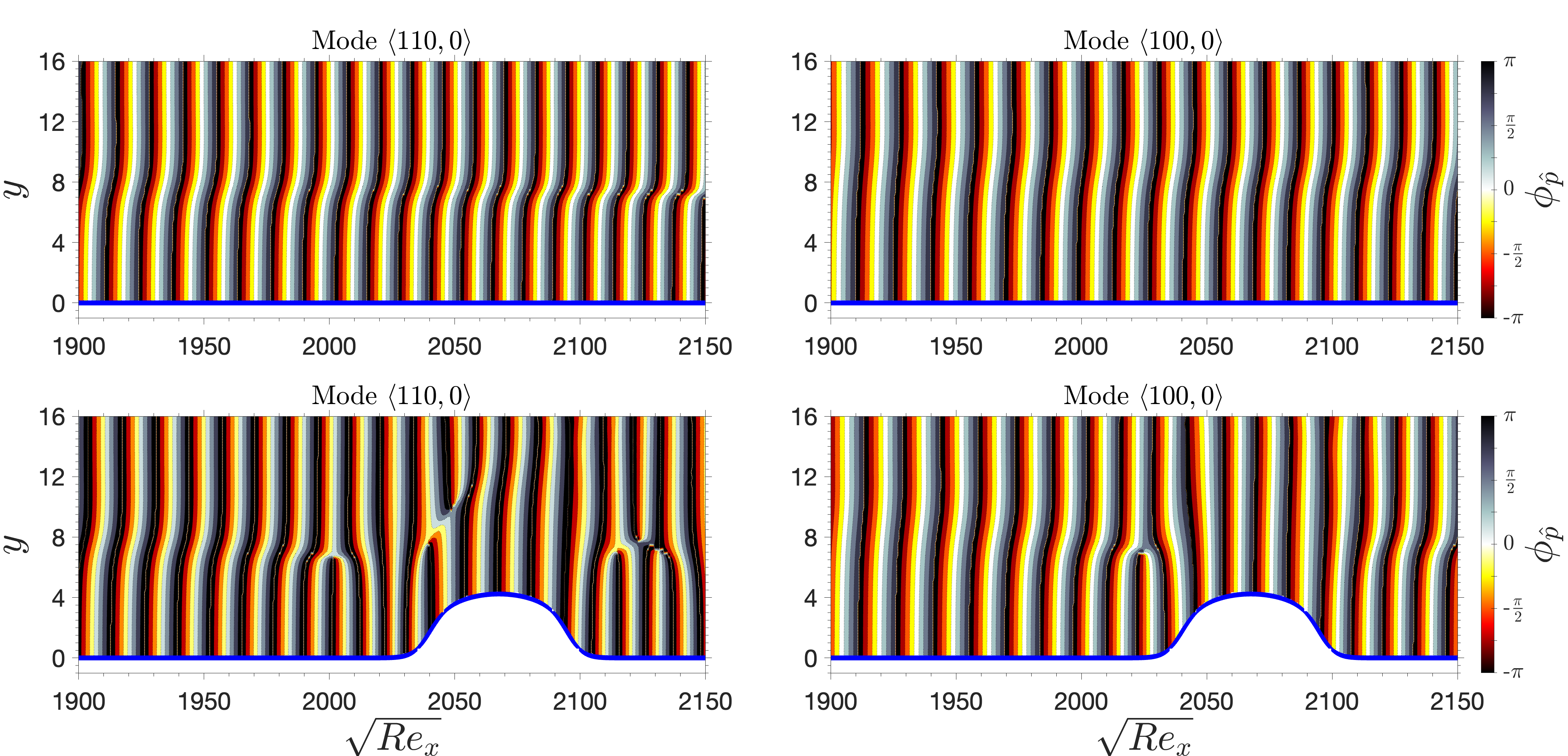}%
\put(-385.0,165.0){$(a)$}
\put(-202.0,165.0){$(b)$}
\put(-385.0,78.0){$(c)$}
\put(-202.0,78.0){$(d)$}
}%
\vspace{2.0mm}
\centerline{%
\includegraphics[trim=0 430 0 0, clip,width=0.99\textwidth] {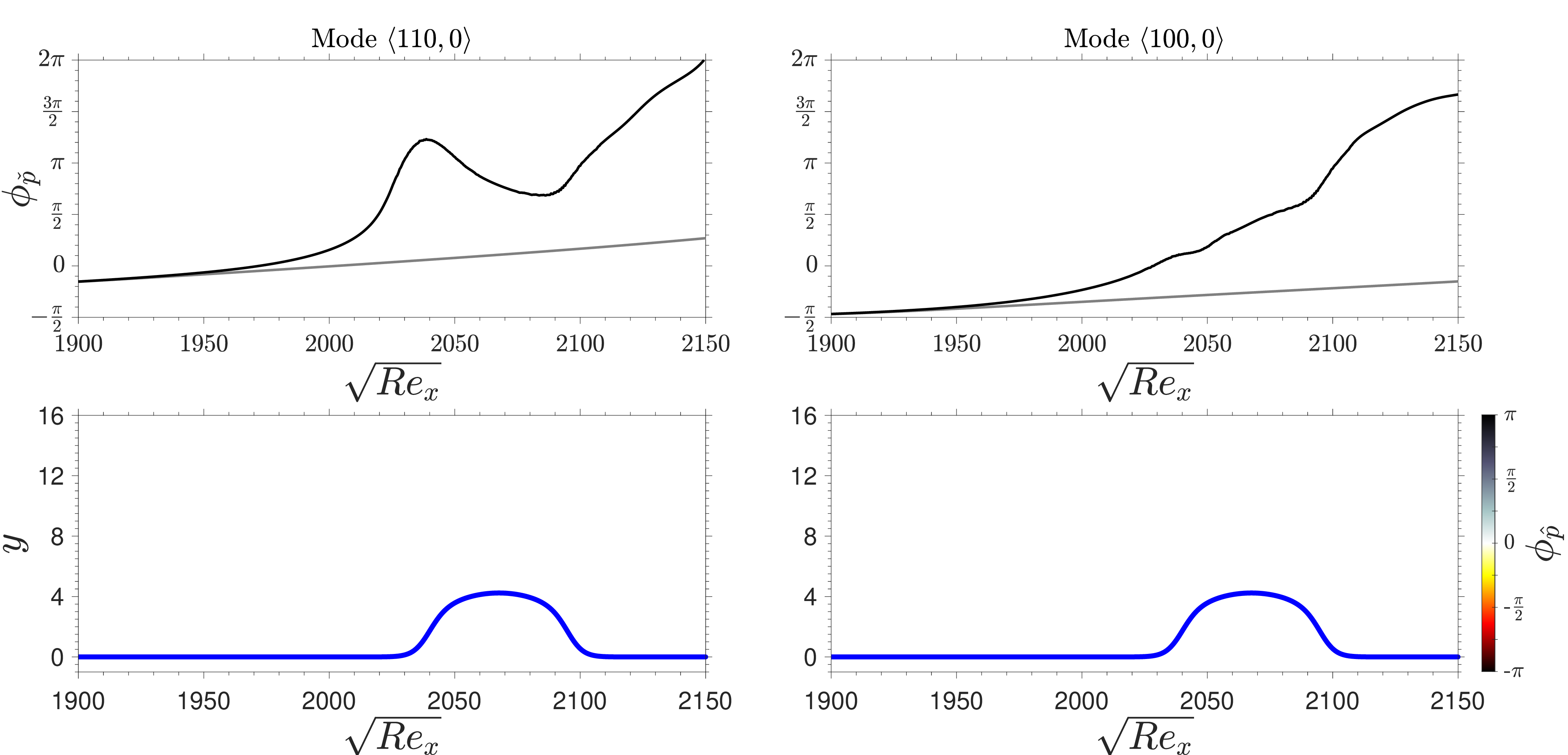}%
\put(-385.0,81.0){$(e)$}
\put(-202.0,81.0){$(f)$}
}%
\caption{($a$-$d$) Contours of the phase of the Fourier representation of pressure, $\phi_{\hat{p}}$: ($a$,$b$) Reference flat-plate configuration; ($c$,$d$) optimized-roughness case. $(e,f)$ Phase of mode-shape of pressure signal, $\phi_{\check{p}}$, at the wall; Gray lines are reference curves for flat-plate case, while the black lines are the results of the boundary layer over the optimal roughness. The left panels show mode $\left< 110 , 0 \right>$ and the right panels show mode $\left< 100 , 0 \right>$.}
\label{FIG:X2OptBump_PressurePhase_2ndModes}
\end{figure}

We now return to the change of phase observed in the near-wall pressure fields in figure \ref{FIG:X2OptBump_midPlane_SchlierenPressure}.  While the interpretation in terms of detuning of the individual instability waves
is plausible, the same visual pattern can be caused by other effects, for example dispersion of the waves that comprise the plotted fluctuating pressure field. 
In order to provide a more precise assessment of the influence of the roughness on the key instabilities, 
we perform a Fourier transform of the pressure signal in time and the span,
\begin{equation}
    p\left(\boldsymbol{x},t\right) = \sum_{F,k_z} \hat{p}_{\left< F , k_z \right>}(x,y) \; \exp\left[{-\textrm{i} \left( \frac{\sqrt{Re_{x_0}} F }{10^6} t + \frac{2\pi k_z}{L_z} z \right) }\right],
\label{eq:PFourierRep}
\end{equation}
where $\hat{p} ( x , y )$ is the complex Fourier coefficient whose phase we will denote as $\phi_{\hat{p}}$. Inspired by stability theory, we introduce the ansatz,  
\begin{equation}
    \hat{p}_{\left< F , k_z \right>} \equiv \check{p}_{\left< F , k_z \right>} \; \exp \left( \int_{x_0}^x \alpha_{\left< F , k_z \right>} dx \right), 
\label{eq:PModalRep}
\end{equation}
where $\alpha (x) = \alpha_r + i \alpha_i$ is a complex streamwise wavenumber, which is evaluated from, 
\begin{equation}
    \alpha_{\left< F , k_z \right>} = \frac{0.5}{ \mathcal{E}_{\left< F , k_z \right>} } \int_{0}^{L_y} \left[ \overline{\rho}  \biggl\{ \boldsymbol{\hat{u}}^* \frac{\partial \boldsymbol{\hat{u}}}{\partial x} \biggl\} + \frac{\overline{\rho} \overline{T}}{(\gamma-1) \gamma M_\infty^2} \Biggl\{ \frac{ \hat{\rho}^* }{\overline{\rho}^2} \frac{\partial \hat{\rho}}{\partial x} + \frac{ \hat{T}^* }{\overline{T}^2} \frac{\partial \hat{T}}{\partial x} \Biggl\} \right]_{\left< F , k_z \right>} dy.
\end{equation}
The mode shape is therefore $\check{p}(x,y)$, and its phase will be denoted $\phi_{\check{p}}$.

The phases of instability modes $\left< 110 , 0 \right>$ and $\left< 100 , 0 \right>$ are reported in figure \ref{FIG:X2OptBump_PressurePhase_2ndModes}.  The contour plots contrast the behaviour of $\phi_{\hat{p}}$ in the reference flat-plate configuration (figures \ref{FIG:X2OptBump_PressurePhase_2ndModes}$a$,$b$) and as the instability waves approach the optimal roughness (figures \ref{FIG:X2OptBump_PressurePhase_2ndModes}$c$,$d$). The results highlight the distinction between the modal pressure fluctuations that are below and above the relative sonic line, the former travelling supersonically relative to the mean flow.
In panels $(a)$ and $(b)$, the phase change across the relative sonic line is maintained at a nearly constant value along the flat plate in the shown region in the figures.
In contrast, panels $(c)$ and $(d)$ clearly show that the relative phase across the sonic line is disrupted as each instability wave approaches the roughness.  Most importantly, the phases of the pressure modes in the trapped superspnic regions, below the relative-sonic lines, are rapidly altered.
For mode $\left< 110 , 0 \right>$ in panel $(c)$, this change occurs at $\sqrt{Re_x} \approx 2000$ which, according to figure \ref{FIG:EnergySpectralDensity}$(c)$, is the location where the amplification of this instability wave is abruptly halted and it starts to decay.
Similarly for mode $\left< 100 , 0 \right>$ in panel $(d)$, detuning of the pressure fluctuations takes place at $\sqrt{Re_x} \approx 2020$, which also corresponds to the location where this instability wave reaches a local maximum amplitude after which it decays in figure \ref{FIG:EnergySpectralDensity}$(d)$.
Figures \ref{FIG:X2OptBump_PressurePhase_2ndModes}($e$,$f$) provide a more quantitative picture.  We report the phases of the pressure in the modal representation (\ref{eq:PModalRep}), or $\phi_{\check{p}}$, along the wall.  The influence of the roughness relative to the flat-pate case is evident, as well as the correlation between the changes in the phases and in the modal spectra from figures \ref{FIG:EnergySpectralDensity}($c$,$d$).

The results presented in \S\ref{sec:OptimalRoughness} support our original physical argument regarding the effect of roughness on the stability behaviour of second-mode instabilities.  As the thickness of the near-wall,  relative supersonic region changes abruptly, the instability waves are altered: their growth rate is modulated, the phase of their acoustic component is ``scrambled", and their net amplification is reduced relative to the flow over a flat plate. As a result, the nonlinear interactions among key instability waves, downstream of the optimal roughness, are effectively weakened and transition to turbulence is nearly suppressed.


\section{Summary}
\label{sec:Summary}

The capacity of isolated roughness to delay laminar-to-turbulent transition in a Mach-4.5 boundary layer is examined.  The study is conducted using direct numerical simulations of boundary layers over flat plate, with isolated roughness elements that have various geometrical parameters mounted on the surface.  
The oncoming disturbance that interacts with the boundary layer in the simulations corresponds to the nonlinearly most dangerous disturbance for the prescribed energy level \citep{Jahanbakhshi2019}, and is comprised primarily of two planar second-mode instabilities and an oblique first-mode instability.

The roughness shape was parameterized to provide control over the location, streamwise wave-number, height, width and abruptness of the roughness.  The influence of these parameters was examined, in particular their impact on the location of laminar-to-turbulence transition relative to the reference smooth-wall configuration.
The effects of location and streamwise wave-number were explored first, and provided the initial estimate of a roughness design that was adopted in subsequent optimization where the height, width and abruptness of the roughness were optimized to achieve maximum transition delay in our computational domain.
The constrained optimization was performed using an ensemble-variational approach, in which a cost function is defined in terms of the skin-friction coefficient and is minimized to ensure the latest possible transition to turbulence.
The optimal roughness that was identified was able to maintain a laminar state throughout the computational domain.

The roughness elements minimally affected the first-mode instability that was part of the inflow disturbance, while the second-mode waves and their nonlinear interactions were appreciably attenuated.  
Whether the second-mode waves were stabilized or destabilized depended on the location and geometrical features of the roughness. Specifically, stability is affected by  
(i) the relative position of the roughness and the synchronization of the slow and fast modes and (ii) the height of the near-wall region where the second-mode instability waves travel supersonically relative to the mean flow.
Pre-synchronization, altering the supersonic region by a protruding roughness destabilizes the second-mode waves, while post-synchronization the net effect is stabilizing.
The change in stability is due to the shift in the phase of the trapped acoustic waves relative to the harmonic vortical and thermal waves that are beyond the relative sonic line.  
The net outcome is an effective delay of the instability growth, mitigation of nonlinear interaction and ultimately transition delay.  The optimized roughness was more slender, slightly taller and more abrupt than the initial design. This roughness was more effective, entirely eliminating transition from the computational domain.

Depending on their shape and locations, roughness elements can have stabilizing, destabilizing, or neutral net effect on the amplification of instability waves. As a result, the impact on the nonlinear stages of transition and the onset of turbulence are difficult to anticipate.
Our EnVar framework provides objective guarantees that the optimized roughness delays transition to turbulence at design conditions, and we did not observe any undesirable or unexpected destabilization of originally benign disturbances. 
The optimized roughness does not, however, guarantee transition delay for other disturbances, away from the design conditions, e.g.\,since new transition mechanisms may be active. For the purpose of this work, we tested a broadband inflow spectrum of instability waves at sufficiently high amplitude to trigger transition within the computational domain over the flat plate (see Appendix \ref{Appendix_B}), and still observed transition delay when the optimal roughness from \S\ref{sec:OptimalRoughness} was introduced.  Despite this additional test, there remains no guarantee that transition would be delayed for different inflow disturbance spectra.  Essentially, the optimized roughness should not be interpreted as optimal for all inflow disturbances, but rather for the inflow condition that was adopted in the optimization procedure.
By considering the nonlinearly most dangerous disturbance, our goal was to demonstrate that the optimization of the roughness can be effective, even in the most aggressive scenario.


\par\bigskip
\noindent
\textbf{Acknowledgements.} The authors are grateful to Professor J. Larsson for sharing the \textit{Hybrid} code.

\par\bigskip
\noindent
\textbf{Funding.}  This work was supported in part by the US Air Force Office of Scientific Research (grant no.\,FA9550-19-1-0230) and by the Office of Naval Research (grants no.\,N00014-17-1-2339, N00014-21-1-2148). Computational resources were provided in part by the Maryland Advanced Research Computing Center (MARCC) and the AI.Panther HPC cluster at Florida Institute of Technology (funded by the National Science Foundation major Research Implementation grant no. MRI-2016818).

\par\bigskip
\noindent
\textbf{Declaration of interests.} 
The authors report no conflict of interest.

\par\bigskip
\noindent
\textbf{Author ORCIDs.} \\
Reza Jahanbakhshi \url{https://orcid.org/0000-0001-7477-8892} \\
Tamer A. Zaki, \url{https://orcid.org/0000-0002-1979-7748}


\appendix

\section{Flow over complex geometries}
\label{Appendix_A}

This appendix provides a brief description of the cut-stencil method implemented in the code \textit{Hybrid} \citep{Johnsen2010} and a validation test.
As shown in the schematic of figure \ref{FIG:Figure0}, the cut-stencil is a sharp interface method where the discretization of the governing equations is modified near the solid body in order to explicitly enforce the no-slip boundary conditions at the fluid-solid interface. The precise location for the solid/fluid interface is adopted in order to ensure local, and therefore also global, conservation. 
The present implementation is similar to that by \citet{Greene2016}.
Therefore, herein, we introduce the cut-stencil method very briefly and refer the readers to \citet{Greene2016} for a more detailed discussion of the method.
We then provide a sample validation study by comparing the results from our code with the published data for a flow simulation that is relevant to the topic of current work.

\subsection{A brief description of the cut-stencil}

\begin{figure}
\centerline{%
\includegraphics[trim=0 0 0 0, clip,width=0.60\textwidth] {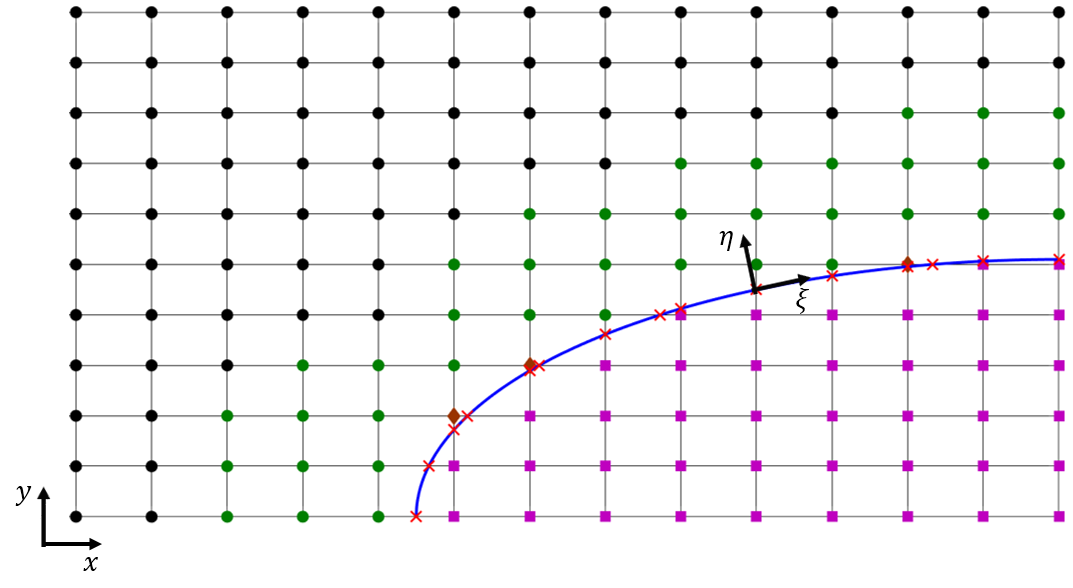}%
\hspace{3.00mm}
\includegraphics[trim=0 -50 0 0, clip,width=0.40\textwidth] {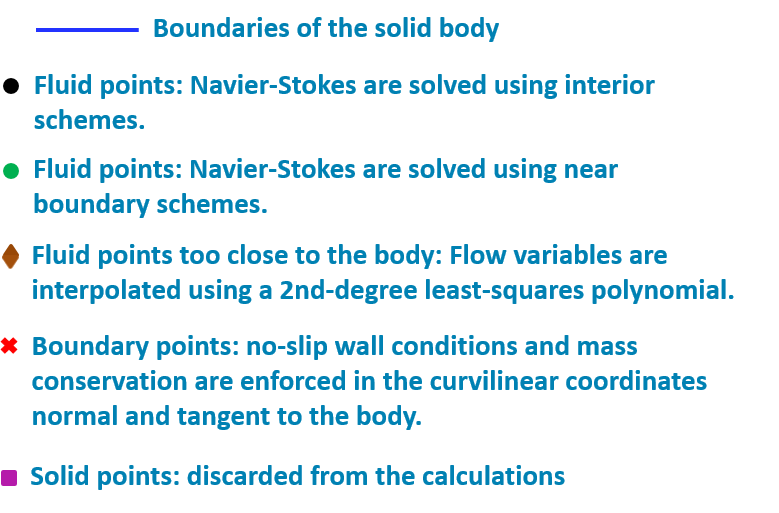}%
\put(-395.0,120.0){$(a)$}
\put(-160.0,120.0){$(b)$}
}%
\caption{$(a)$ Schematic of the grid near the immersed body. $(a)$ Point classification for the cut-stencil implementation in code \textit{Hybrid}.}
\label{FIG:Figure0}
\end{figure}

\begin{figure}
\centerline{%
\includegraphics[trim=80 230 50 220, clip,width=0.999\textwidth] {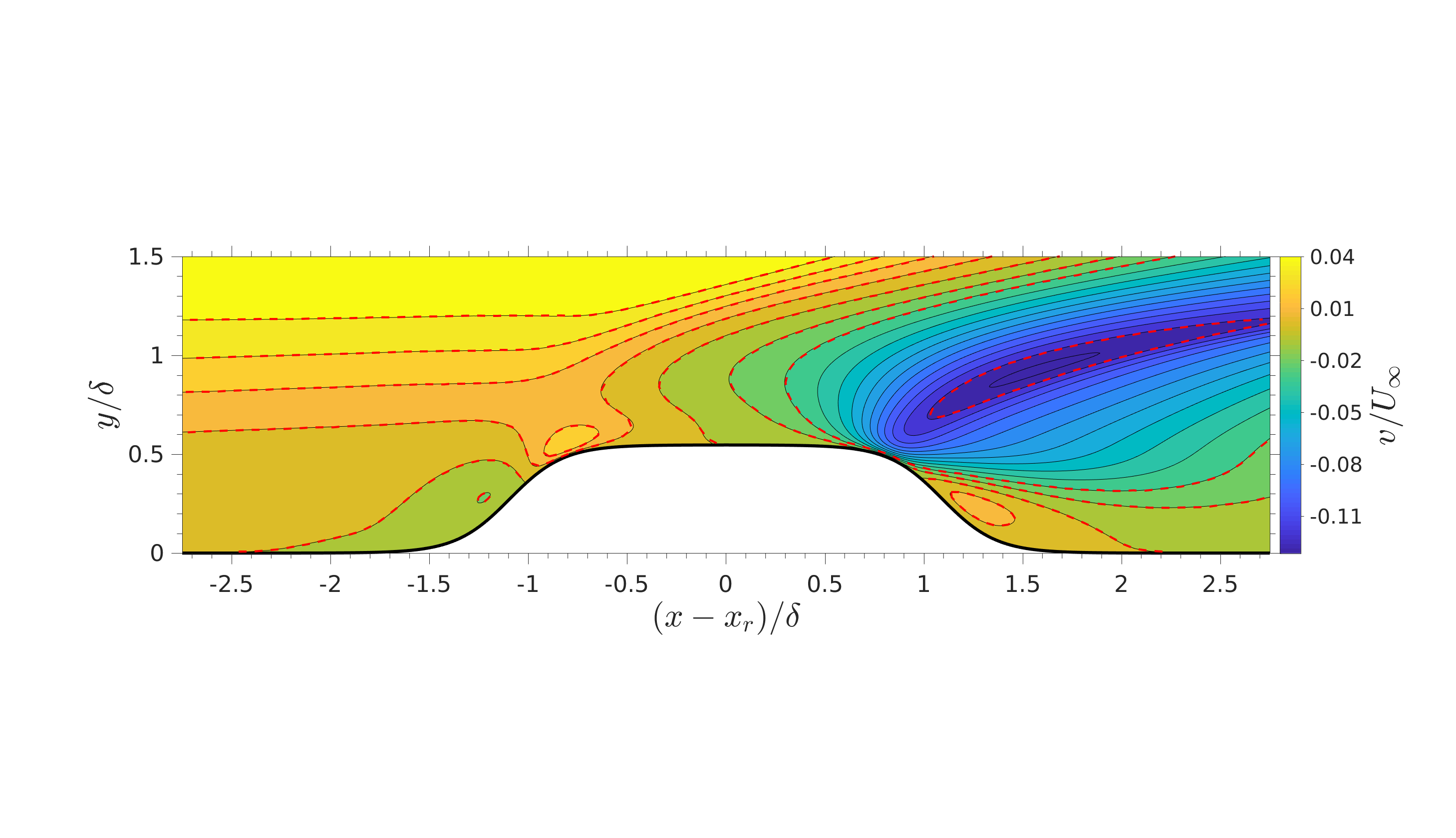}%
\put(-390.0,95.0){$(a)$}
}%
\centerline{%
\includegraphics[trim=80 230 50 220, clip,width=0.999\textwidth] {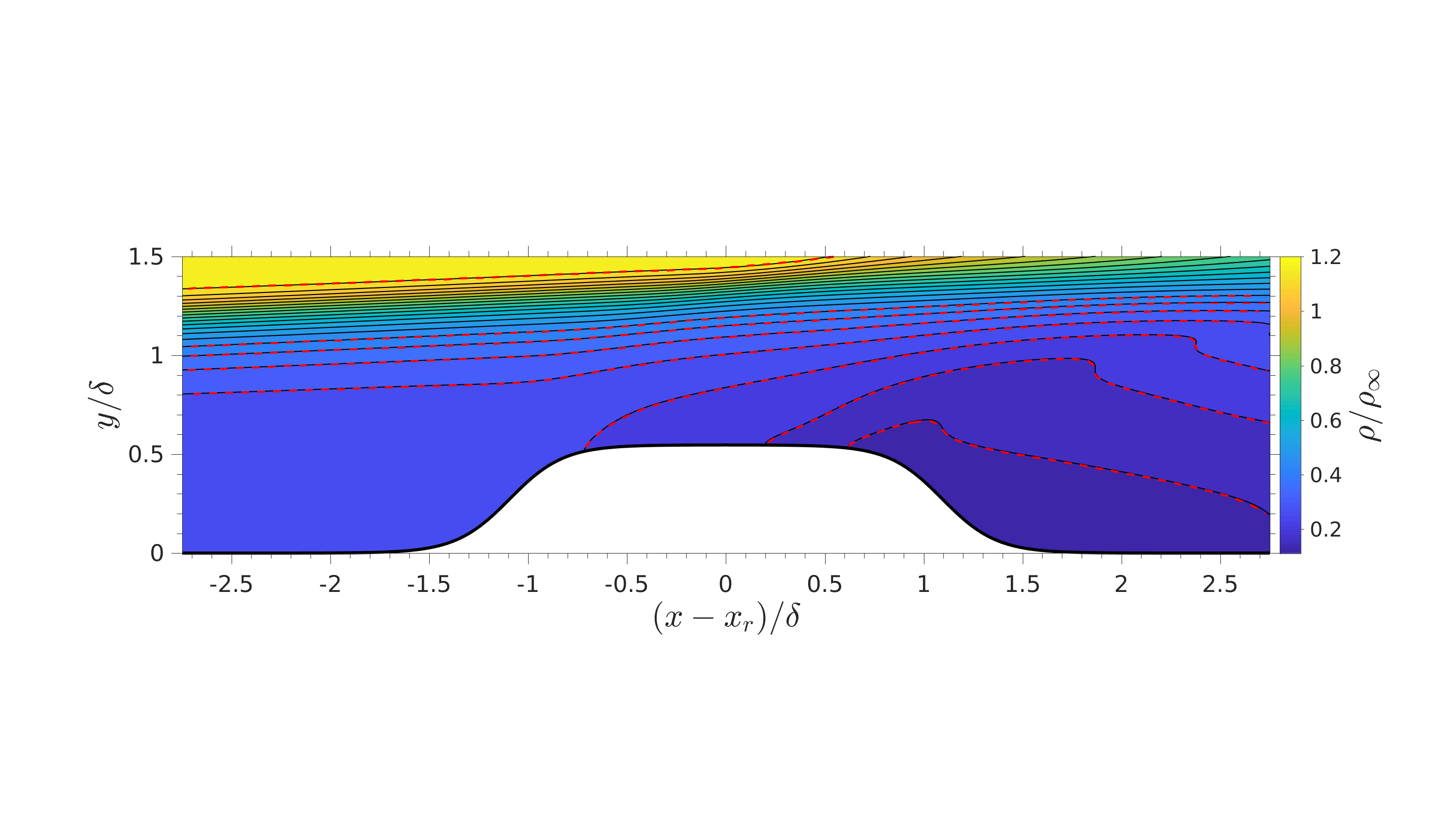}%
\put(-390.0,95.0){$(b)$}
}%
\centerline{%
\includegraphics[trim=80 190 50 220, clip,width=0.999\textwidth] {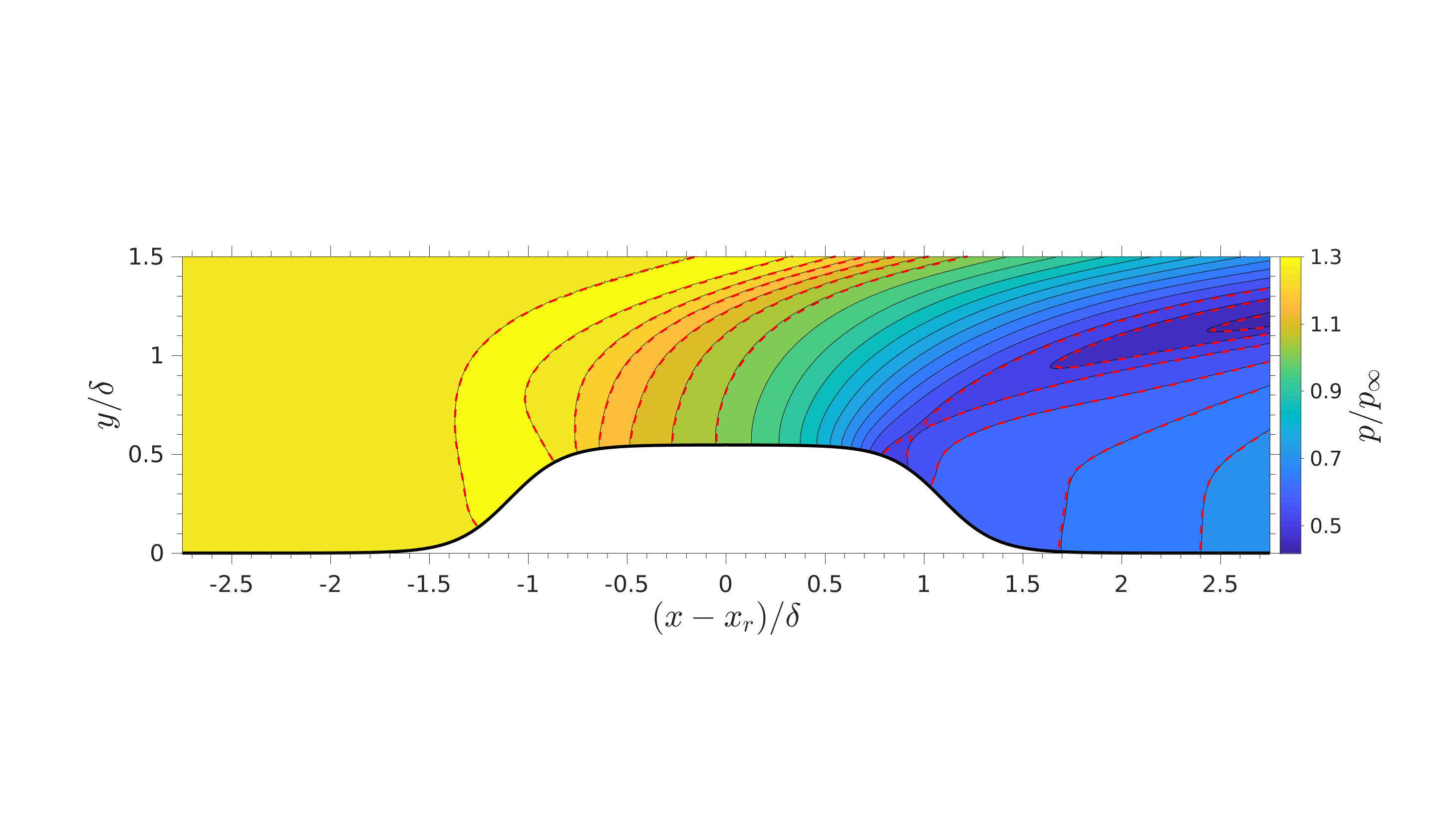}%
\put(-390.0,107.0){$(c)$}
}%
\caption{Contours of $(a)$ wall-normal velocity, $(b)$ density, and $(c)$ pressure. The contour colors and thin gray lines correspond to the current simulation and red-dashed lines are the published results by \citet{Greene2016}.}
\label{FIG:ValidationContoursLines}
\end{figure}

A two-dimensional schematic of the implementation is shown in figure \ref{FIG:Figure0}.
All grid points inside the solid body are discarded, and the cut-cell discretization generates four classes of fluid points that are treated separately: 
\begin{enumerate}
    \item For interior fluid grid points that are far from the body, the Navier-Stokes equations are discretized using the interior schemes.
    \item For the first three fluid grid points near the solid body, sixth-, fourth- and second-order schemes are used to discretize the Navier-Stokes on these points.
    \item Points that are denoted as irregular fluid points in the figure are excluded from the computations along the directions where they reside too close to the solid (within 25\% of the local grid spacing). However, in the other directions the points may be included in the computations of other neighbouring fluid points. Therefore, flow variables are required at these points and are computed via an interpolation scheme from their neighbors including the boundary points.
    \item Boundary points are locations where the body surface intersects the grid, and where the flow variables are also needed. In current implementation, the velocity of the immersed object is computed using no-slip wall condition and the temperature is computed from the thermal boundary condition. The remaining parameters are the velocity gradients and the pressure (or density). Velocity gradients are computed using a fourth-order one-side finite-difference scheme in the direction normal the solid body; The pressure is extrapolated from the neighbouring fluid points, and the equation of state for an ideal gas is used to computed density.
\end{enumerate}

The solver requires interpolation/extrapolation of flow quantities onto the irregular and boundary points, and points within the fluid domain that are not on the grid.
For example, for computing $\partial F / \partial \eta$ at the boundary points, the value of $F$ is required at four points equally spaced from the surface in the direction of $\eta$.
Similar to \citet{Greene2016}, the interpolation/extrapolation function is a second-degree polynomial in $x$, $y$ and $z$, while the coefficients of the polynomial are computed using least-squares fitting of the neighbouring points.

\subsection{A sample validation study}

The current implementation of the cut-stencil algorithm has been validated extensively against published data. The comparisons included multiple configurations for two-dimensional and three-dimensional solid bodies, and a range of Mach numbers from 0.2 to 4.8. 
Here, we provide the results from the validation case most relevant to the present effort, namely a high-Mach-number boundary layer over a two-dimensional isolated roughness.

\begin{figure}
\centerline{%
\includegraphics[trim=0 0 0 0, clip,width=0.999\textwidth] {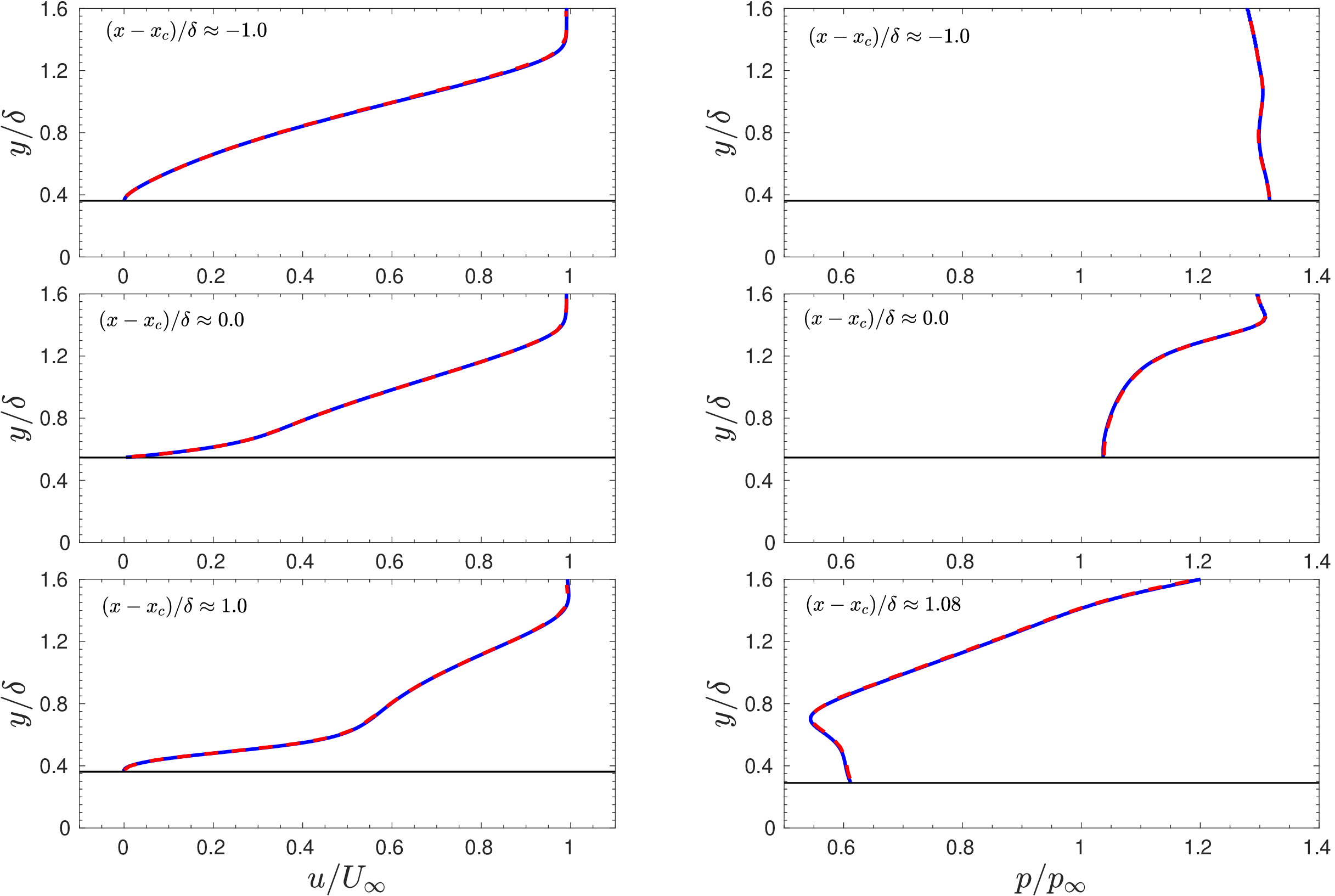}%
\put(-390.0,250.0){$(a)$}
\put(-185.0,250.0){$(b)$}
\put(-390.0,168.0){$(c)$}
\put(-185.0,168.0){$(d)$}
\put(-390.0,85.0){$(e)$}
\put(-185.0,85.0){$(f)$}
}%
\caption{$(a, c, e)$ Streamwise velocity and $(b, d, f)$ pressure profiles at three streamwise positions. Blue lines correspond to current simulation and red-dashed lines are the published results by \citet{Greene2016}. Horizontal black lines mark the height of roughness.}
\label{FIG:ValidationUP_Pointdata}
\end{figure}
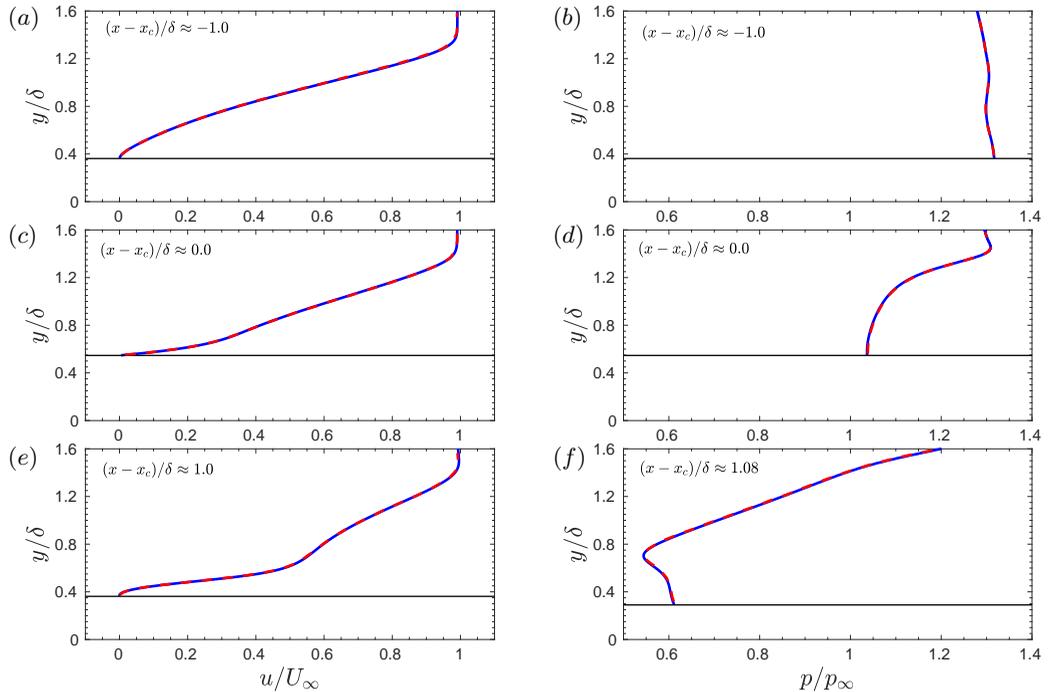

We compare results from our cut-stencil implementation to published data by \citet{Greene2016} from their body-fitted curvilinear solver.
The computational domain starts at $\sqrt{Re_{x_{0}}} = 762 $ and the free-stream Mach number is $Ma_{\infty} = 4.8$.
The Blasius length scale at $x_0$ and free-stream quantities are adopted as the reference scales.  
The geometry of the roughness is a smooth protrusion which is given by two hyperbolic tangent functions as
\begin{equation}
    h(x) = \frac{h_r}{2} \left[ \tanh\left( L_r [x - x_r + W_r] \right) - \tanh \left( L_r [x - x_r - W_r] \right)  \right] ,
\end{equation}
where $h_r = 13.115$, $x_r = 1961.30$, $L_r = 0.152497592 $, and $W_r = 26.23$ are, respectively, the non-dimensionalized height, center location, abruptness, and width.
More information about the simulation parameters can be found in table 4 by \citet{Greene2016}.

Comparisons of our simulation results and those by \citet{Greene2016} are provided in figures \ref{FIG:ValidationContoursLines} and \ref{FIG:ValidationUP_Pointdata}.
In these figures, $\delta \approx 24$ is the undisturbed boundary-layer thickness at $x_r$.
As demonstrated by figure \ref{FIG:ValidationContoursLines}, the velocity, density and pressure contours from the cut-stencil implementation and from the reference body-fitted simulation are indistinguishable. For a more detailed view, profiles of the streamwise velocity and pressure were extracted at three locations and are reported in figure \ref{FIG:ValidationUP_Pointdata}.  Agreement, in particular in the near-wall region, is evident, which verifies the accuracy of the implementation.

\section{Effect of optimized roughness on transition due to broadband spectrum}
\label{Appendix_B}

Additional simulations were carried out, where we tested the performance of the optimized roughness obtained in \S\ref{sec:OptimalRoughness} for an off-design condition.  Specifically, rather than considering the nonlinearly most dangerous inflow disturbance, we adopted a broadband inflow spectrum where the total disturbance energy was distributed equally among the 400 instability waves that span the frequency and wavenumber range in figure \ref{FIG:InflowSpectra}$a$, with randomly assigned phases.

\begin{figure}
\centerline{%
\includegraphics[trim=70 0 100 0, clip,width=0.999\textwidth] {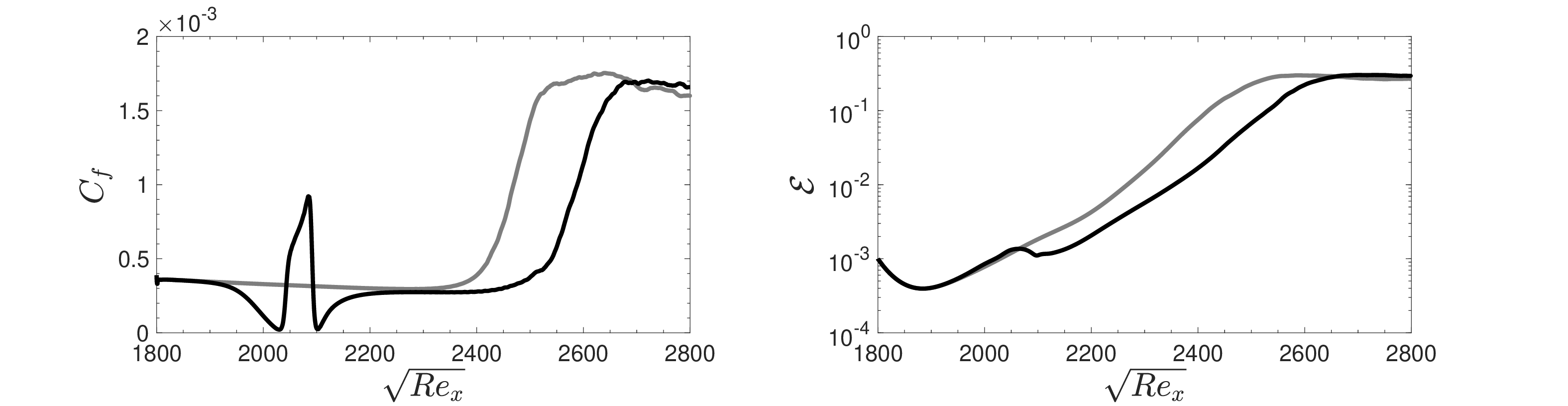}%
\put(-385.0,100.0){$(a)$}
\put(-190.0,100.0){$(b)$}
}%
\caption{Influence of optimal roughness on transition due to high-energy, broadband inflow disturbance spectrum: $(a)$ Skin-friction coefficient and $(b)$ total disturbance energy versus downstream Reynolds number. Gray lines are reference curves for flat-plate case; black lines are the results of the boundary layer over the optimal roughness.}
\label{FIG:BroadbandDisturbance}
\end{figure}

When the total energy of the inflow instability waves was the same as in the main body of this paper, i.e., $\sum_{F,k_z} \mathcal{E}_{\left< F , k_z \right>} = 2 \times 10^{-5}$, transition did not take place within the computational domain.  We therefore increased the inflow total disturbance energy by two orders of magnitude, $\sum_{F,k_z} \mathcal{E}_{\left< F , k_z \right>} = 10^{-3}$.  Under these conditions, transition takes place within the computational domain.  Results that compare flow over a flat plate and the optimized roughness from \S\ref{sec:OptimalRoughness} are shown in figure \ref{FIG:BroadbandDisturbance}, where we report the skin-friction coefficient and the downstream evolution of the total disturbance energy. The results demonstrate that the roughness remains stabilizing for this configuration, shifting the onset of turbulence downstream.  We stress, however, that the influence of the roughness could have been destabilizing since there is no performance guarantee when the inflow disturbance is far from the design condition. For example, the roughness may still promote transition at different (higher or lower) energy levels in case of broadband inflow spectrum, or in case of an entirely different inflow spectrum that interacts with the roughness in a destabilizing fashion that is not featured at design conditions.


\bibliographystyle{jfm}
\bibliography{MainDocument}

\begin{thebibliography}{58}
\expandafter\ifx\csname natexlab\endcsname\relax\def\natexlab#1{#1}\fi
\def\au#1{#1} \def\ed#1{#1} \def\yr#1{#1}\def\at#1{#1}\def\jt#1{\textit{#1}}
  \def\bt#1{#1}\def\bvol#1{\textbf{#1}} \def\vol#1{#1} \def\pg#1{#1}
  \def\publ#1{#1}\def\arxiv#1{#1}\def\org#1{#1}\def\st#1{\textit{#1}}

\bibitem[Bountin {\em et~al.\/}(2013)Bountin, Chimitov, Maslov, Novikov,
  Egorov, Fedorov \& Utyuzhnikov]{Bountin2013}
{\sc \au{Bountin, D.}, \au{Chimitov, T.}, \au{Maslov, A.}, \au{Novikov, A.},
  \au{Egorov, I.}, \au{Fedorov, A.} \& \au{Utyuzhnikov, S.}} \yr{2013}
  \at{Stabilization of a hypersonic boundary layer using a wavy surface}.
  \jt{AIAA J.}  \bvol{51}~(5),  \pg{1203--1210}.

\bibitem[Buchta {\em et~al.\/}(2022)Buchta, Laurence \& Zaki]{Buchta2022}
{\sc \au{Buchta, D.A.}, \au{Laurence, S.J.} \& \au{Zaki, T.A.}} \yr{2022}
  \at{Assimilation of wall-pressure measurements in high-speed flow over a
  cone}.  \jt{J. Fluid Mech}  \bvol{947},  \pg{R2}.

\bibitem[Buchta \& Zaki(2021)]{Buchta2021}
{\sc \au{Buchta, D.A.} \& \au{Zaki, T.A.}} \yr{2021}  \at{Observation-infused
  simulations of high-speed boundary-layer transition}.  \jt{J. Fluid Mech}
  \bvol{916},  \pg{A44}.

\bibitem[Casper {\em et~al.\/}(2016)Casper, Beresh, Henfling, Spillers, Pruett
  \& Schneider]{Casper2016}
{\sc \au{Casper, K.M.}, \au{Beresh, S.J.}, \au{Henfling, J.F.}, \au{Spillers,
  R.W.}, \au{Pruett, B.O.M.} \& \au{Schneider, S.P.}} \yr{2016}  \at{Hypersonic
  wind-tunnel measurements of boundary-layer transition on a slender cone}.
  \jt{AIAA J.}  \bvol{54}~(1),  \pg{1250--1263}.

\bibitem[Cheung \& Zaki(2010)]{Cheung2010}
{\sc \au{Cheung, Lawrence~C} \& \au{Zaki, Tamer~A}} \yr{2010}  \at{Linear and
  nonlinear instability waves in spatially developing two-phase mixing layers}.
   \jt{Phys. Fluids}  \bvol{22}~(5),  \pg{052103}.

\bibitem[Cowley \& Hall(1990)]{cowley1990instability}
{\sc \au{Cowley, Stephen} \& \au{Hall, Philip}} \yr{1990}  \at{On the
  instability of hypersonic flow past a wedge}.  \jt{Journal of Fluid
  Mechanics}  \bvol{214},  \pg{17--42}.

\bibitem[Dong {\em et~al.\/}(2020)Dong, Liu \& Wu]{Dong2020}
{\sc \au{Dong, M.}, \au{Liu, Y.} \& \au{Wu, X.}} \yr{2020}  \at{Receptivity of
  inviscid modes in supersonic boundary layers due to scattering of free-stream
  sound by localised wall roughness}.  \jt{J. Fluid Mech}  \bvol{896},
  \pg{A23}.

\bibitem[Dong \& Zhao(2021)]{Dong2021}
{\sc \au{Dong, M.} \& \au{Zhao, L.}} \yr{2021}  \at{An asymptotic theory of the
  roughness impact on inviscid {M}ack modes in supersonic/hypersonic boundary
  layers}.  \jt{J. Fluid Mech}  \bvol{913},  \pg{A22}.

\bibitem[Duan {\em et~al.\/}(2010)Duan, Wang \& Zhong]{Duan2010}
{\sc \au{Duan, L.}, \au{Wang, X.} \& \au{Zhong, X.}} \yr{2010}  \at{A
  high-order cut-cell method for numerical simulation of hypersonic
  boundary-layer instability with surface roughness}.  \jt{J. Comput. Phys.}
  \bvol{229}~(19),  \pg{7207--7237}.

\bibitem[Ducros {\em et~al.\/}(2000)Ducros, Laporte, Soul{\`e}res, Guinot,
  Moinat \& Caruelle]{Ducros2000}
{\sc \au{Ducros, F.}, \au{Laporte, F.}, \au{Soul{\`e}res, T.H.}, \au{Guinot,
  V.}, \au{Moinat, P.H.} \& \au{Caruelle, B.}} \yr{2000}  \at{High-order fluxes
  for conservative skew-symmetric-like schemes in structured meshes:
  application to compressible flows}.  \jt{J. Comput. Phys.}  \bvol{161}~(1),
  \pg{114--139}.

\bibitem[Ergin \& White(2006)]{Ergin2006}
{\sc \au{Ergin, F.G.} \& \au{White, E.B.}} \yr{2006}  \at{Unsteady and
  transitional flows behind roughness elements}.  \jt{AIAA J.}  \bvol{44}~(11),
   \pg{2504--2514}.

\bibitem[Fedorov(2011)]{Fedorov2011}
{\sc \au{Fedorov, A.}} \yr{2011}  \at{Transition and stability of high-speed
  boundary layers}.  \jt{Annu. Rev. Fluid Mech.}  \bvol{43},  \pg{79--95}.

\bibitem[Fedorov \& Tumin(2011)]{Fedorov2011b}
{\sc \au{Fedorov, A.} \& \au{Tumin, A.}} \yr{2011}  \at{High-speed
  boundary-layer instability: old terminology and a new framework}.  \jt{AIAA
  J.}  \bvol{49}~(8),  \pg{1647--1657}.

\bibitem[Fong {\em et~al.\/}(2015)Fong, Wang, Huang, Zhong, McKiernan, Fisher
  \& Schneider]{Fong2015}
{\sc \au{Fong, K.D.}, \au{Wang, X.}, \au{Huang, Y.}, \au{Zhong, X.},
  \au{McKiernan, G.R.}, \au{Fisher, R.A.} \& \au{Schneider, S.P.}} \yr{2015}
  \at{Second mode suppression in hypersonic boundary layer by roughness: Design
  and experiments}.  \jt{AIAA J.}  \bvol{53}~(10),  \pg{3138--3144}.

\bibitem[Fujii(2006)]{Fujii2006}
{\sc \au{Fujii, K.}} \yr{2006}  \at{Experiment of the two-dimensional roughness
  effect on hypersonic boundary-layer transition}.  \jt{J Spacecr Rockets}
  \bvol{43}~(4),  \pg{731--738}.

\bibitem[Greene {\em et~al.\/}(2016)Greene, Eldredge, Zhong \& Kim]{Greene2016}
{\sc \au{Greene, P.T.}, \au{Eldredge, J.D.}, \au{Zhong, X.} \& \au{Kim, J.}}
  \yr{2016}  \at{A high-order multi-zone cut-stencil method for numerical
  simulations of high-speed flows over complex geometries}.  \jt{J. Comput.
  Phys.}  \bvol{316},  \pg{652--681}.

\bibitem[Groskopf \& Kloker(2016)]{Groskopf2016}
{\sc \au{Groskopf, G.} \& \au{Kloker, M.J.}} \yr{2016}  \at{Instability and
  transition mechanisms induced by skewed roughness elements in a high-speed
  laminar boundary layer}.  \jt{J. Fluid Mech}  \bvol{805},  \pg{262--302}.

\bibitem[Haley \& Zhong(2023)]{haley2023roughness}
{\sc \au{Haley, Christopher} \& \au{Zhong, Xiaolin}} \yr{2023}  \at{Roughness
  effect on hypersonic second mode instability and transition on a cone}.
  \jt{Physics of Fluids}  \bvol{35}~(3),  \pg{034113}.

\bibitem[Harvey(1978)]{Harvey1978}
{\sc \au{Harvey, W.D.}} \yr{1978}  \at{Influence of free-stream disturbances on
  boundary-layer transition}.  \jt{NASA Technical Memorandum 78635} .

\bibitem[Holloway \& Sterrett(1964)]{Holloway1964}
{\sc \au{Holloway, P.F.} \& \au{Sterrett, J.R.}} \yr{1964}  \at{Effect of
  controlled surface roughness on boundary-layer transition and heat transfer
  at {Mach} numbers of 4.8 and 6.0}.  \jt{NASA Technical Note D-2054} .

\bibitem[Jahanbakhshi \& Zaki(2019)]{Jahanbakhshi2019}
{\sc \au{Jahanbakhshi, R.} \& \au{Zaki, T.A.}} \yr{2019}  \at{Nonlinearly most
  dangerous disturbance for high-speed boundary-layer transition}.  \jt{J.
  Fluid Mech}  \bvol{876},  \pg{87--121}.

\bibitem[Jahanbakhshi \& Zaki(2021)]{Jahanbakhshi2021}
{\sc \au{Jahanbakhshi, R.} \& \au{Zaki, T.A.}} \yr{2021}  \at{Optimal heat flux
  for delaying transition to turbulence in a high-speed boundary layer}.
  \jt{J. Fluid Mech}  \bvol{916},  \pg{A46}.

\bibitem[James(1959)]{James1959}
{\sc \au{James, C.S.}} \yr{1959}  \at{Boundary-layer transition on hollow
  cylinders in supersonic free flight as affected by {Mach} number and a
  screwthread type of surface roughness}.  \jt{NASA Technical Memorandum
  1-20-59A} .

\bibitem[Johnsen {\em et~al.\/}(2010)Johnsen, Larsson, Bhagatwala, Cabot, Moin,
  Olson, Rawat, Shankar, Sj{\"o}green, Yee {\em et~al.\/}]{Johnsen2010}
{\sc \au{Johnsen, E.}, \au{Larsson, J.}, \au{Bhagatwala, A.V.}, \au{Cabot,
  W.H.}, \au{Moin, P.}, \au{Olson, B.J.}, \au{Rawat, P.S.}, \au{Shankar, S.K.},
  \au{Sj{\"o}green, B.}, \au{Yee, H.C.} \& \au{others}} \yr{2010}
  \at{Assessment of high-resolution methods for numerical simulations of
  compressible turbulence with shock waves}.  \jt{J. Comput. Phys.}
  \bvol{229}~(4),  \pg{1213--1237}.

\bibitem[Kawai \& Larsson(2012)]{Kawai2012}
{\sc \au{Kawai, S.} \& \au{Larsson, J.}} \yr{2012}  \at{Wall-modeling in large
  eddy simulation: Length scales, grid resolution, and accuracy}.  \jt{Phys.
  Fluids}  \bvol{24}~(1),  \pg{015105}.

\bibitem[Kegerise \& Rufer(2016)]{Kegerise2016}
{\sc \au{Kegerise, M.A.} \& \au{Rufer, S.J.}} \yr{2016}  \at{Unsteady heat-flux
  measurements of second-mode instability waves in a hypersonic flat-plate
  boundary layer}.  \jt{Exp Fluids}  \bvol{57}~(8),  \pg{130}.

\bibitem[Kendall(1975)]{Kendall1975}
{\sc \au{Kendall, J.M.}} \yr{1975}  \at{Wind tunnel experiments relating to
  supersonic and hypersonic boundary-layer transition}.  \jt{AIAA J.}
  \bvol{13}~(3),  \pg{290--299}.

\bibitem[Larsson \& Lele(2009)]{Larsson2009}
{\sc \au{Larsson, J.} \& \au{Lele, S.K.}} \yr{2009}  \at{Direct numerical
  simulation of canonical shock/turbulence interaction}.  \jt{Phys. Fluids}
  \bvol{21}~(12),  \pg{126101}.

\bibitem[Laurence {\em et~al.\/}(2016)Laurence, Wagner \&
  Hannemann]{Laurence2016}
{\sc \au{Laurence, S.J.}, \au{Wagner, A.} \& \au{Hannemann, K.}} \yr{2016}
  \at{Experimental study of second-mode instability growth and breakdown in a
  hypersonic boundary layer using high-speed {Schlieren} visualization}.
  \jt{J. Fluid Mech}  \bvol{797},  \pg{471--503}.

\bibitem[Lees \& Lin(1946)]{Lees1946investigation}
{\sc \au{Lees, L.} \& \au{Lin, C.C.}} \yr{1946}  \bt{Investigation of the
  stability of the laminar boundary layer in a compressible fluid}. {\em Tech.
  Rep.\/}.  \org{National Advisory Committee for Aeronautics}.

\bibitem[Liepmann \& Roshko(2001)]{Liepmann2001elements}
{\sc \au{Liepmann, H.W.} \& \au{Roshko, A.}} \yr{2001} {\em Elements of
  gasdynamics\/}.  \publ{Dover Publication}.

\bibitem[Liu {\em et~al.\/}(2019)Liu, Yi, Xu, Shi, Ouyang \& Xiong]{Liu2019}
{\sc \au{Liu, X.}, \au{Yi, S.}, \au{Xu, X.}, \au{Shi, Y.}, \au{Ouyang, T.} \&
  \au{Xiong, H.}} \yr{2019}  \at{Experimental study of second-mode wave on a
  flared cone at {Mach} 6}.  \jt{Phys. Fluids}  \bvol{31}~(7),  \pg{074108}.

\bibitem[Liu {\em et~al.\/}(2020)Liu, Dong \& Wu]{Liu2020}
{\sc \au{Liu, Y.}, \au{Dong, M.} \& \au{Wu, X.}} \yr{2020}  \at{Generation of
  first {M}ack modes in supersonic boundary layers by slow acoustic waves
  interacting with streamwise isolated wall roughness}.  \jt{J. Fluid Mech}
  \bvol{888},  \pg{A10}.

\bibitem[Lysenko \& Maslov(1984)]{Lysenko1984}
{\sc \au{Lysenko, V.I.} \& \au{Maslov, A.A.}} \yr{1984}  \at{The effect of
  cooling on supersonic boundary-layer stability}.  \jt{J. Fluid Mech}
  \bvol{147},  \pg{39--52}.

\bibitem[Mack(1969)]{Mack1969}
{\sc \au{Mack, L.M.}} \yr{1969}  \bt{Boundary layer stability theory}. {\em
  Tech. Rep.\/}.  \org{California Institute of Technology, Jet Propulsion
  Laboratory}.

\bibitem[Mack(1984)]{Mack1984}
{\sc \au{Mack, L.M.}} \yr{1984}  \bt{Boundary-layer linear stability theory}.
  {\em Tech. Rep.\/}.  \org{California Institute of Technology, Jet Propulsion
  Laboratory}.

\bibitem[Marxen {\em et~al.\/}(2010)Marxen, Iaccarino \& Shaqfeh]{Marxen2010}
{\sc \au{Marxen, O.}, \au{Iaccarino, G.} \& \au{Shaqfeh, E.S.}} \yr{2010}
  \at{Disturbance evolution in a {M}ach 4.8 boundary layer with two-dimensional
  roughness-induced separation and shock}.  \jt{J. Fluid Mech}  \bvol{648},
  \pg{435--469}.

\bibitem[Mons {\em et~al.\/}(2021)Mons, Du \& Zaki]{Mons2021}
{\sc \au{Mons, V.}, \au{Du, Y.} \& \au{Zaki, T.A.}} \yr{2021}
  \at{Ensemble-variational assimilation of statistical data in large-eddy
  simulation}.  \jt{Phys. Rev. Fluids}  \bvol{6},  \pg{104607}.

\bibitem[Mons {\em et~al.\/}(2019)Mons, Wang \& Zaki]{Mons2019}
{\sc \au{Mons, V.}, \au{Wang, Q.} \& \au{Zaki, T.A.}} \yr{2019}
  \at{Kriging-enhanced ensemble variational data assimilation for scalar-source
  identification in turbulent environments}.  \jt{J. Comput. Phys.}
  \bvol{398},  \pg{108856}.

\bibitem[Morkovin(1987)]{Morkovin1987}
{\sc \au{Morkovin, M.V.}} \yr{1987}  \at{Transition at hypersonic speeds}.
  \jt{NASA Contractor Report No. NAS1-18107} .

\bibitem[Park \& Zaki(2019)]{Park2019}
{\sc \au{Park, J.} \& \au{Zaki, T.A.}} \yr{2019}  \at{Sensitivity of high-speed
  boundary-layer stability to base-flow distortion}.  \jt{J. Fluid Mech}
  \bvol{859},  \pg{476--515}.

\bibitem[Radeztsky~Jr {\em et~al.\/}(1999)Radeztsky~Jr, Reibert \&
  Saric]{Radeztsky1999}
{\sc \au{Radeztsky~Jr, R.H.}, \au{Reibert, M.S.} \& \au{Saric, W.S.}} \yr{1999}
   \at{Effect of isolated micron-sized roughness on transition in swept-wing
  flows}.  \jt{AIAA J.}  \bvol{37}~(11),  \pg{1370--1377}.

\bibitem[Riley {\em et~al.\/}(2014)Riley, McNamara \& Johnson]{Riley2014}
{\sc \au{Riley, Z.B.}, \au{McNamara, J.J.} \& \au{Johnson, H.B.}} \yr{2014}
  \at{Assessing hypersonic boundary-layer stability in the presence of
  structural deformation}.  \jt{AIAA J.}  \bvol{52}~(11),  \pg{2547--2558}.

\bibitem[Schneider(1999)]{Schneider1999}
{\sc \au{Schneider, S.P.}} \yr{1999}  \at{Flight data for boundary-layer
  transition at hypersonic and supersonic speeds}.  \jt{J Spacecr Rockets}
  \bvol{36}~(1),  \pg{8--20}.

\bibitem[Smith(1973)]{Smith1973laminar}
{\sc \au{Smith, F.T.}} \yr{1973}  \at{Laminar flow over a small hump on a flat
  plate}.  \jt{J. Fluid Mech}  \bvol{57}~(4),  \pg{803--824}.

\bibitem[Smith(1989)]{smith1989first}
{\sc \au{Smith, FT}} \yr{1989}  \at{On the first-mode instability in subsonic,
  supersonic or hypersonic boundary layers}.  \jt{Journal of Fluid Mechanics}
  \bvol{198},  \pg{127--153}.

\bibitem[Smith \& Brown(1990)]{smith1990inviscid}
{\sc \au{Smith, FT} \& \au{Brown, SN}} \yr{1990}  \at{The inviscid instability
  of a blasius boundary layer at large values of the {M}ach number}.
  \jt{Journal of Fluid Mechanics}  \bvol{219},  \pg{499--518}.

\bibitem[Stetson \& Kimmel(1992)]{Stetson1992}
{\sc \au{Stetson, K.} \& \au{Kimmel, R.}} \yr{1992} On hypersonic
  boundary-layer stability.  \bt{In {\em 30th aerospace sciences meeting and
  exhibit\/}},  \pg{p. 737}.  \publ{American Institute of Aeronautics and
  Astronautics}.

\bibitem[Stewartson(1969)]{Stewartson1969flow}
{\sc \au{Stewartson, K.}} \yr{1969}  \at{On the flow near the trailing edge of
  a flat plate ii}.  \jt{Mathematika}  \bvol{16}~(1),  \pg{106--121}.

\bibitem[Sutherland(1893)]{Sutherland1893}
{\sc \au{Sutherland, W.}} \yr{1893}  \at{Lii. the viscosity of gases and
  molecular force}.  \jt{Lond. Edinb. Dublin philos. mag.}  \bvol{36}~(223),
  \pg{507--531}.

\bibitem[Tumin(2007)]{Tumin2007}
{\sc \au{Tumin, A.}} \yr{2007}  \at{Three-dimensional spatial normal modes in
  compressible boundary layers}.  \jt{J. Fluid Mech}  \bvol{586},
  \pg{295--322}.

\bibitem[Volpiani {\em et~al.\/}(2018)Volpiani, Bernardini \&
  Larsson]{Volpiani2018}
{\sc \au{Volpiani, P.S.}, \au{Bernardini, M.} \& \au{Larsson, J.}} \yr{2018}
  \at{Effects of a nonadiabatic wall on supersonic shock/boundary-layer
  interactions}.  \jt{Phys. Rev. Fluids}  \bvol{3}~(8),  \pg{083401}.

\bibitem[Wu \& Dong(2016)]{Wu2016}
{\sc \au{Wu, Xuesong} \& \au{Dong, Ming}} \yr{2016}  \at{A local scattering
  theory for the effects of isolated roughness on boundary-layer instability
  and transition: transmission coefficient as an eigenvalue}.  \jt{J. Fluid
  Mech}  \bvol{794},  \pg{68--108}.

\bibitem[Xu {\em et~al.\/}(2016)Xu, Sherwin, Hall \& Wu]{Xu2016}
{\sc \au{Xu, H.}, \au{Sherwin, S.J.}, \au{Hall, P.} \& \au{Wu, X.}} \yr{2016}
  \at{The behaviour of tollmien--schlichting waves undergoing small-scale
  localised distortions}.  \jt{Journal of Fluid Mechanics}  \bvol{792},
  \pg{499--525}.

\bibitem[Zaki \& Wang(2021)]{Zaki2021}
{\sc \au{Zaki, T.A.} \& \au{Wang, M.}} \yr{2021}  \at{From limited observations
  to the state of turbulence: Fundamental difficulties of flow reconstruction}.
   \jt{Phys. Rev. Fluids}  \bvol{6},  \pg{100501}.

\bibitem[Zhao {\em et~al.\/}(2019)Zhao, Dong \& Yang]{Zhao2019}
{\sc \au{Zhao, L.}, \au{Dong, M.} \& \au{Yang, Y.}} \yr{2019}  \at{Harmonic
  linearized {Navier-Stokes} equation on describing the effect of surface
  roughness on hypersonic boundary-layer transition}.  \jt{Phys. Fluids}
  \bvol{31}~(3),  \pg{034108}.

\bibitem[Zhao {\em et~al.\/}(2018)Zhao, Wen, Tian, Long \& Yuan]{Zhao2018}
{\sc \au{Zhao, R.}, \au{Wen, C.Y.}, \au{Tian, X.D.}, \au{Long, T.H.} \&
  \au{Yuan, W.}} \yr{2018}  \at{Numerical simulation of local wall heating and
  cooling effect on the stability of a hypersonic boundary layer}.  \jt{Int. J.
  Heat Mass Transf.}  \bvol{121},  \pg{986--998}.

\bibitem[Zhu {\em et~al.\/}(2018)Zhu, Chen, Wu, Chen, Lee \& Gad-el
  Hak]{Zhu2018}
{\sc \au{Zhu, Y.}, \au{Chen, X.}, \au{Wu, J.}, \au{Chen, S.}, \au{Lee, C.} \&
  \au{Gad-el Hak, M.}} \yr{2018}  \at{Aerodynamic heating in transitional
  hypersonic boundary layers: Role of second-mode instability}.  \jt{Phys.
  Fluids}  \bvol{30}~(1),  \pg{011701}.

\end{thebibliography}

\end{document}